\documentclass[aps,prb,twocolumn,longbibliography]{revtex4-2}
\usepackage{graphicx}
\usepackage{color}
\usepackage[usenames,dvipsnames,svgnames,table]{xcolor}
\usepackage{natbib}
\usepackage{epstopdf}
\usepackage{epsfig}
\usepackage{amsmath}
\usepackage{amssymb}
\usepackage{slashed}
\usepackage[colorlinks=true]{hyperref}
\usepackage{comment}
\newcommand{\beq}{\begin{equation}}
\newcommand{\eeq}{\end{equation}}
\newcommand{\bal}{\begin{aligned}}
\newcommand{\eal}{\end{aligned}}
\usepackage{physics}

\usepackage{soul}

\begin{document}

\title{Effective action approach to the filling anomaly in crystalline topological matter}
\author{Pranav Rao}
\affiliation{Department of Physics and Institute for Condensed Matter Theory, University of Illinois at Urbana-Champaign, Urbana IL, 61801-3080, USA}

\author{Barry Bradlyn}
\email{bbradlyn@illinois.edu}
\affiliation{Department of Physics and Institute for Condensed Matter Theory, University of Illinois at Urbana-Champaign, Urbana IL, 61801-3080, USA}

\date{\today}

\begin{abstract}
In two dimensions, magnetic higher-order topological insulators (HOTIs) are characterized by excess boundary charge and a compensating bulk ``filling anomaly.'' At the same time, without additional noncrystalline symmetries, the boundaries of two-dimensional HOTIs are gapped and featureless at low energies, while the bulk of the system is predicted to have a topological response to the insertion of lattice (particularly disclination) defects. 
Until recently, a precise connection between these effects has remained elusive. 
In this work, we point the direction towards a unifying field-theoretic description for the bulk and boundary response of magnetic HOTIs. 
By focusing on the low-energy description of the gapped boundary of a two-dimensional magnetic HOTI with no time-reversing symmetries, we show that the boundary charge and filling anomaly arise from the gravitational ``Gromov-Jensen-Abanov'' (GJA) response action first introduced in [Phys. 
Rev. 
Lett. 116, 126802 (2016)] in the context of the quantum Hall effect. 
As in quantum Hall systems the GJA action cancels apparent anomalies associated with bulk response to disclinations, allowing us to derive a concrete connection between the bulk and boundary theories of HOTIs. 
We show how our results elucidate the connection between higher order topology and geometric response both in band insulators, and point towards a new route to understanding interacting higher order topological phases beyond the simple cases considered here.
\end{abstract}

\maketitle

\section{Introduction}

Traditionally, the most fruitful experimental and theoretical investigations into topological insulators have focused on the presence of gapless edge states~\cite{kane2005quantum,kane2005topological,bernevig2006quantum,konig2007quantum,hsieh2009observation,xu2012observation,xia2009observation,fu2007topological,fu2007topologicala,hsieh2012topological,teo2008surface}. 
However, recent work on topological insulators protected by crystal symmetry has shown that in many cases, topologically nontrivial systems need not have gapless states on surfaces that are not invariant under crystal symmetries~\cite{po2017symmetrybased,song2018quantitative,bradlyn2017topological,song2020realspace,slager2013space,kruthoff2017topological,cano2021band}. 
Instead, many topological crystalline insulators feature gapless corner (in two dimensions) or hinge states (in three dimensions) states at the point where boundary facets meet, at least in highly symmetric finite size geometries~\cite{benalcazar2017quantized,benalcazar2017electric,schindler2018higher,khalaf2018symmetry}. 
Such systems, known as higher-order topological insulators (HOTIs) have attracted great interest since their discovery. 
HOTI phases are now ubiquitous in condensed matter~\cite{schindler2018higherordera,zhou2015topological,li2019pressureinduced,shumiya2022evidence,noguchi2021evidence,wang2019higherorder,he2019symtopo,xu2020highthroughput,wieder2022topological,vergniory2019complete,vergniory2022all,elcoro2021magnetic,tang2019efficient,tang2019comprehensive,xu2019higherorder,xiao2018realization,liu2020robust,jo2020intrinsic,gao2021layer}, metamaterial~\cite{cerjan2020observation,grinberg2020robusta,peterson2020fractional}, and photonic~\cite{xie2018secondorder,ota2019photonic,elhassan2019corner,chen2019direct,zhang2020higherorder,xie2019visualization,wang2021higherorder,li2020higherorder,kim2020recent,proctor2020robustness} systems.

While many three-dimensional HOTIs are topologically nontrivial in the sense that the occupied states cannot support exponentially localized Wannier functions~\cite{schindler2018higher,Po2017,wieder2018axion}, the case of two-dimensional HOTIs is more subtle. 
Simple models of rotationally-symmetric HOTIs in two-dimensions feature fractionally charged midgap corner states and quantized bulk electric multipole moments. 
However, much like with electric polarization in one dimension, midgap corner states in two-dimensional HOTIs can only be pinned in energy by additional noncrystalline symmetries such as charge conjugation or chiral symmetry. 
In the absence of such symmetries, corner modes may increase or decrease in energy, becoming indistinguishable from the continuum of bulk states. 
Nevertheless, even when corner states do not appear within the bulk energy gap, there remains an excess charge on the boundary and a ``filling anomaly,'' where the number of filled electronic states in the bulk is mismatched from the number of unit cells~\cite{wieder2020strong,fang2021filling,benalcazar2019quantization,song2017ensuremath2,wieder2018axion,hwang2019fragile}. 
Thus, much like the one-dimensional SSH chain, two-dimensional HOTIs with only crystalline symmetries are obstructed atomic insulators (or, more generally, fragile topological phases~\cite{Po2017,wieder2018axion,wieder2020strong,bouhon2019wilson,cano2018topology,bradlyn2019disconnected}), which admit exponentially localized Wannier functions (after the appropriate addition of trivial occupied bands), albeit displaced from atomic positions.

At the same time, recent work has indicated that---in addition to quantized multipole moments---two-dimensional HOTIs can host nontrivial excitations on lattice defects in the bulk~\cite{liu2019shift,manjunath2021crystalline,li2020fractional,may2021crystalline,schindler2022topological,geier2021bulk,zhang2022fractional}. 
Particularly relevant for this work, it has been shown that for magnetic (time-reversal breaking) $C_{2n}$-symmetric HOTIs, disclination defects in the bulk with deficit angle $\pi/n$ bind fractional electric charge (fractional fermion number). 
In order to capture this phenomenon in a low-energy effective theory, field-theoretic descriptions of magnetic HOTIs have been proposed that, after integrating out fermionic degrees of freedom, involve couplings between the electromagnetic field and the geometry of the lattice. 
Similar field theories were first discussed by Wen and Zee~\cite{wen1992shift} in the context of the quantum Hall effect. 
Considering these Wen-Zee (WZ) responses to geometry has led to a number of consequences; as a notable example, the WZ response gives rise to the quantized Hall viscosity in rotationally-invariant quantum Hall phases\cite{read2011hall, bradlyn2014low,Gromov20141,Abanov2014,bradlyn2015gcs}. 
Some progress has been made viewing the excess corner and disclination charges that stem from higher-order topology as a (discrete) geometric response. 
For instance, it was recently shown that the charge bound to disclinations in 2D magnetic HOTIs could be captured by a Wen-Zee like response involving the coupling of the electromagnetic field to discrete rotational gauge fields~\cite{manjunath2021crystalline,may2021crystalline,herzog2022interacting,geier2021bulk,zhang2022fractional}. 

At first sight, the bulk Wen-Zee response does not capture the filling anomaly, which is present even in systems with vanishing curvature (and no disclinations). 
The close geometric analogy between corners and disclinations suggests that it may be possible to capture corner charge and filling anomaly in HOTIs via Wen-Zee like geometric response. 
Steps in this direction were taken in Ref.~\cite{may2021crystalline}. 
That work derived the bulk Wen-Zee action for a model of a fourfold ($C_4$) rotationally-symmetric magnetic HOTI, and argued how to connect this to the fractional charge bound to corners. 
However, Ref.~\cite{may2021crystalline} also showed that the bulk Wen-Zee action predicts an anomaly on the boundary of a magnetic HOTI that raises questions about how to describe the gapped boundary.

A similar puzzle was encountered in the study of the geometric response of quantum Hall states, where it was shown by Gromov, Jensen, and Abanov that the nonconservation of charge at the boundary of a system with a Wen-Zee response could be cancelled by a local counterterm~\cite{gromov2016boundary}. 
In that sense, the Wen-Zee term does not result in a boundary anomaly~\cite{callan1985anomalies,bardeen1984consistent} and does not require gapless edge modes; rather, it implies a nontrivial coupling between electromagnetism and \emph{extrinsic} curvature in the (possibly gapped) boundary theory. 

In this work, we show how the anomaly-cancelling Gromov-Jensen-Abanov (GJA) boundary action $S_\mathrm{GJA}$ arises in two-dimensional magnetic HOTIs, focusing on systems with no time-reversing symmetries (i.e. in type-I magnetic space groups~\cite{bradley1972mathematical}). 
We will show that $S_\mathrm{GJA}$ cancels the inflow of current from the bulk predicted in Ref.~\cite{may2021crystalline}, rendering the total bulk-plus-boundary theory of the magnetic HOTI anomaly-free. 
At the same time, $S_\mathrm{GJA}$ also necessitates an excess charge on the boundary of the magnetic HOTI even in flat space, due to the nontrivial Euler characteristic (extrinsic curvature) of a closed boundary. 
We will validate these phenomenological calculations by deriving $S_\mathrm{GJA}$ explicitly from the boundary theory in a microscopic model of a magnetic HOTI, following the approach of Refs.~\cite{hwang2019fragile,wieder2020strong}. 
We will emphasize the important role of discrete (in contrast to continuous) rotational symmetry in determining the matching conditions between the boundary filling anomaly and the bulk disclination response.

The structure of this paper is as follows: First, in Sec.~\ref{sec:WZreview} we will review the induced action formalism for geometric response in 2D magnetic systems. 
We will introduce the Wen-Zee action, show how it captures charge bound to disclination defects in the bulk, and review how it appears in the study of magnetic HOTIs. 
Then, we will introduce the Gromov-Jensen-Abanov boundary action, which cancels the apparent anomaly of the Wen-Zee theory and allows a system described by the Wen-Zee action in the bulk to be consistent with a gapped boundary. 
We will show how the Gromov-Jensen-Abanov term determines the filling anomaly and corner charge of a magnetic HOTI, thus closing the conceptual gap between bulk disclination response, corner charge, and the filling anomaly. 

To justify these observations, we will in Sec.~\ref{sec:microscopic} introduce a microscopic model of a $p4m$-symmetric magnetic HOTI adapted from Ref.~\cite{wieder2020strong}. 
We will derive the low-energy theory for the gapped boundary in terms of a two-component Dirac fermion with a $p4m$-symmetric but space-dependent mass term, paying particular attention to the role of the extrinsic geometry of the boundary. 
In Sec.~\ref{sec:charge} we will bosonize the theory and determine the boundary current density and excess charge. 
We will thus see how the filling anomaly of the HOTI arises due to the chiral anomaly in the boundary theory via the Goldstone-Wilczek mechanism. 
Since the boundary mass depends on the extrinsic geometry of the boundary, we will show how the current density can be expressed in terms of the extrinsic curvature. 
Building on this, in Sec.~\ref{sec:action} we will integrate our expression for the current to obtain the boundary Gromov-Jensen-Abanov induced action for the magnetic HOTI. 
We will then show how the Gromov-Jensen-Abanov boundary action necessitates the existence of a bulk Wen-Zee action, highlighting the ambiguities that arise due to the presence of discrete rotational symmetry. 
Finally, in Sec.~\ref{sec:discussion} we will discuss the implications of our results for more general $C_{2n}$-symmetric magnetic insulators and gappable boundaries of interacting topological phases. 
We will discuss how our result cements the Wen-Zee and Gromov-Jensen-Abanov actions as the hallmarks of obstructed atomic insulators in magnetic two-dimensional systems.   

\section{The Wen-Zee and Gromov-Jensen-Abanov Actions}\label{sec:WZreview}

We will begin by reviewing the induced action approach to geometric response in $2+1$-dimensional magnetic topological phases (i.e. topological phases with broken time-reversal symmetry). 
We imagine our system is defined on a spacetime manifold $\mathcal{M}=\Gamma\times\mathbb{R}$, where $\Gamma$ is a two-dimensional spatial manifold, and $\mathbb{R}$ represents time. 
The spatial manifold $\Gamma$ may be either a closed manifold, or a manifold with boundary $\partial\Gamma$; the boundary $\partial\Gamma$ separates the system of interest from a topologically trivial region which we can treat as the vacuum. 
We note that in this case the boundary of $\mathcal{M}$ is $\partial\mathcal{M}=\partial\Gamma\times\mathbb{R}$. 
We will be interested in gapped systems coupled to both an external electromagnetic field $A=A_\mu dx^\mu$ (where $\mu=0,1,2$ is a spacetime index, and $\mu=0$ is the time direction), and to deformations of the geometry of the spatial manifold $\Gamma$ (we will for simplicity neglect deformations that mix space and time in this work). 
The geometry of $\Gamma$ can be specified in terms of a set of orthonormal frames $e_i^a$, where $i=1,2$ indexes spatial directions in the manifold and $a=1,2$ specifies one of the two frames. 
We can also introduce the inverse frames $E^i_a$ satisfying
\begin{align}
 E^i_a e_i^b &= \delta_a^b, \\
 E^i_a e_j^a &= \delta_j^i,
\end{align} 
where $\delta$ is the Kronecker delta symbol. 
Here and throughout this work we use the Einstein summation convention for repeated indices. 
While a full accounting of this nonrelativistic geometry can be found in, e.g., Refs.~\cite{bradlyn2014low,Gromov20141,gromov2016boundary,gromov2015thermal,son2013newton}, we will for our purposes only highlight that the frames determine the metric
\begin{equation}
g_{ij} = e^a_i e^a_j 
\end{equation}
on the manifold $\Gamma$, and that the change in the frames between nearby points on the manifold is determined by the spin connection $\omega=\omega_\mu dx^\mu$. 
In terms of the frames and the metric, we have the explicit expression~\cite{bradlyn2014low,Gromov20141}
\begin{align}\label{eq:spinconnection}
\omega_0 &= \frac{1}{2}\epsilon^a_{\hphantom{a}b}E^i_a\partial_0 e_{i}^b, \nonumber  \\
\omega_k &= \frac{1}{2}\epsilon^a_{\hphantom{a}b}E^i_a\partial_i e_{i}^b + \frac{1}{2}\epsilon^{ab}E^i_a E^j_b\partial_i g_{jk}
\end{align}
for the components of the spin connection, where $\epsilon^{ab}$ is the antisymmetric symbol, and indices $a,b$ are lowered with the identity matrix. 

Given a gapped system of fermions coupled to the electromagnetic field and background geometry, we can integrate out the microscopic degrees of freedom to determine the induced action (generating functional) $S[A,e,\omega]$. 
Functional derivatives of $S[A,e,\omega]$ with respect to the background fields determine ground state averages and correlation functions (response coefficients). 
For a gapped system, we expect $S[A,e,\omega]$ to be a local functional of the fields,
\begin{align}
S[A,e,\omega] &\equiv S_{loc}[e,F,R] + S_{top}[A,e,\omega] \\
&= \int_{\mathcal{M}} \mathcal{L}_{loc}(e,F,R) + \mathcal{L}_{top},
\end{align} 
where the scalar function (or, more precisely, the three-form) $\mathcal{L}_{loc}(e,F,R)$ is a function of the frames, the electromagnetic field strength $F=dA$, and the curvature $R=d\omega$; by construction $\mathcal{L}_{loc}(e,F,R)$ is invariant under electromagnetic ($U(1)$) gauge transformations and local changes of frame [which are themselves $\mathrm{SO}(2)$ local rotations (gauge transformations)]. 
On the other hand, $\mathcal{L}_{top}$ is invariant under gauge transformations only up to boundary terms. 
The non-invariance of $\mathcal{L}_{top}$ under gauge transformations implies that coupling constants for terms in $\mathcal{L}_{top}$ cannot be locally renormalized in a space-dependent way by perturbations of the microscopic Hamiltonian that respect the symmetries of the problem.

The prototypical example of a contribution to $S_{top}$ is the electromagnetic Chern-Simons action
\begin{equation}
S_\mathrm{ECS} = \frac{\nu}{4\pi}\int_\mathcal{M} A \wedge dA.
\end{equation}
By varying $S_\mathrm{ECS}$ with respect to $A_\mu$, we find that the coefficient $\nu$ determines the Hall conductivity of the sample. 
At the same time, the fact that $S_\mathrm{ECS}$ changes by a boundary term under electromagnetic gauge transformations necessitates the existence of gapless boundary modes when the system has a boundary.

In this work, we will be focusing on magnetic topological phases with $\nu=0$, so that the Hall conductivity vanishes. 
In this case, the leading contribution to $S_{top}$ involves mixed coupling between electromagnetism and background geometry, first introduced by Wen and Zee in the context of the quantum Hall effect~\cite{wen1992shift}. 
This Wen-Zee action is a mixed Chern-Simons term coupling the electromagnetic vector potential $A$ and the spin connection $\omega$,
\begin{equation}
    \label{eq:wenzee}
    S_{\mathrm{WZ}}[A,\omega,\bar{s}] = \frac{\bar{s}}{2\pi} \int_{\mathcal{M}} A \wedge d\omega 
\end{equation}
We have made explicit the dependence of $S_{WZ}$ on the coupling constant $\bar{s}$, known as the orbital spin per particle (working in units where $\hbar=c=e=1$. 
Note also that $\bar{s}$ as defined here differs from the similar quantity introduced in the Hall effect by a factor of $\nu$). 

The Wen-Zee action determines the response of a topological phase to variations in background geometry. 
In particular, varying Eq.~\eqref{eq:wenzee} with respect to the frames using Eq.~\eqref{eq:spinconnection} determines the ground state stress tensor, and $\bar{s}/{2\pi}$ determines the Hall viscosity (per unit of magnetic flux through the system)~\cite{read2011hall,bradlyn2012kubo,hoyos2014hall,Abanov2014,Hoyos2012}.  
Additionally, we can vary Eq.~\eqref{eq:wenzee} with respect to the vector potential to determine the excess charge density
\begin{equation}\label{eq:disclinationcharge}
\Delta \rho_R = \frac{\delta S_{\mathrm{WZ}}}{\delta A_0} = \frac{\bar{s}}{2\pi} d\omega
\end{equation}
bound to sources of curvature $R=d\omega$. 
We see that the Wen-Zee action determines the charge bound to disclination defects $R=\alpha\delta(\mathbf{x}-\mathbf{x_0})$ of deficit angle $\alpha$ at position $\mathbf{x}_0$ in otherwise flat space. 
Furthermore, we can integrate Eq.~\eqref{eq:disclinationcharge} over space to find the shift in the total charge on a closed spatial manifold ($\partial\Gamma=0)$,
\begin{equation}\label{eq:bulkcharge}
\Delta Q_{R} = \frac{\bar{s}}{2\pi}\int_{\Gamma} R = \bar{s}\chi_{\Gamma},
\end{equation}
where $\chi_\Gamma$ is the Euler characteristic of the spatial manifold $\Gamma$.

In Ref.~\cite{may2021crystalline}, the authors showed that when coupled to nontrivial geometry the low-energy bulk behavior of a model for a 2D magnetic HOTI with $C_4$ rotation symmetry is governed by the Wen-Zee action Eq.~\eqref{eq:wenzee} with $\bar{s}=2$. 
This leads to the prediction that their $C_4$-symmetric HOTI will feature fractional $1/2$ charge bound to bulk $\pi/2$ disclination defects. 
However, in flat geometries where $\omega=0$, the Wen-Zee action does not lead to an accumulation of charge in the ground state, and so alone cannot explain the filling anomaly in rotationally-symmetric magnetic HOTIs.

The Wen-Zee action Eq.~\eqref{eq:wenzee} is invariant with respect to $\mathrm{SO}(2)$ gauge transformations of the spin connection (i.e., local rotations of the frames). 
However, under $U(1)$ (electromagnetic) gauge transformations of $A$, Eq.~\eqref{eq:wenzee} is only invariant up to a boundary term. 
In particular, when the spacetime $\mathcal{M}$ has a boundary $\partial\mathcal{M}=\partial\Gamma\times\mathbb{R}$, then under the $U(1)$ gauge transformation $A\rightarrow A + df$ we have
\begin{align}
S_{\mathrm{WZ}}&[A,\omega,\bar{s}]\rightarrow S_{\mathrm{WZ}}[A+df,\omega,\bar{s}] \\
&=S_{\mathrm{WZ}}[A,\omega,\bar{s}] + \int_\mathcal{M} df\wedge d\omega \\
&=S_{\mathrm{WZ}}[A,\omega,\bar{s}] + \frac{\bar{s}}{2\pi}\int_\mathcal{M}d^3x \epsilon^{\mu\nu\lambda}\partial_\mu(f\partial_\nu\omega_\lambda) \\
& = S_{\mathrm{WZ}}[A,\omega,\bar{s}] +\frac{\bar{s}}{2\pi} \int_{\partial\mathcal{M}}d^2x \epsilon^{\mu\nu\lambda}\hat{n}_\mu f\partial_\nu\omega_\lambda\label{eq:swzvarintermediate} \\
&=S_{\mathrm{WZ}}[A,\omega,\bar{s}]- \frac{\bar{s}}{2\pi}\int_{\partial\mathcal{M}}fd\omega\label{eq:swzvar},
\end{align} 
where $\hat{n}_i$ is an outward-directed unit one-form on the boundary, and $\epsilon^{\mu\nu\lambda}$ is the Levi-Civita symbol. 
A minus sign arises in going from Eq.~\eqref{eq:swzvarintermediate} to Eq.~\eqref{eq:swzvar} due to the spacetime orientation of $\partial\mathcal{M}$, following the conventions of Ref.~\cite{gromov2016boundary}.

Naively, the non-invariance Eq.~\eqref{eq:swzvar} seems to imply the existence of an anomaly in the current on the boundary of the system~\cite{callan1985anomalies}. 
This poses a conceptual problem for the application of Eq.~\eqref{eq:wenzee} to 2D HOTIs: since the boundary of a 2D HOTI is generically gapped, charge on the boundary must be conserved. 
This conundrum was first highlighted in Ref.~\cite{may2021crystalline} in the context of a fourfold symmetric magnetic HOTI.

To resolve this paradox, we note that---as first shown in Ref.~\cite{gromov2016boundary}---there exists a local action that can be added to the boundary theory to cancel the non-invariance Eq.~\eqref{eq:swzvar} of the Wen-Zee action. 
Introducing the extrinsic curvature $K$ of the boundary $\partial M$, We can write the Gromov-Jensen-Abanov action
\begin{equation}
	\label{eq:gja}
    S_{\mathrm{GJA}}[A,K,\bar{s}] = \frac{\bar{s}}{2\pi} \int_{\partial \mathcal{M}} A \wedge K.
\end{equation}
The extrinsic curvature $K$ differs from the spin connection projected to the boundary $\omega|_{\partial\mathcal{M}}$ by a closed form
\begin{equation}\label{eq:omegaKrel}
d\alpha = \omega|_{\partial \mathcal{M}} + K.
\end{equation}
Geometrically, the form $d\alpha$ corresponds to the derivative of the angle between the boundary normal and tangent vectors and the bulk frames evaluated on the boundary. 
Under a $U(1)$ gauge transformation, $S_{\mathrm{GJA}}$ transforms as
\begin{align}\label{eq:gjavar}
S_{\mathrm{GJA}}&[A,K,\bar{s}] \rightarrow S_{\mathrm{GJA}}[A+df,K,\bar{s}]\nonumber \\
&= S_{\mathrm{GJA}}[A,K,\bar{s}] + \frac{\bar{s}}{2\pi}\int_{\partial \mathcal{M}} df\wedge K \nonumber \\
&=S_{\mathrm{GJA}}[A,K,\bar{s}] + \frac{\bar{s}}{2\pi}\int_{\partial \mathcal{M}} df\wedge (d\alpha-\omega)\nonumber \\
&=S_{\mathrm{GJA}}[A,K,\bar{s}] +\frac{\bar{s}}{2\pi}\int_{\partial \mathcal{M}} fd\omega,
\end{align}
which exactly compensates the boundary variation Eq.~\eqref{eq:swzvar} of the Wen-Zee action. 
Thus, the total action $S_\mathrm{total} = S_{\mathrm{WZ}} +S_{\mathrm{GJA}} $ is gauge invariant:
\begin{equation}
    \label{eq:totalaction}
    S_\mathrm{total} = \frac{\bar{s}}{2\pi} \left(\int_\mathcal{M} A \wedge d\omega + \int_{\partial \mathcal{M}} A \wedge K\right)
\end{equation}
The GJA action gives rise viscous forces at the boundary $\partial \mathcal{M}$, which can be seen by considering the stress tensor associated to variations of $S_\mathrm{GJA}$ with respect to the boundary tangent and normal frames\cite{rao2021resolving,abanov2019free,ganeshan2017odd}. 
In the hydrodynamic context, this results in a modification of the boundary conditions for fluid flow. 
For our present purposes, however, we are more interested in the electromagnetic response captured by $S_\mathrm{GJA}$. 
By varying $S_\mathrm{GJA}$ with respect to the vector potential, we find that it quantifies a ground state average current density proportional to the extrinsic curvature,
\begin{equation}
j^{\bar{\mu}}_{FA} = \frac{\bar{s}}{2\pi} \epsilon^{\bar{\mu}\bar{\nu}}K_{\bar{\nu}},
\end{equation} 
where $\bar{\mu}=0,1$ indexes spacetime directions on $\partial\mathcal{M}$. 
Focusing on the $\bar{\mu}=0$ component and integrating over the spatial boundary $\partial\Gamma$ given by
\begin{equation}
\label{eq:extracharge}
\Delta Q_\mathrm{GJA}  = \int_{\partial\Gamma} \frac{\delta S_\mathrm{GJA}}{\delta A_0} =  \frac{\bar{s}}{2\pi} \int_{\partial\Gamma} K. 
\end{equation}

Eqs.~\eqref{eq:swzvar} and \eqref{eq:gjavar} show that $S_\mathrm{GJA}$ is necessary for a system with bulk Wen-Zee response to be consistent with a gapped boundary; hence, by the results of Refs.~\cite{may2021crystalline,liu2019shift,manjunath2021crystalline,geier2021bulk,zhang2022fractional,li2020fractional} we can deduce that a magnetic HOTI where disclinations bind fractional charge in the bulk \emph{must} have a boundary described by $S_\mathrm{GJA}$, when the boundary is gapped. 
This is one of the central results of this work, and in Secs.~\ref{sec:microscopic}--\ref{sec:action} we will derive $S_\mathrm{GJA}$ from a microscopic model of a magnetic HOTI boundary. 
Before moving on however, let us note that 
Eqs.~\eqref{eq:extracharge} and \eqref{eq:disclinationcharge} shows that there is a concrete connection between charge bound to disclinations in the bulk, and the excess charge at the boundary of the system. 
This allows us to understand the filling anomaly in the context of the induced action.  
Combining Eq.~\eqref{eq:extracharge} for the boundary charge due to the GJA action with Eq.~\eqref{eq:bulkcharge} for the excess charge due to the bulk Wen-Zee term, and using the Gauss-Bonnet theorem~\cite{eguchi1980gravitation} we find
\begin{align}\label{eq:extrachargedisk}
\Delta Q &= \frac{\bar{s}}{2\pi}\left(\int_\Gamma d\omega + \int_{\partial\Gamma} K\right) \nonumber\\
&=\frac{\bar{s}}{2\pi}\int_{\partial\Gamma} \omega|_{\partial\mathcal{M}} + K \nonumber\\
&= \frac{\bar{s}}{2\pi}\int_{\partial\Gamma} d\alpha \nonumber\\ 
&=\bar{s}\chi_{d},
\end{align} 
where $\chi_d$ is the generalized Euler characteristic for the manifold with boundary. 
In the absence of disclinations, $\chi_d=1$ for a system with a single closed boundary, and we recover the filling anomaly of an excess charge $\bar{s}$ above charge neutrality in a magnetic HOTI. 
For a flat disc, Eq.~\eqref{eq:extrachargedisk} shows that the excess charge is entirely localized on the boundary, consistent with known results on magnetic HOTIs in Refs.~\cite{benalcazar2019quantization,may2021crystalline,wieder2020strong}; in the presence of corners, $d\alpha=0$ everywhere except at the corners, meaning the excess charge will be localized on the corners modulo locally invariant modifications to the actions. 
Finally we see that Eq.~\eqref{eq:extrachargedisk} also incorporates the charge bound to bulk disclinations as computed from the Wen-Zee action Eq.~\eqref{eq:disclinationcharge}. 
Eq.~\eqref{eq:extrachargedisk} in particular shows that when a (uncharged) disclination of deficit angle $\Delta\alpha$ is dragged adiabatically from infinity, through the boundary of the system, and deep into the bulk of the system, it acquires a charge $\bar{s}\Delta\alpha$ in the bulk at the singularity of $d\omega$, and leaves behind an image charge of magnitude $-\bar{s}\Delta\alpha$ on the boundary due to the  deficit in $K$. 

So far, we have deduced the presence of $S_\mathrm{GJA}$ in magnetic HOTIs based on the conclusions of Ref.~\cite{may2021crystalline,liu2019shift,manjunath2021crystalline,geier2021bulk,zhang2022fractional,li2020fractional} on fractional disclination charge in magnetic HOTIs. 
Already we have seen how a careful study of $S_\mathrm{GJA}$ yields a simple view of the connection between bulk geometric response, boundary corner charge, and the filling anomaly. 
In what follows, we will introduce a specific model for a magnetic HOTI with $p4m$ symmetry, and will derive $S_\mathrm{GJA}$ from the microscopic theory for the gapped boundary. 
In doing so, we will see that there is an additional subtlety in the connection between $S_{WZ}$ and $S_\mathrm{GJA}$ in systems with only discrete rotational symmetry, related to our ability to add charge symmetrically to the boundary of the system.

\section{A Microscopic Model for the HOTI Boundary}\label{sec:microscopic}

We consider a microscopic model of a $p4m$-symmetric HOTI first introduced in Ref.~\cite{wieder2020strong}, and related to the model considered in Refs.~\cite{benalcazar2017quantized,may2021crystalline,hwang2019fragile}. 
To begin, we will review the derivation of the boundary theory for this model. 
We start with a 2D TI in layer group $p4/mmm1'$ with hybridized $p_z$ and $d_{x^2 - y^2}$ orbitals located at the origin of the unit cell. 
We consider the low-energy $k\cdot p$ description of this 2D TI expanded about the $\Gamma$ point in the Brillouin Zone (BZ), where the Bloch Hamiltonian is
\begin{equation}\label{eq:kphambulk}
H_\Gamma({\bf k}) = m \tau^z + v k_x \tau^x \sigma^y + v k_y \tau^x \sigma^x.
\end{equation}
Here the $\tau_i$ are Pauli matrices acting in the $\{p_z,d_{x^2-y^2}\}$ orbital space, and $\sigma_i$ are Pauli matrices acting on spin. 
The Hamiltonian Eq.~\eqref{eq:kphambulk} is invariant under fourfold rotation $C_4$, time-reversal $\mathcal{T}$, mirror $M_x$, and inversion $\mathcal{I}$ symmetry represented as
\begin{align}\label{eq:symmetry}
\mathcal{T}&: \sigma_y H^*(\mathbf{k})\sigma_y = H(-\mathbf{k}) \\
\mathcal{I}&: \tau_z H(\mathbf{k})\tau_z = H(-\mathbf{k}) \\
M_x&: \tau_y\sigma_z H(\mathbf{k})\tau_z\sigma_z = H(-k_x,k_y) \\
C_4&: \tau_ze^{-i\pi/4\sigma_z} H(\mathbf{k})\tau_ze^{i\pi/4\sigma_z} = H(-k_y,k_x) 
\end{align}
While the $\mathbf{k}\cdot\mathbf{p}$ Hamiltonian Eq.~\eqref{eq:kphambulk} has an enhanced emergent continuous rotational symmetry, we focus here only on those symmetries that are relevant for the magnetic HOTI phase. 
However, the low-energy rotational symmetry will allow us to more easily discuss coupling to curvature in Secs.~\ref{sec:charge} and \ref{sec:action}. 
The spectrum of $H$ is gapped for all $m\neq 0$; when $m>0$ filling the negative energy states gives a trivial insulator, while when $m<0$ filling the negative energy states gives a two-dimensional topological insulator.

In order to ultimately analyze edge modes of this model, we can consider a circular domain wall between trivial and topological insulating phases. 
To model this, we consider a spatially varying mass $m(r)$ in polar coordinates $r^2=x^2+y^2$. 
We let $m(r<R)\rightarrow -|M_0|$ be large and negative within a circle of radius $R>>a$, where $a$ is the lattice constant. 
Similarly, we take $m(r>R)\rightarrow +|M_0|$ for $r>R$. 
We take $m(r)$ to be independent of the polar angle $\theta=\arctan(y/x)$. 
In polar coordinates the Hamiltonian operator takes the form
\begin{equation}
\label{eq:edgecartesian}
H_\Gamma(r,\theta) = m(r) \tau^z - i v \tau^x \left[\sigma^1(\theta) \partial_r + \frac{1}{r} \sigma^2(\theta) \partial_\theta\right],
\end{equation}
where the $\theta$-dependent Pauli matrices are defined by
\begin{align}
\sigma^1(\theta) &= \sin\theta \sigma_x + \cos\theta\sigma_y\label{pauli1} \\
\sigma^2(\theta) &= -\sin\theta \sigma_y + \cos\theta\sigma_x. \label{pauli2}
\end{align}
For $r\approx R$, the mass $m(r)$ passes through zero, representing the domain wall between trivial and topological insulating phases. 
Choosing the ordered basis $\ket{\tau_z\sigma_z} = \left(\ket{\uparrow \uparrow},\ket{\uparrow\downarrow},
\ket{\downarrow\uparrow},\ket{\downarrow\downarrow}\right)$ for the four-state Hilbert space, we can introduce two Jackiw-Rebbi zero modes of the radial part of Eq.~\eqref{eq:edgecartesian},
\begin{align}\label{eq:zeromodes}
\ket{\phi_1}&=\frac{1}{\sqrt{2}}e^{-\frac{1}{v}\int_R^r m(r')dr'}\begin{pmatrix}
-e^{i\theta} \\ 0 \\ 0 \\ 1
\end{pmatrix} \\
\ket{\phi_2}&=\frac{1}{\sqrt{2}}e^{-\frac{1}{v}\int_R^r m(r')dr'}\begin{pmatrix}
0 \\ e^{-i\theta} \\ 1 \\ 0.
\end{pmatrix}
\end{align} 
We can obtain the edge Hamiltonian $H_\mathrm{TI}$ for the counterpropagating boundary modes of our system by projecting Eq.~\eqref{eq:edgecartesian} into the basis of zero modes $\ket{\phi_i}$ Introducing a set of Pauli matrices $s_i$ acting in the subspace of $\{\ket{\phi_1},\ket{\phi_2}\}$, we find
\begin{equation}\label{eq:TIedge}
H_\mathrm{TI} = \frac{v}{R}\left(\frac{1}{2} + is_z\partial_\theta\right).
\end{equation}

We now consider perturbations that break time-reversal and inversion symmetry in the bulk while preserving $C_4$ and $M_x$ symmetries. 
This causes a transition from a two-dimensional topological insulator to a magnetic HOTI. 
The most general bulk magnetic potential $U$ consistent with $C_4$ and $M_x$ symmetries and guaranteed to enlarge the bulk and edge gaps [i.e., anticommuting with the angular terms in Eq.~\eqref{eq:kphambulk}] takes the form 
\begin{align}\label{eq:bulkmagmass}
U_{\mathrm{bulk}} &= \sum_{n=0}^\infty \tau_x\sigma_z m^-_n\sin((4n+2)\theta) + \tau_y m^+_n\cos((4n+2)\theta) \nonumber \\
&+\sum_{n=0}^\infty \tau_x\sigma^1(\theta)m^1_n\cos(4n\theta) + \tau_y\sigma^2(\theta)m^2_n\sin(4n\theta)
\end{align}
We can obtain the microscopic boundary theory of the magnetic $p4m$-symmetric HOTI by projecting the magnetic mass Eq.~\eqref{eq:bulkmagmass} into the basis $\{\ket{\phi_1},\ket{\phi_2}\}$ of Jackiw-Rebbi boundary states of the two-dimensional topological insulator. 
Doing so, we find
\begin{widetext}
\begin{align}\label{eq:hotiedge}
H_\mathrm{edge} &= \frac{iv s^z}{R} \left(-\frac{i}{2} s^z + \partial_\theta\right) +\sum\limits_n \left[s^1(\theta)m^-_n\sin((4n+2)\theta)+ s^2(\theta)m^+_n\cos((4n+2)\theta)\right],
\end{align}
\end{widetext}
where we have introduced the rotated boundary Pauli matrices
\begin{align}
s^1(\theta)&=\cos\theta s_x -\sin\theta s_y \\
s^2(\theta)&=\sin\theta s_x + \cos\theta s_y.
\end{align}

Eq.~\eqref{eq:hotiedge} will be the starting point for our analysis of the HOTI boundary theory. 
Ref.~\cite{wieder2020strong} showed that when only a single mass term $m_n^\pm\neq 0$ was nonvanishing, the boundary has a chiral symmetry protecting the existence of exact zero-energy modes. 
In this limit, Eq.~\eqref{eq:hotiedge} hosts a set of $4n+2$ charge-$1/2$ zero modes localized at the zeros of the oscillating mass term. 
A perturbative analysis then showed how weakly breaking chiral symmetry causes all zero modes to gain the same energy, resulting in the filling anomaly: either all four zero modes are filled or all are empty, changing the ground state charge by $\Delta Q=\pm 2$. 
Here, using a field-theoretic treatment we will show how the filling anomaly arises from Eq.~\eqref{eq:hotiedge} generally, as a result of the coupling of the low-energy fermions to the extrinsic geometry of the HOTI boundary.

To see this, let us first recast Eq.~\eqref{eq:hotiedge} into a second-quantized form. 
First, note that the constant $v/{2R}$ term in the edge Hamiltonians Eq.~\eqref{eq:TIedge} and \eqref{eq:hotiedge} indicate the absence of exact zero energy modes of the boundary fermions when the mass $U$ is zero. 
This constant term arises because the local spin quantization axis for the fermions on the boundary is determined from the bulk Pauli matrices in polar coordinates, Eqs.~\eqref{pauli1} and \eqref{pauli2}. 
This forces the spin of the boundary modes in Eq.~\eqref{eq:zeromodes} to precess along the boundary as a function of $\theta$. 
More generally, this means that the Pauli matrices that appear in the boundary theory are rotated into a frame aligned with the boundary tangent and normal vectors.  
We note that this is an \emph{extrinsic} geometric effect, due to the embedding of the one-dimensional domain wall into two-dimensional space. 
For the same reason, the Pauli matrices in the boundary mass term in the second line of Eq.~\eqref{eq:hotiedge} are also dependent on the polar angle $\theta$. 

To account for these extrinsic effects, we can introduce a second-quantized antiperiodic fermion field
\begin{equation}\label{eq:fermions}
\psi(\theta) = \sum_n e^{in\theta}\begin{pmatrix}e^{i\theta/2}c_{n1} \\
e^{-i\theta/2}c_{n2}
\end{pmatrix}
\end{equation}
where $c_{ni}$ destroys an electron in the eigenstate $\ket{\psi_{ni}} = e^{in\theta}\ket{\phi_i}$ of Eq.~\eqref{eq:TIedge}. 
In terms of $\psi_{i}(\theta)$, we can second quantize the Hamiltonian Eq.~\eqref{eq:hotiedge} as
\begin{widetext}
\begin{align}
H_\mathrm{edge}\rightarrow\mathcal{H}_{edge} &=\int d\theta \psi^\dag\left[i\frac{v}{R}s_z\partial_\theta + \sum_n \left[s^x m^-_n\sin((4n+2)\theta)+ s^y m^+_n\cos((4n+2)\theta)\right]\right]\psi.\label{eq:2quantedge0}
\end{align}
\end{widetext}
Finally, we can introduce a coordinate $x$ along the boundary such that $\theta=\theta(x)$, a set of Dirac matrices
\begin{equation}\label{eq:gammas}
\gamma_0=s_y,\;\; \gamma_1=-is_x, \gamma_5=\gamma_0\gamma_1=s_z,
\end{equation}
and the Dirac adjoint $\bar\psi = \psi^\dag\gamma_0$.  
These allow us to rewrite Eq.~\eqref{eq:2quantedge0} in the suggestive form
\begin{equation}
\mathcal{H}_{edge} = \int dx \bar{\psi}\left[iv\gamma_1\partial_x + M_0(x) -i M_5(x)\gamma_5\right]\psi,\label{eq:edgediracgeneral}
\end{equation}
with
\begin{align}\label{eq:edgemassdefs}
M_0(x) &= \sum_n m^+_n\cos((4n+2)\theta(x)), \\
M_5(x) &= \sum_n m^-_n\sin((4n+2)\theta(x)),
\end{align}
where for a circular boundary $\theta(x)=x/R$.

We see that the boundary theory for a magnetic HOTI is given by a two-component Dirac fermion with a space-dependent mass $M_0(x)$ and a space-dependent chiral mass $M_5(x)$. 
The spatial profiles of $M_0(x)$ and $M_5(x)$ are dictated by the rotational and mirror symmetries of the bulk HOTI. 
In particular, we can examine the representation of $C_4$ and $M_x$ inherited from the bulk. 
Noting that $C_4$ takes $\theta\rightarrow \theta+\pi/2$ (and hence $x\rightarrow x+ \pi R/2$), we can use Eq.~\eqref{eq:symmetry} and the definition of the edge basis Eq.~\eqref{eq:zeromodes} to see that
\begin{equation}
C_4: \psi(x)\rightarrow -i\gamma_5\psi(x-\pi R/2).\label{eq:c4edge}
\end{equation}
Similarly, mirror symmetry $M_x$ maps $\theta\rightarrow \pi-\theta$ (and hence $x\rightarrow~\pi~R~-~x$), and so is represented by
\begin{equation}
M_x: \psi(x)\rightarrow i\gamma_2\psi(\pi R - x).\label{eq:mxedge}
\end{equation}
Note that due to the antiperiodic boundary conditions on $\psi(x)$, the representative Eq.~\eqref{eq:c4edge} of $C_4$ satisfies $(i\gamma_2)^4=+1$. 
In the edge theory, the fourfold rotation $C_4$ acts as translation by one quarter the circumference of the system; Eq.~\eqref{eq:c4edge} shows that the spinor basis of edge states transforms nontrivially (by $i\gamma_2$) under this translation. 
This means that even though the $\gamma$ matrices in Eq.~\eqref{eq:edgediracgeneral} appear independent of position, the mass term in the Hamiltonian must be consistent with rotational symmetry; hence $M_0(x)$ and $M_5(x)$ contain only angular harmonics $e^{im x/R}$ with $m=2 \mod 4$. 
In other words, the quantization axis for the boundary pseudospin $s_i$ matrices is determined from the extrinsic geometry of the embedding of the boundary in two-dimensional space, and consistent with the $C_4$ symmetry of the boundary. 
We can view the spatial dependence of the masses $M_0(x)$ and $M_5(x)$ as arising due to a competition between the two types of frames in the system: the boundary tangent and normal vector, and the bulk space-independent Cartesian frames evaluated on the boundary.

Finally, although we derived Eq.~\eqref{eq:edgediracgeneral} by assuming a rotationally-invariant bulk mass $m(r)$, the symmetry considerations of Eqs.~\eqref{eq:edgediracgeneral}--\eqref{eq:mxedge} allow us to generalize Eq.~\eqref{eq:edgediracgeneral} to other fourfold-symmetric boundaries. 
Provided the boundary is a single closed curve, then $x\in [0,2\pi R]$ gives a coordinate along the boundary, while the polar angle $\theta(x)$ can be written as
\begin{equation}\label{eq:alphadef}
\theta(x) = \alpha(x) + f(x),
\end{equation}
 where $\alpha(x)$ is the angle between the two-dimensional x-axis and the outward normal to the boundary, and $f(x)$ is a periodic function. 
 Enforcing $p4m$ symmetry on the boundary requires that
\begin{align}
\alpha(x+\pi R/2) &= \alpha(x) + \pi/2 \\
\alpha(\pi R - x) & = \pi - \alpha(x) \\
f(x+\pi R/2) &= f(x) \\
f(\pi R - x) & = - f(x)
\end{align}
With this parameterization, we note from Eq.~\eqref{eq:omegaKrel} that since we are working in a system where the bulk spin connection vanishes, the extrinsic curvature of the boundary is given by
\begin{equation}\label{eq:kdef}
K_x = \partial_x\alpha.
\end{equation}
Since the boundary $\partial\Gamma\times\mathbb{R}$ is embedded in flat spacetime and $\partial\Gamma$ forms a simple closed curve, the Gauss-Bonnet theorem implies that
\begin{equation}
\frac{1}{2\pi}\int dx K_x = 1 = \frac{1}{2\pi}\int dx \partial_x\theta
\end{equation}
For the remainder of this work, we will consider this general boundary. 
We will show that we can integrate out the gapped boundary fermions and derive the geometric effective action for a $p4m$-symmetric magnetic HOTI. 

\section{Anomalous Charge}\label{sec:charge}

Given the HOTI boundary Hamiltonian Eq.~\eqref{eq:edgediracgeneral}, we would like to integrate out the fermions to obtain the induced action (generating functional). 
To do so, we will first derive the expectation value of the current on the boundary as a function of extrinsic curvature. 
We can then (functionally) integrate the current to obtain the boundary geometric action. 
To begin, in this section, we derive the Goldstone-Wilczek expression for the filling anomaly current $j^{\bar{\mu}}_\mathrm{FA}$ by bosonizing the Hamiltonian Eq.~\eqref{eq:edgediracgeneral}. 

Following Refs.~\cite{fradkin2013field,von1998bosonization,kane1992transport,kane1992transmission}, we can introduce bosonic fields $\eta(x)$ and $\beta(x)$ to rewrite our second-quantized boundary fermions in Eq.~\eqref{eq:fermions} as
\begin{equation}
\begin{pmatrix}
\psi_1(x) \\
\psi_2(x)
\end{pmatrix}
= \frac{1}{\sqrt{2\pi R}}\begin{pmatrix}
F_1 :e^{i(\eta+\beta- x/2R)}: \\
F_2 :e^{i(\eta-\beta+x/2R)}: 
\end{pmatrix}
,\label{eq:fermiondict}
\end{equation}
where $:\cdot:$ denotes normal ordering with respect to the ground state of Eq.~\eqref{eq:TIedge} at charge neutrality (zero chemical potential) and $F_i$ are anticommuting Klein factors connecting states with different total fermion number. 
This is well-defined since Eq.~\eqref{eq:TIedge} has no zero modes. 
The bosons $\eta$ and $\beta$ are compact. 

The normal-ordered fermion density (i.e., the fermion density measured in excess of the charge neutral ground state) is given by
\begin{equation}
\rho(x) = :\bar{\psi}\gamma_0\psi: = \frac{1}{\pi}\partial_x \beta.\label{eq:densityformula}
\end{equation}
Since Eq.~\eqref{eq:TIedge} has linear dispersion, we can rewrite $H_\mathrm{TI}$ in terms of density fluctuations to find
\begin{equation}
H_\mathrm{TI}\equiv \int dx \frac{v}{2\pi} :(\partial_x\beta)^2:.
\end{equation}
Next, using Eq.~\eqref{eq:fermiondict} and the definition Eq.~\eqref{eq:gammas} for our $\gamma$ matrices, we find that
\begin{equation}\label{eq:m0boson}
\bar{\psi}\psi = i(\psi_2^\dag\psi_1-\psi_1^\dag\psi_2) = \frac{1}{\pi\mu}:\sin 2\beta:\end{equation}
and
\begin{equation}\label{eq:m5boson}
\bar{\psi}\gamma_5\psi = i(\psi_2^\dag\psi_1+\psi_1^\dag\psi_2) = \frac{i}{\pi\mu}:\cos 2\beta:,
\end{equation}
where $\mu$ is a short-distance cutoff scale. 
We thus have that $H_\mathrm{edge}$ in Eq.~\eqref{eq:edgediracgeneral} is equivalent to the bosonic Hamiltonian
\begin{equation}
H_\mathrm{edge}\rightarrow \int dx \frac{v}{2\pi}:(\partial_x\beta)^2 + \frac{M_0(x)}{\pi\mu}\sin2\beta + \frac{M_5(x)}{\pi\mu}\cos2\beta:
\end{equation}
Note that our choice of $\gamma$ matrices has resulted in the sine and cosine in Eqs.~\eqref{eq:m0boson} and \eqref{eq:m5boson} being reversed with respect to the bosonization formulas of Ref.~\cite{goldstone1981fractional}; this does not affect any observables of the theory, and merely reflects a choice of representation for the Klein factors~\cite{senechal2004introduction}.

In the limit of large $|M_0(x)|^2+|M_5(x)|^2$, the boson field $\beta$ will be pinned to the minimum of the classical potential
\begin{equation}
\pi\mu V(x,\beta)= M_0(x)\sin2\beta + M_5(x)\cos2\beta,
\end{equation}
which occurs when
\begin{equation}
\langle\beta(x)\rangle =\beta_{cl}(x)= -\frac{\pi}{4} - \frac{1}{2}\arctan\frac{M_5}{M_0}
\end{equation}
Using the bosonization relation Eq.~\eqref{eq:densityformula}, this implies that the edge of the HOTI has an excess charge density
\begin{equation}
\langle\rho(x)\rangle = \frac{1}{\pi}\partial_x\beta_{cl} = \frac{1}{2\pi}\partial_\mu \mathrm{Im}\log (M_0(x) - i M_5(x)).\label{eq:gwdensity}
\end{equation}
Similar results for the boundary charge density were first obtained in Ref.~\cite{hwang2019fragile}. 
Now, by Eq.~\eqref{eq:edgemassdefs}, we know that $M_0(x)$ and $M_5(x)$ consist only of Fourier harmonics with winding number $2$ modulo $4$. 
We can thus write
\begin{equation}
M_0(x) - i M_5(x) = e^{2i\theta}\sum_z z_n e^{4in\theta} \equiv e^{2i\theta}z(\theta),
\end{equation}
where we have reintroduced the polar angle $\theta(x)$. 
Since $z(\theta)=z(\theta+\pi/2)$ and $M_0(x)^2+M_5(x)^2$ is, by hypothesis, large, we can write
\begin{equation}
z(\theta)= r(\theta) e^{iq(\theta)}
\end{equation}
with
\begin{equation}
|r(\theta)|>0,\;\; q(\theta) = 4n^*\theta + g(\theta)
\end{equation}
with $g(\theta)$ a periodic function. 
Inserting this into Eq.~\eqref{eq:gwdensity} we find
\begin{align}
\langle \rho(x) \rangle &= \frac{1}{2\pi}(4n^*+2)\partial_x\theta + \frac{1}{2\pi}\partial_x g\label{eq:rho1} \\
& =  \frac{1}{2\pi}(4n^*+2)K_x + \frac{1}{2\pi}\partial_x f'.\label{eq:rho2}
\end{align}
In going from Eq.~\eqref{eq:rho1} to \eqref{eq:rho2} we used Eqs.~\eqref{eq:alphadef} and \eqref{eq:kdef} to introduce the periodic function $f'(x) = g(x) + (\theta(x)-\alpha(x))$. 
The two expressions Eq.~\eqref{eq:rho1} and \eqref{eq:rho2} for the ground state excess density in the HOTI phase are equivalent, and each make clear different features of the result. 
Integrating Eq.~\eqref{eq:rho1} over space and using the periodicity of $g$ immediately gives the filling anomaly
\begin{equation}
\Delta Q = \int dx \langle\rho(x)\rangle = 4n^*+2,
\end{equation}
which shows that for all choices of the bulk mass $U$ in Eq.~\eqref{eq:bulkmagmass}, the magnetic HOTI has excess boundary charge $\Delta Q \mod 4 = 2$. 
This represents a field-theoretic derivation of the filling anomaly. 

Our result for the filling anomaly $\Delta Q$ is consistent with Ref.~\cite{wieder2020strong} in the nearly-chiral-symmetric considered in that reference. 
The nearly-chiral-symmetric limit corresponds in our language to the case where $M_0$ and $M_5$ have a single nonvanishing Fourier component 
\begin{align}
M_0&\rightarrow m_0\cos((4n^*+2)\theta), \\
M_5&\rightarrow m_5\sin((4n^*+2)\theta),
\end{align}
with $m_0\gg m_5$ (or $m_5\gg m_0$, however we choose $m_0$ to be the larger mass for concreteness). 
Ref.~\cite{wieder2020strong} treated $m_5$ to first order in perturbation theory, and showed using a (nested) Jackiw-Rebbi analysis that there exist charge-$1/2$ modes localized at the $8n^*+4$ zeros of $\cos(4n^*+2)\theta)$, which are either all occupied (for $m_5<0$) or all unoccupied (for $m_5>0$). 
Thus, there is an excess charge of $|\Delta Q| = |4n^*+2|$ on the boundary relative to the charge neutral ground state with $m_0=m_5=0$. 
Our field-theoretic derivation generalizes this result to generic $p4m$-symmetric mass configurations $M_0(x), M_5(x)$, showing concretely how the filling anomaly results from the coupling of the boundary fermions to extrinsic geometry.

Our Goldstone-Wilczek analysis shows that the connection between chiral symmetry breaking and boundary curvature is a direct result of the chiral anomaly. 
To see this, let us first rewrite our expression Eq.~\eqref{eq:rho2} for the charge density in the reparameterization-covariant form
\begin{align}\label{eq:FAcurrent}
j_\mathrm{FA}^{\bar{\mu}} &= \frac{4n^*+2}{2\pi}\epsilon^{\bar{\mu}\bar{\nu}} K_{\bar{\nu}} + \frac{1}{2 \pi}\epsilon^{\bar{\mu}\bar{\nu}}\partial_{\bar{\nu}} f'.
\end{align}
Although we derived this result using bosonization, it can also be derived directly from the fermionic path integral corresponding to Eq.~\eqref{eq:edgediracgeneral}. 
Following Refs.~\cite{goldstone1981fractional,fujikawa2004path}, we can perform a space-dependent chiral transformation to eliminate the chiral mass $M_5$. 
Doing so results in an anomalous change in the path integral measure due to the chiral anomaly, yielding Eq.~\eqref{eq:FAcurrent} for the ground state average current. 
Thus, we see that breaking chiral symmetry in the bulk of the magnetic HOTI gaps out the corner modes on the boundary and leads to the emergence of the filling anomaly and geometric response via the boundary chiral anomaly.

\section{Implications for the Bulk Action}\label{sec:action}

Finally, we can connect the anomalous charge with the Gromov-Jensen-Abanov boundary action. 
Writing the filling anomaly current from Eq.~\eqref{eq:FAcurrent} as
\begin{equation}
j^{\bar{\mu}}_{FA} = \frac{\partial S[A,K]}{\partial A_{\bar{\mu}}},
\end{equation}
we can integrate with respect to $A_{\bar{\mu}}$ to obtain the generating functional
\begin{align}
S[A,K] &= \int_{\partial \mathcal{M}} \frac{4n^*+2}{2\pi} A \wedge K - \frac{1}{2\pi} f'(x) F \\
&= S_\mathrm{GJA}[A,K,(4n^*+2)] - \frac{1}{2\pi} \int_{\partial \mathcal{M}} f'(x) F,\label{eq:action}
\end{align}
where $F=dA$ is the electromagnetic field strength tensor. 
Eq.~\eqref{eq:action} allows us to identify two contributions to the boundary induced action. 
The first is the GJA boundary term proportional to $A\wedge K$, with orbital spin per particle $\bar{s}=(4n^*+2)$. 
This captures the filling anomaly $\Delta Q \mod 4 = 2$. 
The second term is the integral of a scalar function that is locally invariant under gauge and reparameterization transformations, and can thus arise from integrating out gapped fermions in any 1D system. 

The decomposition Eq.~\eqref{eq:action} of the action into $S_\mathrm{GJA}$ and locally-invariant terms is not unique, however. 
In particular, defining
\begin{equation}
q'(x) = (4n^*-4m^*)\theta - f',
\end{equation}
we can rewrite Eq.~\eqref{eq:action} as 
\begin{align}
S[A,K] &= \int_{\partial \mathcal{M}} \frac{4m^*+2}{2\pi} A \wedge K + \frac{1}{2\pi} q'(x) F \\
&= S_\mathrm{GJA}[A,K,4m^*+2] + S_{q'}
\end{align}
with 
\begin{equation}
S_{q'} =\frac{1}{2\pi} \int_{\partial \mathcal{M}} q'(x) F.
\end{equation}
$S_{q'}$ is naively a locally-invariant function, and is furthermore consistent with $C_4$ symmetry. 
For a compact boundary spacetime, we have that a $C_4$ transformation takes $q(x)\rightarrow q(x+\pi R) = q(x)+2\pi (n^*-m^*)$ and so 
\begin{align}
S_q &\rightarrow \frac{1}{2\pi}\int_{\partial \mathcal{M}} q(x)F + n^*\int_{\partial \mathcal{M}} F\\
& = S_q + 2\pi (n^*-m^*).
\end{align}
Thus, the propagator $e^{iS_{q'}}$ is invariant under $C_4$. 
Physically, $S_{q'}$ corresponds to our ability to add an integer number of electrons in each quarter of the 1D boundary without closing a bulk gap. 
For example, adding one filled single-particle orbital to each quarter of the boundary in a $p4m$ symmetric way will shift $q'(x)\rightarrow q'(x)+4\theta(x)$. 
Nevertheless, $C_4$ symmetry combined with Eq.~\eqref{eq:action} shows that the coefficient of $S_\mathrm{GJA}$ for the $p4m$-symmetric HOTI cannot be equal to zero, reflecting the filling anomaly.

Above, we have examined how the filling anomaly in higher-order topological insulators can be explained by notions of geometric response and specifically a proper consideration of the GJA boundary action $S_\mathrm{GJA}$. 
The existence of this term on the boundary 
\textit{necessitates} a bulk WZ response Eq.~\eqref{eq:wenzee}. 
In turn, there must be a universal disclination response in these systems as dictated by the bulk WZ response. 
To determine the coefficient of the bulk WZ term, we note that the presence of $S_{q'}$ presents an ambiguity in linking the bulk and boundary induced actions that is not present in systems with continuous rotational symmetry. 
Continuous rotational symmetry requires $\partial_x q=0$, and so all boundary charge is due to the GJA action. 
However,  $n^*$ (and hence $q(x)$) can change by an integer when a boundary gap closes, while the bulk gap remains open. 
This means that the boundary filling anomaly determines the coefficient of the bulk Wen-Zee term only modulo 4. 
Putting it all together, we deduce that in the most general case the bulk plus boundary action for a $p4m$-symmetric HOTI with gapped boundary can be written
\begin{widetext}
\begin{equation}
S[e,A,\omega]=\frac{(4n^*+2)}{2\pi}\int_{\mathcal{M}} A\wedge d\omega + \frac{4n^*+2}{2\pi}\int_{\partial \mathcal{M}} A\wedge K + \frac{1}{2\pi}\int_{\partial \mathcal{M}}(4m^*)\theta(x) F + \dots,  
\end{equation}
\end{widetext}
where the omitted terms on the boundary do not contribute to $\Delta Q$. 
The bulk spin per particle $\bar{s}_\mathrm{bulk}=4n^*+2$ determines the charge bound to disclinations. 
Gauge-non-invariance of the bulk Wen-Zee action requires the presence of the boundary GJA action with coefficient $\bar{s}=\bar{s}_\mathrm{bulk}=4n^*+2$. 
Finally, the boundary charge is given by $\Delta Q = \bar{s}_{bulk} + 4m^*$. 
Only $\Delta Q \mod 4$, and hence only $\bar{s}_\mathrm{bulk} \mod 4$, is a universal property of the bulk HOTI phase. 
This ambiguity corresponds to the fact that angular momentum $\bar{s}=4$ per particle cannot be distinguished by a fourfold rotation, which has eigenvalues $\lambda_{C_4} = e^{i\pi\bar{s}/2}$.

Our result also highlights the importance of chiral symmetry breaking to understanding the filling anomaly. 
The Gromov-Jensen-Abanov boundary term explicitly breaks chiral symmetry: Eq.~\eqref{eq:extracharge} shows that a system with $\bar{s}\neq 0$ cannot have a charge neutral (chiral-symmetric) ground state in the presence of a curved boundary. 
Similarly, the bulk Wen-Zee actions shows that a bulk system with $\bar{s}\neq 0$ cannot have a charge neutral ground state in the presence of disclination defects.

\section{Discussion and Future Directions}\label{sec:discussion}

In this work, we have considered the geometric response of higher-order topological insulators that exhibit a filling anomaly. 
By focusing on the low-energy theory for the (generically gapped) boundary of two-dimensional magnetic HOTIs, we have shown that the induced action contains the Gromov-Jensen-Abanov action Eq.~\eqref{eq:gja}. 
We demonstrated this both at the phenomenological level in Sec.~\ref{sec:WZreview}, and for a concrete microscopic model of a magnetic HOTI with $p4m$ symmetry in Secs.~\ref{sec:charge} and \ref{sec:action}. 
The variation of $S_\mathrm{GJA}$ with respect to the electromagnetic potential $A_0$ in flat space gives the excess charge density on the boundary of the HOTI in the ground state, which integrates to the filling anomaly. 
At the same time, $S_{GJA}$ couples electromagnetism to extrinsic curvature on the boundary, cancelling the anomalous inflow of current due to the bulk Wen-Zee term in the presence of disclinations in the bulk (or other smooth sources of curvature). 

Although we paid particular attention to $p4m$ (and hence $C_4$) symmetric magnetic HOTIs, our results generalize straightforwardly to magnetic HOTIs with $C_{2n}$ symmetry (note that the case of $C_3$-symmetry alone is rather subtle, as discussed in Ref.~\cite{van2020topological,kempkes2019robust,xue2019acoustic,ni2019observation}). 
In this case, as was argued in Refs.~\cite{proctor2020robustness,wieder2020strong,hwang2019fragile}, we can start with our same $p-d$ hybridized topological insulator from Eq.~\eqref{eq:kphambulk}, and add a $C_{2n}$-symmetric magnetic potential. 
Furthermore, we can consider models where the bulk tight-binding orbitals are localized at Wyckoff positions other than the origin of the unit cell, allowing for more general (nonzero angular momentum) representations of rotations in the subspace of low-energy modes. 
We will find that the boundary theory describes a two-component Dirac Fermion with spatially-dependent masses $M_0(\theta)$ and $M_5(\theta)$ as in Eq.~\eqref{eq:edgemassdefs}, although now the allowed Fourier harmonics in the masses will be of the form $e^{im\theta}$ with $m=2n\ell+k$, where $2n$ is the order of the $C_{2n}$ rotational symmetry, $\ell$ is an integer, and $k$ is an integer modulo $2n$ that depends on the occupied Wyckoff positions~\cite{fang2021filling,benalcazar2019quantization}. 
Carrying out the same Goldstone-Wilczek calculation as in Sec.~\ref{sec:charge}, we expect to find the total induced action
\begin{widetext}
\begin{equation}\label{eq:generalc2naction}
S_{2n}[e,A,\omega] = \frac{2nn^*+k}{2\pi}\left(\int_\mathcal{M} A\wedge d\omega + \int_{\partial\mathcal{M}} A\wedge K\right) + \frac{2nm^*}{2\pi}\int_{\partial\mathcal{M}}\theta(x) F +\dots
\end{equation}  
\end{widetext}
We see that in the $C_{2n}$-symmetric magnetic HOTI, the spin per particle $\bar{s}_{2n} = k \mod 2n$. 
From the Gromov-Jensen-Abanov term we recover the observation that for a system with boundary given by a $C_{2n}$ symmetric regular $2n$-gon, each corner will bind a fractional $k/{2n}$ charge (modulo $1$), although crucially the charge can be shifted away from the corners by locally invariant terms in the action. 

Additionally, we emphasize that in the $C_{2n}$-symmetric case the spin per particle $\bar{s}$---and hence the filling anomaly and disclination charge---are only invariant modulo $2n$. 
This reflects our ability to add $2n$ electrons to the system in a $C_{2n}$-symmetric fashion. 
In this way, discrete $2n$-fold rotational symmetry leads to a reduction in our ability to deduce the boundary excess charge (filling anomaly) directly from the bulk disclination charge. 
Locally invariant terms of the form
\begin{equation}\label{eq:bddlocterm}
\mathcal{L}_{loc} = 2nm^*\theta(x)F
\end{equation}
are allowed by $2n$-fold rotational symmetry and add $2n$ electrons to the boundary. 
We also saw in Sec.~\ref{sec:action} that we can interpret this ambiguity in terms of angular momentum eigenvalues: breaking continuous rotational symmetry down to $C_{2n}$ breaks the conservation of angular momentum: in a $C_{2n}$-symmetric system angular momentum is only conserved modulo $2n$. 
Thus, the spin per particle $\bar{s}$ is only robust modulo $2n$. 
By tuning the Fourier components of the bulk mass $U$ in Eq.~\eqref{eq:bulkmagmass} (and its appropriate $C_{2n}$-symmetric generalization), we can change $\bar{s}$ by multiples of $2n$ without closing a bulk gap. 
This leads to an analogous shift in the filling anomaly due to the Goldstone-Wilczek mechanism. 
At the same time, we can also perturb the edge theory independent of the bulk and change the filling anomaly by multiples of $2n$ without changing the bulk spin $\bar{s}$. 
Perturbations in the edge theory that change the filling anomaly by $2n$ require closing a gap in the edge theory, and correspond to shifting sets of $2n$ corner modes from the bulk conduction band manifold to the bulk valence band manifold; this changes the integer $m^*$ in the locally-invariant Lagrangian Eq.~\eqref{eq:bddlocterm}. 
This can be alternatively reinterpreted in terms of crystalline gauge fields, by viewing the role of the bulk magnetic mass as ``Higgsing'' the $\mathrm{SO}(2)$ gauge group of the spin connection down to $\mathbb{Z}_{2n}$. 
In this language the ambiguity of the filling anomaly and the disclination charge is connected to the fact that a $2\pi$ disclination is trivial.

Our work opens up several intriguing directions for future research. 
First, our microscopic edge theory approach to HOTIs can allow for an examination of how crystalline gauge fields emerge in effective descriptions of topological crystalline phases~\cite{huang2022effective,gioia2021unquantized,else2021quantum,else2021topological,dubinkin2021higher,manjunath2021crystalline,may2021crystalline,geier2021bulk,zhang2022fractional,han2019generalized}, by completing the analogy between rotational symmetry-breaking mass terms and the Higgs mechanism described in the previous paragraph. 
More generally, our work shows that focusing on the boundary theory offers a new window into \emph{bulk} geometric response. 
This indirect approach can be useful for geometric response beyond the Wen-Zee action Eq.~\eqref{eq:wenzee}, for instance in probing the second Wen-Zee action
\begin{equation}
S_{\mathrm{WZ},2} = \frac{\overline{s^2}}{4\pi}\int_\mathcal{M} \omega\wedge d\omega
\end{equation}
in microscopic models of HOTIs. 
As argued in Ref.~\cite{gromov2016boundary}, the second Wen-Zee action requires a local counterterm
\begin{equation}
S_{\mathrm{GJA},2} = \frac{\overline{s^2}}{4\pi}\int_{\partial\mathcal{M}} \omega\wedge K.\label{eq:gja2}
\end{equation}
The second Wen-Zee term has primarily been investigated only in quantum Hall systems, which feature gapless boundary modes. 
In these systems, it is difficult to distinguish the second Wen-Zee and gravitational Chern-Simons induced actions~\cite{stone2012gravitational,read2000paired,gromov2015thermal,cappelli2016multipole}. 
By applying the formalism we developed here to magnetic HOTIs, the second Wen-Zee response can be probed via Eq.~\eqref{eq:gja2} in the gapped boundary theory.

Our point of view also offers a new perspective on crystalline gauge fields\cite{manjunath2021crystalline,geier2021bulk,zhang2022fractional} and the geometric response of topological phases that has promise for broader studies. 
By probing the boundary physics, we are able to make statements about the bulk geometric response without needing to directly gauge a discrete spatial symmetry. 
This indirect approach can be useful for other geometric responses, for example considering the analog of the second WZ term, or even more nuanced translation responses. 
Furthermore, this approach could shed light on what ingredients are necessary to consistently treat the bulk response, in this instance by inferring a consistent way to define the discrete rotation gauge field from the boundary details, which we leave for future work.

Next, although we considered noninteracting magnetic HOTIs in this work, our Goldstone-Wilczek derivation in Sec.~\ref{sec:charge} can be applied equally well to interacting boundary theories, such as those that may arise on the boundary of non-chiral (Abelian) quantum Hall systems. 
There the boundary theory would consist of several counterpropagating bosonic modes. 
While many non-chiral quantum Hall edges are known to be gappable~\cite{levin2013protected}, it is possible that discrete rotational (or other point group) symmetries can place constraints on the space dependence of the interaction between modes. 
This would lead to the emergence of a fractionalized filling anomaly and fractional HOTI phases, via a generalization of our arguments in Sec.~\ref{sec:charge}. 

Additionally, although we focused in this work on 2D HOTIs with no time-reversing symmetries (i.e. with type-I magnetic space group symmetries~\cite{bradley1972mathematical}), our approach based on geometric response can be extended to systems with more complicated symmetries.
First, our work can also be extended to (interacting and noninteracting) time-reversal invariant HOTI phases in 2D (i.e., type-II magnetic space group symmetries), where we expect to find spin-resolved responses~\cite{lin2022spin,lange2022projected,hwang2022magnetic} and charge-neutral defect states~\cite{wieder2018axion,schindler2022topological}. Such an extension would be particularly relevant to experimental searches for non-magnetic obstructed atomic limit phases~\cite{xu2021filling,xu2021three,nie2021application}. 
Second, it would be interesting to apply our approach to 2D HOTIs with type-III and IV magnetic symmetries, where a richer interplay between boundary geometry and symmetry could be expected.
Third, our work can be extended to three-dimensional systems with ``$R\wedge F$'' responses~\cite{may2022topological}, and topological semimetals with unquantized anomalies~\cite{gioia2021unquantized}. 

Finally, we conclude by emphasizing that our work further supports the point made recently in the literature that the Wen-Zee action is a ``geometric'', rather than topological action. 
As we have seen here, a nonvanishing $\bar{s}$ is a signature of an obstructed atomic limit (or fragile topological) phase. 
This is consistent with recent work on disclination charge in noninteracting~\cite{may2021crystalline,benalcazar2019quantization} and interacting~\cite{herzog2022interacting} fragile topological and obstructed atomic limit phases.

\begin{acknowledgments}
The authors thank O. 
Dubinkin, J. 
May-Mann, G. Palumbo, and B. Wieder for helpful discussions. 
This work was supported by the National Science Foundation under grant DMR-1945058.
\end{acknowledgments}

\bibliography{refs}

\begin{thebibliography}{119}%
\makeatletter
\providecommand \@ifxundefined [1]{%
 \@ifx{#1\undefined}
}%
\providecommand \@ifnum [1]{%
 \ifnum #1\expandafter \@firstoftwo
 \else \expandafter \@secondoftwo
 \fi
}%
\providecommand \@ifx [1]{%
 \ifx #1\expandafter \@firstoftwo
 \else \expandafter \@secondoftwo
 \fi
}%
\providecommand \natexlab [1]{#1}%
\providecommand \enquote  [1]{``#1''}%
\providecommand \bibnamefont  [1]{#1}%
\providecommand \bibfnamefont [1]{#1}%
\providecommand \citenamefont [1]{#1}%
\providecommand \href@noop [0]{\@secondoftwo}%
\providecommand \href [0]{\begingroup \@sanitize@url \@href}%
\providecommand \@href[1]{\@@startlink{#1}\@@href}%
\providecommand \@@href[1]{\endgroup#1\@@endlink}%
\providecommand \@sanitize@url [0]{\catcode `\\12\catcode `\$12\catcode
  `\&12\catcode `\#12\catcode `\^12\catcode `\_12\catcode `\%12\relax}%
\providecommand \@@startlink[1]{}%
\providecommand \@@endlink[0]{}%
\providecommand \url  [0]{\begingroup\@sanitize@url \@url }%
\providecommand \@url [1]{\endgroup\@href {#1}{\urlprefix }}%
\providecommand \urlprefix  [0]{URL }%
\providecommand \Eprint [0]{\href }%
\providecommand \doibase [0]{https://doi.org/}%
\providecommand \selectlanguage [0]{\@gobble}%
\providecommand \bibinfo  [0]{\@secondoftwo}%
\providecommand \bibfield  [0]{\@secondoftwo}%
\providecommand \translation [1]{[#1]}%
\providecommand \BibitemOpen [0]{}%
\providecommand \bibitemStop [0]{}%
\providecommand \bibitemNoStop [0]{.\EOS\space}%
\providecommand \EOS [0]{\spacefactor3000\relax}%
\providecommand \BibitemShut  [1]{\csname bibitem#1\endcsname}%
\let\auto@bib@innerbib\@empty
\bibitem [{\citenamefont {Kane}\ and\ \citenamefont
  {Mele}(2005{\natexlab{a}})}]{kane2005quantum}%
  \BibitemOpen
  \bibfield  {author} {\bibinfo {author} {\bibfnamefont {C.~L.}\ \bibnamefont
  {Kane}}\ and\ \bibinfo {author} {\bibfnamefont {E.~J.}\ \bibnamefont
  {Mele}},\ }\bibfield  {title} {\bibinfo {title} {Quantum spin hall effect in
  graphene},\ }\href {https://doi.org/10.1103/PhysRevLett.95.226801} {\bibfield
   {journal} {\bibinfo  {journal} {Physical Review Letters}\ }\textbf {\bibinfo
  {volume} {95}},\ \bibinfo {pages} {226801} (\bibinfo {year}
  {2005}{\natexlab{a}})}\BibitemShut {NoStop}%
\bibitem [{\citenamefont {Kane}\ and\ \citenamefont
  {Mele}(2005{\natexlab{b}})}]{kane2005topological}%
  \BibitemOpen
  \bibfield  {author} {\bibinfo {author} {\bibfnamefont {C.~L.}\ \bibnamefont
  {Kane}}\ and\ \bibinfo {author} {\bibfnamefont {E.~J.}\ \bibnamefont
  {Mele}},\ }\bibfield  {title} {\bibinfo {title} {{{Z}}{$_2$} topological
  order and the quantum spin hall effect},\ }\href
  {https://doi.org/10.1103/PhysRevLett.95.146802} {\bibfield  {journal}
  {\bibinfo  {journal} {Physical Review Letters}\ }\textbf {\bibinfo {volume}
  {95}},\ \bibinfo {pages} {146802} (\bibinfo {year}
  {2005}{\natexlab{b}})}\BibitemShut {NoStop}%
\bibitem [{\citenamefont {Bernevig}\ \emph {et~al.}(2006)\citenamefont
  {Bernevig}, \citenamefont {Hughes},\ and\ \citenamefont
  {Zhang}}]{bernevig2006quantum}%
  \BibitemOpen
  \bibfield  {author} {\bibinfo {author} {\bibfnamefont {B.~A.}\ \bibnamefont
  {Bernevig}}, \bibinfo {author} {\bibfnamefont {T.~L.}\ \bibnamefont
  {Hughes}},\ and\ \bibinfo {author} {\bibfnamefont {S.-C.}\ \bibnamefont
  {Zhang}},\ }\bibfield  {title} {\bibinfo {title} {Quantum spin hall effect
  and topological phase transition in {{HgTe}} quantum wells},\ }\href
  {https://doi.org/10.1126/science.1133734} {\bibfield  {journal} {\bibinfo
  {journal} {Science (New York, N.Y.)}\ }\textbf {\bibinfo {volume} {314}},\
  \bibinfo {pages} {1757} (\bibinfo {year} {2006})}\BibitemShut {NoStop}%
\bibitem [{\citenamefont {K{\"o}nig}\ \emph {et~al.}(2007)\citenamefont
  {K{\"o}nig}, \citenamefont {Wiedmann}, \citenamefont {Br{\"u}ne},
  \citenamefont {Roth}, \citenamefont {Buhmann}, \citenamefont {Molenkamp},
  \citenamefont {Qi},\ and\ \citenamefont {Zhang}}]{konig2007quantum}%
  \BibitemOpen
  \bibfield  {author} {\bibinfo {author} {\bibfnamefont {M.}~\bibnamefont
  {K{\"o}nig}}, \bibinfo {author} {\bibfnamefont {S.}~\bibnamefont {Wiedmann}},
  \bibinfo {author} {\bibfnamefont {C.}~\bibnamefont {Br{\"u}ne}}, \bibinfo
  {author} {\bibfnamefont {A.}~\bibnamefont {Roth}}, \bibinfo {author}
  {\bibfnamefont {H.}~\bibnamefont {Buhmann}}, \bibinfo {author} {\bibfnamefont
  {L.~W.}\ \bibnamefont {Molenkamp}}, \bibinfo {author} {\bibfnamefont {X.-L.}\
  \bibnamefont {Qi}},\ and\ \bibinfo {author} {\bibfnamefont {S.-C.}\
  \bibnamefont {Zhang}},\ }\bibfield  {title} {\bibinfo {title} {Quantum spin
  hall insulator state in {{HgTe}} quantum wells},\ }\href
  {https://doi.org/10.1126/science.1148047} {\bibfield  {journal} {\bibinfo
  {journal} {Science (New York, N.Y.)}\ }\textbf {\bibinfo {volume} {318}},\
  \bibinfo {pages} {766} (\bibinfo {year} {2007})}\BibitemShut {NoStop}%
\bibitem [{\citenamefont {Hsieh}\ \emph {et~al.}(2009)\citenamefont {Hsieh},
  \citenamefont {Xia}, \citenamefont {Wray}, \citenamefont {Qian},
  \citenamefont {Pal}, \citenamefont {Dil}, \citenamefont {Osterwalder},
  \citenamefont {Meier}, \citenamefont {Bihlmayer}, \citenamefont {Kane},
  \citenamefont {Hor}, \citenamefont {Cava},\ and\ \citenamefont
  {Hasan}}]{hsieh2009observation}%
  \BibitemOpen
  \bibfield  {author} {\bibinfo {author} {\bibfnamefont {D.}~\bibnamefont
  {Hsieh}}, \bibinfo {author} {\bibfnamefont {Y.}~\bibnamefont {Xia}}, \bibinfo
  {author} {\bibfnamefont {L.}~\bibnamefont {Wray}}, \bibinfo {author}
  {\bibfnamefont {D.}~\bibnamefont {Qian}}, \bibinfo {author} {\bibfnamefont
  {A.}~\bibnamefont {Pal}}, \bibinfo {author} {\bibfnamefont {J.~H.}\
  \bibnamefont {Dil}}, \bibinfo {author} {\bibfnamefont {J.}~\bibnamefont
  {Osterwalder}}, \bibinfo {author} {\bibfnamefont {F.}~\bibnamefont {Meier}},
  \bibinfo {author} {\bibfnamefont {G.}~\bibnamefont {Bihlmayer}}, \bibinfo
  {author} {\bibfnamefont {C.~L.}\ \bibnamefont {Kane}}, \bibinfo {author}
  {\bibfnamefont {Y.~S.}\ \bibnamefont {Hor}}, \bibinfo {author} {\bibfnamefont
  {R.~J.}\ \bibnamefont {Cava}},\ and\ \bibinfo {author} {\bibfnamefont
  {M.~Z.}\ \bibnamefont {Hasan}},\ }\bibfield  {title} {\bibinfo {title}
  {Observation of unconventional quantum spin textures in topological
  insulators},\ }\href {https://doi.org/10.1126/science.1167733} {\bibfield
  {journal} {\bibinfo  {journal} {Science (New York, N.Y.)}\ }\textbf {\bibinfo
  {volume} {323}},\ \bibinfo {pages} {919} (\bibinfo {year}
  {2009})}\BibitemShut {NoStop}%
\bibitem [{\citenamefont {Xu}\ \emph {et~al.}(2012)\citenamefont {Xu},
  \citenamefont {Liu}, \citenamefont {Alidoust}, \citenamefont {Neupane},
  \citenamefont {Qian}, \citenamefont {Belopolski}, \citenamefont {Denlinger},
  \citenamefont {Wang}, \citenamefont {Lin}, \citenamefont {Wray},
  \citenamefont {Landolt}, \citenamefont {Slomski}, \citenamefont {Dil},
  \citenamefont {Marcinkova}, \citenamefont {Morosan}, \citenamefont {Gibson},
  \citenamefont {Sankar}, \citenamefont {Chou}, \citenamefont {Cava},
  \citenamefont {Bansil},\ and\ \citenamefont {Hasan}}]{xu2012observation}%
  \BibitemOpen
  \bibfield  {author} {\bibinfo {author} {\bibfnamefont {S.-Y.}\ \bibnamefont
  {Xu}}, \bibinfo {author} {\bibfnamefont {C.}~\bibnamefont {Liu}}, \bibinfo
  {author} {\bibfnamefont {N.}~\bibnamefont {Alidoust}}, \bibinfo {author}
  {\bibfnamefont {M.}~\bibnamefont {Neupane}}, \bibinfo {author} {\bibfnamefont
  {D.}~\bibnamefont {Qian}}, \bibinfo {author} {\bibfnamefont {I.}~\bibnamefont
  {Belopolski}}, \bibinfo {author} {\bibfnamefont {J.~D.}\ \bibnamefont
  {Denlinger}}, \bibinfo {author} {\bibfnamefont {Y.~J.}\ \bibnamefont {Wang}},
  \bibinfo {author} {\bibfnamefont {H.}~\bibnamefont {Lin}}, \bibinfo {author}
  {\bibfnamefont {L.~A.}\ \bibnamefont {Wray}}, \bibinfo {author}
  {\bibfnamefont {G.}~\bibnamefont {Landolt}}, \bibinfo {author} {\bibfnamefont
  {B.}~\bibnamefont {Slomski}}, \bibinfo {author} {\bibfnamefont {J.~H.}\
  \bibnamefont {Dil}}, \bibinfo {author} {\bibfnamefont {A.}~\bibnamefont
  {Marcinkova}}, \bibinfo {author} {\bibfnamefont {E.}~\bibnamefont {Morosan}},
  \bibinfo {author} {\bibfnamefont {Q.}~\bibnamefont {Gibson}}, \bibinfo
  {author} {\bibfnamefont {R.}~\bibnamefont {Sankar}}, \bibinfo {author}
  {\bibfnamefont {F.~C.}\ \bibnamefont {Chou}}, \bibinfo {author}
  {\bibfnamefont {R.~J.}\ \bibnamefont {Cava}}, \bibinfo {author}
  {\bibfnamefont {A.}~\bibnamefont {Bansil}},\ and\ \bibinfo {author}
  {\bibfnamefont {M.~Z.}\ \bibnamefont {Hasan}},\ }\bibfield  {title} {\bibinfo
  {title} {Observation of a topological crystalline insulator phase and
  topological phase transition in {{Pb1-xSnxTe}}},\ }\href@noop {} {\bibfield
  {journal} {\bibinfo  {journal} {Nature communications}\ }\textbf {\bibinfo
  {volume} {3}},\ \bibinfo {pages} {1192} (\bibinfo {year} {2012})}\BibitemShut
  {NoStop}%
\bibitem [{\citenamefont {Xia}\ \emph {et~al.}(2009)\citenamefont {Xia},
  \citenamefont {Qian}, \citenamefont {Hsieh}, \citenamefont {Wray},
  \citenamefont {Pal}, \citenamefont {Lin}, \citenamefont {Bansil},
  \citenamefont {Grauer}, \citenamefont {Hor}, \citenamefont {Cava},\ and\
  \citenamefont {Hasan}}]{xia2009observation}%
  \BibitemOpen
  \bibfield  {author} {\bibinfo {author} {\bibfnamefont {Y.}~\bibnamefont
  {Xia}}, \bibinfo {author} {\bibfnamefont {D.}~\bibnamefont {Qian}}, \bibinfo
  {author} {\bibfnamefont {D.}~\bibnamefont {Hsieh}}, \bibinfo {author}
  {\bibfnamefont {L.}~\bibnamefont {Wray}}, \bibinfo {author} {\bibfnamefont
  {A.}~\bibnamefont {Pal}}, \bibinfo {author} {\bibfnamefont {H.}~\bibnamefont
  {Lin}}, \bibinfo {author} {\bibfnamefont {A.}~\bibnamefont {Bansil}},
  \bibinfo {author} {\bibfnamefont {D.}~\bibnamefont {Grauer}}, \bibinfo
  {author} {\bibfnamefont {Y.~S.}\ \bibnamefont {Hor}}, \bibinfo {author}
  {\bibfnamefont {R.~J.}\ \bibnamefont {Cava}},\ and\ \bibinfo {author}
  {\bibfnamefont {M.~Z.}\ \bibnamefont {Hasan}},\ }\bibfield  {title} {\bibinfo
  {title} {Observation of a large-gap topological-insulator class with a single
  {{Dirac}} cone on the surface},\ }\href {https://doi.org/10.1038/nphys1274}
  {\bibfield  {journal} {\bibinfo  {journal} {Nature physics}\ }\textbf
  {\bibinfo {volume} {5}},\ \bibinfo {pages} {398} (\bibinfo {year}
  {2009})}\BibitemShut {NoStop}%
\bibitem [{\citenamefont {Fu}\ and\ \citenamefont
  {Kane}(2007)}]{fu2007topological}%
  \BibitemOpen
  \bibfield  {author} {\bibinfo {author} {\bibfnamefont {L.}~\bibnamefont
  {Fu}}\ and\ \bibinfo {author} {\bibfnamefont {C.~L.}\ \bibnamefont {Kane}},\
  }\bibfield  {title} {\bibinfo {title} {Topological insulators with inversion
  symmetry},\ }\href {https://doi.org/10.1103/PhysRevB.76.045302} {\bibfield
  {journal} {\bibinfo  {journal} {Physical Review B}\ }\textbf {\bibinfo
  {volume} {76}},\ \bibinfo {pages} {045302} (\bibinfo {year}
  {2007})}\BibitemShut {NoStop}%
\bibitem [{\citenamefont {Fu}\ \emph {et~al.}(2007)\citenamefont {Fu},
  \citenamefont {Kane},\ and\ \citenamefont {Mele}}]{fu2007topologicala}%
  \BibitemOpen
  \bibfield  {author} {\bibinfo {author} {\bibfnamefont {L.}~\bibnamefont
  {Fu}}, \bibinfo {author} {\bibfnamefont {C.~L.}\ \bibnamefont {Kane}},\ and\
  \bibinfo {author} {\bibfnamefont {E.~J.}\ \bibnamefont {Mele}},\ }\bibfield
  {title} {\bibinfo {title} {Topological {{Insulators}} in {{Three
  Dimensions}}},\ }\href {https://doi.org/10.1103/PhysRevLett.98.106803}
  {\bibfield  {journal} {\bibinfo  {journal} {Physical Review Letters}\
  }\textbf {\bibinfo {volume} {98}},\ \bibinfo {pages} {106803} (\bibinfo
  {year} {2007})}\BibitemShut {NoStop}%
\bibitem [{\citenamefont {Hsieh}\ \emph {et~al.}(2012)\citenamefont {Hsieh},
  \citenamefont {Lin}, \citenamefont {Liu}, \citenamefont {Duan}, \citenamefont
  {Bansil},\ and\ \citenamefont {Fu}}]{hsieh2012topological}%
  \BibitemOpen
  \bibfield  {author} {\bibinfo {author} {\bibfnamefont {T.~H.}\ \bibnamefont
  {Hsieh}}, \bibinfo {author} {\bibfnamefont {H.}~\bibnamefont {Lin}}, \bibinfo
  {author} {\bibfnamefont {J.}~\bibnamefont {Liu}}, \bibinfo {author}
  {\bibfnamefont {W.}~\bibnamefont {Duan}}, \bibinfo {author} {\bibfnamefont
  {A.}~\bibnamefont {Bansil}},\ and\ \bibinfo {author} {\bibfnamefont
  {L.}~\bibnamefont {Fu}},\ }\bibfield  {title} {\bibinfo {title} {Topological
  crystalline insulators in the {{SnTe}} material class},\ }\href
  {https://doi.org/10.1038/ncomms1969} {\bibfield  {journal} {\bibinfo
  {journal} {Nature Communications}\ }\textbf {\bibinfo {volume} {3}},\
  \bibinfo {pages} {982} (\bibinfo {year} {2012})}\BibitemShut {NoStop}%
\bibitem [{\citenamefont {Teo}\ \emph {et~al.}(2008)\citenamefont {Teo},
  \citenamefont {Fu},\ and\ \citenamefont {Kane}}]{teo2008surface}%
  \BibitemOpen
  \bibfield  {author} {\bibinfo {author} {\bibfnamefont {J.~C.~Y.}\
  \bibnamefont {Teo}}, \bibinfo {author} {\bibfnamefont {L.}~\bibnamefont
  {Fu}},\ and\ \bibinfo {author} {\bibfnamefont {C.~L.}\ \bibnamefont {Kane}},\
  }\bibfield  {title} {\bibinfo {title} {Surface states and topological
  invariants in three-dimensional topological insulators: {{Application}} to
  \$\textbackslash{{textBi}}\_1\textbackslash
  ensuremath-{{X}}\textbackslash{{textSb}}\_x\$},\ }\href
  {https://doi.org/10.1103/PhysRevB.78.045426} {\bibfield  {journal} {\bibinfo
  {journal} {Physical Review B}\ }\textbf {\bibinfo {volume} {78}},\ \bibinfo
  {pages} {045426} (\bibinfo {year} {2008})}\BibitemShut {NoStop}%
\bibitem [{\citenamefont {Po}\ \emph {et~al.}(2017)\citenamefont {Po},
  \citenamefont {Vishwanath},\ and\ \citenamefont
  {Watanabe}}]{po2017symmetrybased}%
  \BibitemOpen
  \bibfield  {author} {\bibinfo {author} {\bibfnamefont {H.~C.}\ \bibnamefont
  {Po}}, \bibinfo {author} {\bibfnamefont {A.}~\bibnamefont {Vishwanath}},\
  and\ \bibinfo {author} {\bibfnamefont {H.}~\bibnamefont {Watanabe}},\
  }\bibfield  {title} {\bibinfo {title} {Symmetry-based indicators of band
  topology in the 230 space groups},\ }\href
  {https://doi.org/10.1038/s41467-017-00133-2} {\bibfield  {journal} {\bibinfo
  {journal} {Nature Communications}\ }\textbf {\bibinfo {volume} {8}},\
  \bibinfo {pages} {50} (\bibinfo {year} {2017})}\BibitemShut {NoStop}%
\bibitem [{\citenamefont {Song}\ \emph {et~al.}(2018)\citenamefont {Song},
  \citenamefont {Zhang}, \citenamefont {Fang},\ and\ \citenamefont
  {Fang}}]{song2018quantitative}%
  \BibitemOpen
  \bibfield  {author} {\bibinfo {author} {\bibfnamefont {Z.}~\bibnamefont
  {Song}}, \bibinfo {author} {\bibfnamefont {T.}~\bibnamefont {Zhang}},
  \bibinfo {author} {\bibfnamefont {Z.}~\bibnamefont {Fang}},\ and\ \bibinfo
  {author} {\bibfnamefont {C.}~\bibnamefont {Fang}},\ }\bibfield  {title}
  {\bibinfo {title} {Quantitative mappings between symmetry and topology in
  solids},\ }\href {https://doi.org/10.1038/s41467-018-06010-w} {\bibfield
  {journal} {\bibinfo  {journal} {Nature Communications}\ }\textbf {\bibinfo
  {volume} {9}},\ \bibinfo {pages} {3530} (\bibinfo {year} {2018})}\BibitemShut
  {NoStop}%
\bibitem [{\citenamefont {Bradlyn}\ \emph {et~al.}(2017)\citenamefont
  {Bradlyn}, \citenamefont {Elcoro}, \citenamefont {Cano}, \citenamefont
  {Vergniory}, \citenamefont {Wang}, \citenamefont {Felser}, \citenamefont
  {Aroyo},\ and\ \citenamefont {Bernevig}}]{bradlyn2017topological}%
  \BibitemOpen
  \bibfield  {author} {\bibinfo {author} {\bibfnamefont {B.}~\bibnamefont
  {Bradlyn}}, \bibinfo {author} {\bibfnamefont {L.}~\bibnamefont {Elcoro}},
  \bibinfo {author} {\bibfnamefont {J.}~\bibnamefont {Cano}}, \bibinfo {author}
  {\bibfnamefont {M.~G.}\ \bibnamefont {Vergniory}}, \bibinfo {author}
  {\bibfnamefont {Z.}~\bibnamefont {Wang}}, \bibinfo {author} {\bibfnamefont
  {C.}~\bibnamefont {Felser}}, \bibinfo {author} {\bibfnamefont {M.~I.}\
  \bibnamefont {Aroyo}},\ and\ \bibinfo {author} {\bibfnamefont {B.~A.}\
  \bibnamefont {Bernevig}},\ }\bibfield  {title} {\bibinfo {title} {Topological
  quantum chemistry},\ }\href {https://doi.org/10.1038/nature23268} {\bibfield
  {journal} {\bibinfo  {journal} {Nature}\ }\textbf {\bibinfo {volume} {547}},\
  \bibinfo {pages} {298} (\bibinfo {year} {2017})}\BibitemShut {NoStop}%
\bibitem [{\citenamefont {Song}\ \emph {et~al.}(2020)\citenamefont {Song},
  \citenamefont {Fang},\ and\ \citenamefont {Qi}}]{song2020realspace}%
  \BibitemOpen
  \bibfield  {author} {\bibinfo {author} {\bibfnamefont {Z.}~\bibnamefont
  {Song}}, \bibinfo {author} {\bibfnamefont {C.}~\bibnamefont {Fang}},\ and\
  \bibinfo {author} {\bibfnamefont {Y.}~\bibnamefont {Qi}},\ }\bibfield
  {title} {\bibinfo {title} {Real-space recipes for general topological
  crystalline states},\ }\href@noop {} {\bibfield  {journal} {\bibinfo
  {journal} {Nature communications}\ }\textbf {\bibinfo {volume} {11}},\
  \bibinfo {pages} {1} (\bibinfo {year} {2020})}\BibitemShut {NoStop}%
\bibitem [{\citenamefont {Slager}\ \emph {et~al.}(2013)\citenamefont {Slager},
  \citenamefont {Mesaros}, \citenamefont {Juricic},\ and\ \citenamefont
  {Zaanen}}]{slager2013space}%
  \BibitemOpen
  \bibfield  {author} {\bibinfo {author} {\bibfnamefont {R.-J.}\ \bibnamefont
  {Slager}}, \bibinfo {author} {\bibfnamefont {A.}~\bibnamefont {Mesaros}},
  \bibinfo {author} {\bibfnamefont {V.}~\bibnamefont {Juricic}},\ and\ \bibinfo
  {author} {\bibfnamefont {J.}~\bibnamefont {Zaanen}},\ }\bibfield  {title}
  {\bibinfo {title} {The space group classification of topological band
  insulators},\ }\href@noop {} {\bibfield  {journal} {\bibinfo  {journal}
  {Nature Physics}\ }\textbf {\bibinfo {volume} {9}},\ \bibinfo {pages} {98}
  (\bibinfo {year} {2013})}\BibitemShut {NoStop}%
\bibitem [{\citenamefont {Kruthoff}\ \emph {et~al.}(2017)\citenamefont
  {Kruthoff}, \citenamefont {{de Boer}}, \citenamefont {{van Wezel}},
  \citenamefont {Kane},\ and\ \citenamefont
  {Slager}}]{kruthoff2017topological}%
  \BibitemOpen
  \bibfield  {author} {\bibinfo {author} {\bibfnamefont {J.}~\bibnamefont
  {Kruthoff}}, \bibinfo {author} {\bibfnamefont {J.}~\bibnamefont {{de Boer}}},
  \bibinfo {author} {\bibfnamefont {J.}~\bibnamefont {{van Wezel}}}, \bibinfo
  {author} {\bibfnamefont {C.~L.}\ \bibnamefont {Kane}},\ and\ \bibinfo
  {author} {\bibfnamefont {R.-J.}\ \bibnamefont {Slager}},\ }\bibfield  {title}
  {\bibinfo {title} {Topological {{Classification}} of {{Crystalline
  Insulators}} through {{Band Structure Combinatorics}}},\ }\href
  {https://doi.org/10.1103/PhysRevX.7.041069} {\bibfield  {journal} {\bibinfo
  {journal} {Physical Review X}\ }\textbf {\bibinfo {volume} {7}},\ \bibinfo
  {pages} {041069} (\bibinfo {year} {2017})}\BibitemShut {NoStop}%
\bibitem [{\citenamefont {Cano}\ and\ \citenamefont
  {Bradlyn}(2021)}]{cano2021band}%
  \BibitemOpen
  \bibfield  {author} {\bibinfo {author} {\bibfnamefont {J.}~\bibnamefont
  {Cano}}\ and\ \bibinfo {author} {\bibfnamefont {B.}~\bibnamefont {Bradlyn}},\
  }\bibfield  {title} {\bibinfo {title} {Band {{Representations}} and
  {{Topological Quantum Chemistry}}},\ }\href
  {https://doi.org/10.1146/annurev-conmatphys-041720-124134} {\bibfield
  {journal} {\bibinfo  {journal} {Annual Reviews of Condensed Matter Physics}\
  }\textbf {\bibinfo {volume} {12}},\ \bibinfo {pages} {225} (\bibinfo {year}
  {2021})}\BibitemShut {NoStop}%
\bibitem [{\citenamefont {Benalcazar}\ \emph
  {et~al.}(2017{\natexlab{a}})\citenamefont {Benalcazar}, \citenamefont
  {Bernevig},\ and\ \citenamefont {Hughes}}]{benalcazar2017quantized}%
  \BibitemOpen
  \bibfield  {author} {\bibinfo {author} {\bibfnamefont {W.~A.}\ \bibnamefont
  {Benalcazar}}, \bibinfo {author} {\bibfnamefont {B.~A.}\ \bibnamefont
  {Bernevig}},\ and\ \bibinfo {author} {\bibfnamefont {T.~L.}\ \bibnamefont
  {Hughes}},\ }\bibfield  {title} {\bibinfo {title} {Quantized electric
  multipole insulators},\ }\href@noop {} {\bibfield  {journal} {\bibinfo
  {journal} {Science (New York, N.Y.)}\ }\textbf {\bibinfo {volume} {357}},\
  \bibinfo {pages} {61} (\bibinfo {year} {2017}{\natexlab{a}})}\BibitemShut
  {NoStop}%
\bibitem [{\citenamefont {Benalcazar}\ \emph
  {et~al.}(2017{\natexlab{b}})\citenamefont {Benalcazar}, \citenamefont
  {Bernevig},\ and\ \citenamefont {Hughes}}]{benalcazar2017electric}%
  \BibitemOpen
  \bibfield  {author} {\bibinfo {author} {\bibfnamefont {W.~A.}\ \bibnamefont
  {Benalcazar}}, \bibinfo {author} {\bibfnamefont {B.~A.}\ \bibnamefont
  {Bernevig}},\ and\ \bibinfo {author} {\bibfnamefont {T.~L.}\ \bibnamefont
  {Hughes}},\ }\bibfield  {title} {\bibinfo {title} {Electric multipole
  moments, topological multipole moment pumping, and chiral hinge states in
  crystalline insulators},\ }\href {https://doi.org/10.1103/PhysRevB.96.245115}
  {\bibfield  {journal} {\bibinfo  {journal} {Physical Review B}\ }\textbf
  {\bibinfo {volume} {96}},\ \bibinfo {pages} {245115} (\bibinfo {year}
  {2017}{\natexlab{b}})}\BibitemShut {NoStop}%
\bibitem [{\citenamefont {Schindler}\ \emph
  {et~al.}(2018{\natexlab{a}})\citenamefont {Schindler}, \citenamefont {Cook},
  \citenamefont {Vergniory}, \citenamefont {Wang}, \citenamefont {Parkin},
  \citenamefont {Bernevig},\ and\ \citenamefont
  {Neupert}}]{schindler2018higher}%
  \BibitemOpen
  \bibfield  {author} {\bibinfo {author} {\bibfnamefont {F.}~\bibnamefont
  {Schindler}}, \bibinfo {author} {\bibfnamefont {A.~M.}\ \bibnamefont {Cook}},
  \bibinfo {author} {\bibfnamefont {M.~G.}\ \bibnamefont {Vergniory}}, \bibinfo
  {author} {\bibfnamefont {Z.}~\bibnamefont {Wang}}, \bibinfo {author}
  {\bibfnamefont {S.~S.}\ \bibnamefont {Parkin}}, \bibinfo {author}
  {\bibfnamefont {B.~A.}\ \bibnamefont {Bernevig}},\ and\ \bibinfo {author}
  {\bibfnamefont {T.}~\bibnamefont {Neupert}},\ }\bibfield  {title} {\bibinfo
  {title} {Higher-order topological insulators},\ }\href@noop {} {\bibfield
  {journal} {\bibinfo  {journal} {Science advances}\ }\textbf {\bibinfo
  {volume} {4}},\ \bibinfo {pages} {eaat0346} (\bibinfo {year}
  {2018}{\natexlab{a}})}\BibitemShut {NoStop}%
\bibitem [{\citenamefont {Khalaf}\ \emph {et~al.}(2018)\citenamefont {Khalaf},
  \citenamefont {Po}, \citenamefont {Vishwanath},\ and\ \citenamefont
  {Watanabe}}]{khalaf2018symmetry}%
  \BibitemOpen
  \bibfield  {author} {\bibinfo {author} {\bibfnamefont {E.}~\bibnamefont
  {Khalaf}}, \bibinfo {author} {\bibfnamefont {H.~C.}\ \bibnamefont {Po}},
  \bibinfo {author} {\bibfnamefont {A.}~\bibnamefont {Vishwanath}},\ and\
  \bibinfo {author} {\bibfnamefont {H.}~\bibnamefont {Watanabe}},\ }\bibfield
  {title} {\bibinfo {title} {Symmetry {{Indicators}} and {{Anomalous Surface
  States}} of {{Topological Crystalline Insulators}}},\ }\href
  {https://doi.org/10.1103/PhysRevX.8.031070} {\bibfield  {journal} {\bibinfo
  {journal} {Physical Review X}\ }\textbf {\bibinfo {volume} {8}},\ \bibinfo
  {pages} {031070} (\bibinfo {year} {2018})}\BibitemShut {NoStop}%
\bibitem [{\citenamefont {Schindler}\ \emph
  {et~al.}(2018{\natexlab{b}})\citenamefont {Schindler}, \citenamefont {Wang},
  \citenamefont {Vergniory}, \citenamefont {Cook}, \citenamefont {Murani},
  \citenamefont {Sengupta}, \citenamefont {Kasumov}, \citenamefont {Deblock},
  \citenamefont {Jeon}, \citenamefont {Drozdov}, \citenamefont {Bouchiat},
  \citenamefont {Gu{\'e}ron}, \citenamefont {Yazdani}, \citenamefont
  {Bernevig},\ and\ \citenamefont {Neupert}}]{schindler2018higherordera}%
  \BibitemOpen
  \bibfield  {author} {\bibinfo {author} {\bibfnamefont {F.}~\bibnamefont
  {Schindler}}, \bibinfo {author} {\bibfnamefont {Z.}~\bibnamefont {Wang}},
  \bibinfo {author} {\bibfnamefont {M.~G.}\ \bibnamefont {Vergniory}}, \bibinfo
  {author} {\bibfnamefont {A.~M.}\ \bibnamefont {Cook}}, \bibinfo {author}
  {\bibfnamefont {A.}~\bibnamefont {Murani}}, \bibinfo {author} {\bibfnamefont
  {S.}~\bibnamefont {Sengupta}}, \bibinfo {author} {\bibfnamefont {A.~Y.}\
  \bibnamefont {Kasumov}}, \bibinfo {author} {\bibfnamefont {R.}~\bibnamefont
  {Deblock}}, \bibinfo {author} {\bibfnamefont {S.}~\bibnamefont {Jeon}},
  \bibinfo {author} {\bibfnamefont {I.}~\bibnamefont {Drozdov}}, \bibinfo
  {author} {\bibfnamefont {H.}~\bibnamefont {Bouchiat}}, \bibinfo {author}
  {\bibfnamefont {S.}~\bibnamefont {Gu{\'e}ron}}, \bibinfo {author}
  {\bibfnamefont {A.}~\bibnamefont {Yazdani}}, \bibinfo {author} {\bibfnamefont
  {B.~A.}\ \bibnamefont {Bernevig}},\ and\ \bibinfo {author} {\bibfnamefont
  {T.}~\bibnamefont {Neupert}},\ }\bibfield  {title} {\bibinfo {title}
  {Higher-order topology in bismuth},\ }\href
  {https://doi.org/10.1038/s41567-018-0224-7} {\bibfield  {journal} {\bibinfo
  {journal} {Nature Physics}\ }\textbf {\bibinfo {volume} {14}},\ \bibinfo
  {pages} {918} (\bibinfo {year} {2018}{\natexlab{b}})}\BibitemShut {NoStop}%
\bibitem [{\citenamefont {Zhou}\ \emph {et~al.}(2015)\citenamefont {Zhou},
  \citenamefont {Feng}, \citenamefont {Liu},\ and\ \citenamefont
  {Yao}}]{zhou2015topological}%
  \BibitemOpen
  \bibfield  {author} {\bibinfo {author} {\bibfnamefont {J.-J.}\ \bibnamefont
  {Zhou}}, \bibinfo {author} {\bibfnamefont {W.}~\bibnamefont {Feng}}, \bibinfo
  {author} {\bibfnamefont {G.-B.}\ \bibnamefont {Liu}},\ and\ \bibinfo {author}
  {\bibfnamefont {Y.}~\bibnamefont {Yao}},\ }\bibfield  {title} {\bibinfo
  {title} {Topological edge states in single-and multi-layer {{Bi4Br4}}},\
  }\href@noop {} {\bibfield  {journal} {\bibinfo  {journal} {New Journal of
  Physics}\ }\textbf {\bibinfo {volume} {17}},\ \bibinfo {pages} {015004}
  (\bibinfo {year} {2015})}\BibitemShut {NoStop}%
\bibitem [{\citenamefont {Li}\ \emph {et~al.}(2019)\citenamefont {Li},
  \citenamefont {Chen}, \citenamefont {Jin}, \citenamefont {Ma}, \citenamefont
  {Ge}, \citenamefont {Sun}, \citenamefont {Guo}, \citenamefont {Sun},
  \citenamefont {Han}, \citenamefont {Xiao} \emph
  {et~al.}}]{li2019pressureinduced}%
  \BibitemOpen
  \bibfield  {author} {\bibinfo {author} {\bibfnamefont {X.}~\bibnamefont
  {Li}}, \bibinfo {author} {\bibfnamefont {D.}~\bibnamefont {Chen}}, \bibinfo
  {author} {\bibfnamefont {M.}~\bibnamefont {Jin}}, \bibinfo {author}
  {\bibfnamefont {D.}~\bibnamefont {Ma}}, \bibinfo {author} {\bibfnamefont
  {Y.}~\bibnamefont {Ge}}, \bibinfo {author} {\bibfnamefont {J.}~\bibnamefont
  {Sun}}, \bibinfo {author} {\bibfnamefont {W.}~\bibnamefont {Guo}}, \bibinfo
  {author} {\bibfnamefont {H.}~\bibnamefont {Sun}}, \bibinfo {author}
  {\bibfnamefont {J.}~\bibnamefont {Han}}, \bibinfo {author} {\bibfnamefont
  {W.}~\bibnamefont {Xiao}}, \emph {et~al.},\ }\bibfield  {title} {\bibinfo
  {title} {Pressure-induced phase transitions and superconductivity in a
  {{Quasi}}\textendash 1-{{Dimensional}} topological crystalline insulator
  {{$\alpha$}}-{{Bi4Br4}}},\ }\href@noop {} {\bibfield  {journal} {\bibinfo
  {journal} {Proceedings of the National Academy of Sciences}\ }\textbf
  {\bibinfo {volume} {116}},\ \bibinfo {pages} {17696} (\bibinfo {year}
  {2019})}\BibitemShut {NoStop}%
\bibitem [{\citenamefont {Shumiya}\ \emph {et~al.}(2022)\citenamefont
  {Shumiya}, \citenamefont {Hossain}, \citenamefont {Yin}, \citenamefont
  {Wang}, \citenamefont {Litskevich}, \citenamefont {Yoon}, \citenamefont {Li},
  \citenamefont {Yang}, \citenamefont {Jiang}, \citenamefont {Cheng},
  \citenamefont {Lin}, \citenamefont {Zhang}, \citenamefont {Cheng},
  \citenamefont {Cochran}, \citenamefont {Multer}, \citenamefont {Yang},
  \citenamefont {Casas}, \citenamefont {Chang}, \citenamefont {Neupert},
  \citenamefont {Yuan}, \citenamefont {Jia}, \citenamefont {Lin}, \citenamefont
  {Yao}, \citenamefont {Balicas}, \citenamefont {Zhang}, \citenamefont {Yao},\
  and\ \citenamefont {Hasan}}]{shumiya2022evidence}%
  \BibitemOpen
  \bibfield  {author} {\bibinfo {author} {\bibfnamefont {N.}~\bibnamefont
  {Shumiya}}, \bibinfo {author} {\bibfnamefont {M.~S.}\ \bibnamefont
  {Hossain}}, \bibinfo {author} {\bibfnamefont {J.-X.}\ \bibnamefont {Yin}},
  \bibinfo {author} {\bibfnamefont {Z.}~\bibnamefont {Wang}}, \bibinfo {author}
  {\bibfnamefont {M.}~\bibnamefont {Litskevich}}, \bibinfo {author}
  {\bibfnamefont {C.}~\bibnamefont {Yoon}}, \bibinfo {author} {\bibfnamefont
  {Y.}~\bibnamefont {Li}}, \bibinfo {author} {\bibfnamefont {Y.}~\bibnamefont
  {Yang}}, \bibinfo {author} {\bibfnamefont {Y.-X.}\ \bibnamefont {Jiang}},
  \bibinfo {author} {\bibfnamefont {G.}~\bibnamefont {Cheng}}, \bibinfo
  {author} {\bibfnamefont {Y.-C.}\ \bibnamefont {Lin}}, \bibinfo {author}
  {\bibfnamefont {Q.}~\bibnamefont {Zhang}}, \bibinfo {author} {\bibfnamefont
  {Z.-J.}\ \bibnamefont {Cheng}}, \bibinfo {author} {\bibfnamefont {T.~A.}\
  \bibnamefont {Cochran}}, \bibinfo {author} {\bibfnamefont {D.}~\bibnamefont
  {Multer}}, \bibinfo {author} {\bibfnamefont {X.~P.}\ \bibnamefont {Yang}},
  \bibinfo {author} {\bibfnamefont {B.}~\bibnamefont {Casas}}, \bibinfo
  {author} {\bibfnamefont {T.-R.}\ \bibnamefont {Chang}}, \bibinfo {author}
  {\bibfnamefont {T.}~\bibnamefont {Neupert}}, \bibinfo {author} {\bibfnamefont
  {Z.}~\bibnamefont {Yuan}}, \bibinfo {author} {\bibfnamefont {S.}~\bibnamefont
  {Jia}}, \bibinfo {author} {\bibfnamefont {H.}~\bibnamefont {Lin}}, \bibinfo
  {author} {\bibfnamefont {N.}~\bibnamefont {Yao}}, \bibinfo {author}
  {\bibfnamefont {L.}~\bibnamefont {Balicas}}, \bibinfo {author} {\bibfnamefont
  {F.}~\bibnamefont {Zhang}}, \bibinfo {author} {\bibfnamefont
  {Y.}~\bibnamefont {Yao}},\ and\ \bibinfo {author} {\bibfnamefont {M.~Z.}\
  \bibnamefont {Hasan}},\ }\bibfield  {title} {\bibinfo {title} {Evidence of a
  room-temperature quantum spin {{Hall}} edge state in a higher-order
  topological insulator},\ }\bibfield  {journal} {\bibinfo  {journal} {Nature
  Materials}\ }\href {https://doi.org/10.1038/s41563-022-01304-3}
  {10.1038/s41563-022-01304-3} (\bibinfo {year} {2022})\BibitemShut {NoStop}%
\bibitem [{\citenamefont {Noguchi}\ \emph {et~al.}(2021)\citenamefont
  {Noguchi}, \citenamefont {Kobayashi}, \citenamefont {Jiang}, \citenamefont
  {Kuroda}, \citenamefont {Takahashi}, \citenamefont {Xu}, \citenamefont {Lee},
  \citenamefont {Hirayama}, \citenamefont {Ochi}, \citenamefont {Shirasawa},
  \citenamefont {Zhang}, \citenamefont {Lin}, \citenamefont {Bareille},
  \citenamefont {Sakuragi}, \citenamefont {Tanaka}, \citenamefont {Kunisada},
  \citenamefont {Kurokawa}, \citenamefont {Yaji}, \citenamefont {Harasawa},
  \citenamefont {Kandyba}, \citenamefont {Giampietri}, \citenamefont {Barinov},
  \citenamefont {Kim}, \citenamefont {Cacho}, \citenamefont {Hashimoto},
  \citenamefont {Lu}, \citenamefont {Shin}, \citenamefont {Arita},
  \citenamefont {Lai}, \citenamefont {Sasagawa},\ and\ \citenamefont
  {Kondo}}]{noguchi2021evidence}%
  \BibitemOpen
  \bibfield  {author} {\bibinfo {author} {\bibfnamefont {R.}~\bibnamefont
  {Noguchi}}, \bibinfo {author} {\bibfnamefont {M.}~\bibnamefont {Kobayashi}},
  \bibinfo {author} {\bibfnamefont {Z.}~\bibnamefont {Jiang}}, \bibinfo
  {author} {\bibfnamefont {K.}~\bibnamefont {Kuroda}}, \bibinfo {author}
  {\bibfnamefont {T.}~\bibnamefont {Takahashi}}, \bibinfo {author}
  {\bibfnamefont {Z.}~\bibnamefont {Xu}}, \bibinfo {author} {\bibfnamefont
  {D.}~\bibnamefont {Lee}}, \bibinfo {author} {\bibfnamefont {M.}~\bibnamefont
  {Hirayama}}, \bibinfo {author} {\bibfnamefont {M.}~\bibnamefont {Ochi}},
  \bibinfo {author} {\bibfnamefont {T.}~\bibnamefont {Shirasawa}}, \bibinfo
  {author} {\bibfnamefont {P.}~\bibnamefont {Zhang}}, \bibinfo {author}
  {\bibfnamefont {C.}~\bibnamefont {Lin}}, \bibinfo {author} {\bibfnamefont
  {C.}~\bibnamefont {Bareille}}, \bibinfo {author} {\bibfnamefont
  {S.}~\bibnamefont {Sakuragi}}, \bibinfo {author} {\bibfnamefont
  {H.}~\bibnamefont {Tanaka}}, \bibinfo {author} {\bibfnamefont
  {S.}~\bibnamefont {Kunisada}}, \bibinfo {author} {\bibfnamefont
  {K.}~\bibnamefont {Kurokawa}}, \bibinfo {author} {\bibfnamefont
  {K.}~\bibnamefont {Yaji}}, \bibinfo {author} {\bibfnamefont {A.}~\bibnamefont
  {Harasawa}}, \bibinfo {author} {\bibfnamefont {V.}~\bibnamefont {Kandyba}},
  \bibinfo {author} {\bibfnamefont {A.}~\bibnamefont {Giampietri}}, \bibinfo
  {author} {\bibfnamefont {A.}~\bibnamefont {Barinov}}, \bibinfo {author}
  {\bibfnamefont {T.~K.}\ \bibnamefont {Kim}}, \bibinfo {author} {\bibfnamefont
  {C.}~\bibnamefont {Cacho}}, \bibinfo {author} {\bibfnamefont
  {M.}~\bibnamefont {Hashimoto}}, \bibinfo {author} {\bibfnamefont
  {D.}~\bibnamefont {Lu}}, \bibinfo {author} {\bibfnamefont {S.}~\bibnamefont
  {Shin}}, \bibinfo {author} {\bibfnamefont {R.}~\bibnamefont {Arita}},
  \bibinfo {author} {\bibfnamefont {K.}~\bibnamefont {Lai}}, \bibinfo {author}
  {\bibfnamefont {T.}~\bibnamefont {Sasagawa}},\ and\ \bibinfo {author}
  {\bibfnamefont {T.}~\bibnamefont {Kondo}},\ }\bibfield  {title} {\bibinfo
  {title} {Evidence for a higher-order topological insulator in a
  three-dimensional material built from van der {{Waals}} stacking of
  bismuth-halide chains},\ }\href {https://doi.org/10.1038/s41563-020-00871-7}
  {\bibfield  {journal} {\bibinfo  {journal} {Nature Materials}\ }\textbf
  {\bibinfo {volume} {20}},\ \bibinfo {pages} {473} (\bibinfo {year}
  {2021})}\BibitemShut {NoStop}%
\bibitem [{\citenamefont {Wang}\ \emph {et~al.}(2019)\citenamefont {Wang},
  \citenamefont {Wieder}, \citenamefont {Li}, \citenamefont {Yan},\ and\
  \citenamefont {Bernevig}}]{wang2019higherorder}%
  \BibitemOpen
  \bibfield  {author} {\bibinfo {author} {\bibfnamefont {Z.}~\bibnamefont
  {Wang}}, \bibinfo {author} {\bibfnamefont {B.~J.}\ \bibnamefont {Wieder}},
  \bibinfo {author} {\bibfnamefont {J.}~\bibnamefont {Li}}, \bibinfo {author}
  {\bibfnamefont {B.}~\bibnamefont {Yan}},\ and\ \bibinfo {author}
  {\bibfnamefont {B.~A.}\ \bibnamefont {Bernevig}},\ }\bibfield  {title}
  {\bibinfo {title} {Higher-{{Order Topology}}, {{Monopole Nodal Lines}}, and
  the {{Origin}} of {{Large Fermi Arcs}} in {{Transition Metal
  Dichalcogenides}} \${{X}}\textbackslash{{mathrmTe}}\_2\$
  (\${{X}}=\textbackslash{{mathrmMo}},\textbackslash{{mathrmW}}\$)},\ }\href
  {https://doi.org/10.1103/PhysRevLett.123.186401} {\bibfield  {journal}
  {\bibinfo  {journal} {Physical Review Letters}\ }\textbf {\bibinfo {volume}
  {123}},\ \bibinfo {pages} {186401} (\bibinfo {year} {2019})}\BibitemShut
  {NoStop}%
\bibitem [{\citenamefont {He}\ \emph {et~al.}(2019)\citenamefont {He},
  \citenamefont {Jiang}, \citenamefont {Zhang}, \citenamefont {Huang},
  \citenamefont {Fang},\ and\ \citenamefont {Jin}}]{he2019symtopo}%
  \BibitemOpen
  \bibfield  {author} {\bibinfo {author} {\bibfnamefont {Y.}~\bibnamefont
  {He}}, \bibinfo {author} {\bibfnamefont {Y.}~\bibnamefont {Jiang}}, \bibinfo
  {author} {\bibfnamefont {T.}~\bibnamefont {Zhang}}, \bibinfo {author}
  {\bibfnamefont {H.}~\bibnamefont {Huang}}, \bibinfo {author} {\bibfnamefont
  {C.}~\bibnamefont {Fang}},\ and\ \bibinfo {author} {\bibfnamefont
  {Z.}~\bibnamefont {Jin}},\ }\bibfield  {title} {\bibinfo {title}
  {{{SymTopo}}: {{An}} automatic tool for calculating topological properties of
  nonmagnetic crystalline materials},\ }\href@noop {} {\bibfield  {journal}
  {\bibinfo  {journal} {Chinese Physics B}\ }\textbf {\bibinfo {volume} {28}},\
  \bibinfo {pages} {087102} (\bibinfo {year} {2019})}\BibitemShut {NoStop}%
\bibitem [{\citenamefont {Xu}\ \emph {et~al.}(2020)\citenamefont {Xu},
  \citenamefont {Elcoro}, \citenamefont {Song}, \citenamefont {Wieder},
  \citenamefont {Vergniory}, \citenamefont {Regnault}, \citenamefont {Chen},
  \citenamefont {Felser},\ and\ \citenamefont
  {Bernevig}}]{xu2020highthroughput}%
  \BibitemOpen
  \bibfield  {author} {\bibinfo {author} {\bibfnamefont {Y.}~\bibnamefont
  {Xu}}, \bibinfo {author} {\bibfnamefont {L.}~\bibnamefont {Elcoro}}, \bibinfo
  {author} {\bibfnamefont {Z.-D.}\ \bibnamefont {Song}}, \bibinfo {author}
  {\bibfnamefont {B.~J.}\ \bibnamefont {Wieder}}, \bibinfo {author}
  {\bibfnamefont {M.~G.}\ \bibnamefont {Vergniory}}, \bibinfo {author}
  {\bibfnamefont {N.}~\bibnamefont {Regnault}}, \bibinfo {author}
  {\bibfnamefont {Y.}~\bibnamefont {Chen}}, \bibinfo {author} {\bibfnamefont
  {C.}~\bibnamefont {Felser}},\ and\ \bibinfo {author} {\bibfnamefont {B.~A.}\
  \bibnamefont {Bernevig}},\ }\bibfield  {title} {\bibinfo {title}
  {High-throughput calculations of magnetic topological materials},\ }\href
  {https://doi.org/10.1038/s41586-020-2837-0} {\bibfield  {journal} {\bibinfo
  {journal} {Nature}\ }\textbf {\bibinfo {volume} {586}},\ \bibinfo {pages}
  {702} (\bibinfo {year} {2020})}\BibitemShut {NoStop}%
\bibitem [{\citenamefont {Wieder}\ \emph {et~al.}(2022)\citenamefont {Wieder},
  \citenamefont {Bradlyn}, \citenamefont {Cano}, \citenamefont {Wang},
  \citenamefont {Vergniory}, \citenamefont {Elcoro}, \citenamefont {Soluyanov},
  \citenamefont {Felser}, \citenamefont {Neupert}, \citenamefont {Regnault},\
  and\ \citenamefont {Bernevig}}]{wieder2022topological}%
  \BibitemOpen
  \bibfield  {author} {\bibinfo {author} {\bibfnamefont {B.~J.}\ \bibnamefont
  {Wieder}}, \bibinfo {author} {\bibfnamefont {B.}~\bibnamefont {Bradlyn}},
  \bibinfo {author} {\bibfnamefont {J.}~\bibnamefont {Cano}}, \bibinfo {author}
  {\bibfnamefont {Z.}~\bibnamefont {Wang}}, \bibinfo {author} {\bibfnamefont
  {M.~G.}\ \bibnamefont {Vergniory}}, \bibinfo {author} {\bibfnamefont
  {L.}~\bibnamefont {Elcoro}}, \bibinfo {author} {\bibfnamefont {A.~A.}\
  \bibnamefont {Soluyanov}}, \bibinfo {author} {\bibfnamefont {C.}~\bibnamefont
  {Felser}}, \bibinfo {author} {\bibfnamefont {T.}~\bibnamefont {Neupert}},
  \bibinfo {author} {\bibfnamefont {N.}~\bibnamefont {Regnault}},\ and\
  \bibinfo {author} {\bibfnamefont {B.~A.}\ \bibnamefont {Bernevig}},\
  }\bibfield  {title} {\bibinfo {title} {Topological materials discovery from
  crystal symmetry},\ }\href {https://doi.org/10.1038/s41578-021-00380-2}
  {\bibfield  {journal} {\bibinfo  {journal} {Nature Reviews Materials}\
  }\textbf {\bibinfo {volume} {7}},\ \bibinfo {pages} {196} (\bibinfo {year}
  {2022})}\BibitemShut {NoStop}%
\bibitem [{\citenamefont {Vergniory}\ \emph {et~al.}(2019)\citenamefont
  {Vergniory}, \citenamefont {Elcoro}, \citenamefont {Felser}, \citenamefont
  {Regnault}, \citenamefont {Bernevig},\ and\ \citenamefont
  {Wang}}]{vergniory2019complete}%
  \BibitemOpen
  \bibfield  {author} {\bibinfo {author} {\bibfnamefont {M.}~\bibnamefont
  {Vergniory}}, \bibinfo {author} {\bibfnamefont {L.}~\bibnamefont {Elcoro}},
  \bibinfo {author} {\bibfnamefont {C.}~\bibnamefont {Felser}}, \bibinfo
  {author} {\bibfnamefont {N.}~\bibnamefont {Regnault}}, \bibinfo {author}
  {\bibfnamefont {B.~A.}\ \bibnamefont {Bernevig}},\ and\ \bibinfo {author}
  {\bibfnamefont {Z.}~\bibnamefont {Wang}},\ }\bibfield  {title} {\bibinfo
  {title} {A complete catalogue of high-quality topological materials},\
  }\href@noop {} {\bibfield  {journal} {\bibinfo  {journal} {Nature}\ }\textbf
  {\bibinfo {volume} {566}},\ \bibinfo {pages} {480} (\bibinfo {year}
  {2019})}\BibitemShut {NoStop}%
\bibitem [{\citenamefont {Vergniory}\ \emph {et~al.}(2022)\citenamefont
  {Vergniory}, \citenamefont {Wieder}, \citenamefont {Elcoro}, \citenamefont
  {Parkin}, \citenamefont {Felser}, \citenamefont {Bernevig},\ and\
  \citenamefont {{Nicolas Regnault}}}]{vergniory2022all}%
  \BibitemOpen
  \bibfield  {author} {\bibinfo {author} {\bibfnamefont {M.~G.}\ \bibnamefont
  {Vergniory}}, \bibinfo {author} {\bibfnamefont {B.~J.}\ \bibnamefont
  {Wieder}}, \bibinfo {author} {\bibfnamefont {L.}~\bibnamefont {Elcoro}},
  \bibinfo {author} {\bibfnamefont {S.~S.~P.}\ \bibnamefont {Parkin}}, \bibinfo
  {author} {\bibfnamefont {C.}~\bibnamefont {Felser}}, \bibinfo {author}
  {\bibfnamefont {B.~A.}\ \bibnamefont {Bernevig}},\ and\ \bibinfo {author}
  {\bibnamefont {{Nicolas Regnault}}},\ }\bibfield  {title} {\bibinfo {title}
  {All topological bands of all nonmagnetic stoichiometric materials},\ }\href
  {https://doi.org/10.1126/science.abg9094} {\bibfield  {journal} {\bibinfo
  {journal} {Science (New York, N.Y.)}\ }\textbf {\bibinfo {volume} {376}},\
  \bibinfo {pages} {eabg9094} (\bibinfo {year} {2022})}\BibitemShut {NoStop}%
\bibitem [{\citenamefont {Elcoro}\ \emph {et~al.}(2021)\citenamefont {Elcoro},
  \citenamefont {Wieder}, \citenamefont {Song}, \citenamefont {Xu},
  \citenamefont {Bradlyn},\ and\ \citenamefont
  {Bernevig}}]{elcoro2021magnetic}%
  \BibitemOpen
  \bibfield  {author} {\bibinfo {author} {\bibfnamefont {L.}~\bibnamefont
  {Elcoro}}, \bibinfo {author} {\bibfnamefont {B.~J.}\ \bibnamefont {Wieder}},
  \bibinfo {author} {\bibfnamefont {Z.}~\bibnamefont {Song}}, \bibinfo {author}
  {\bibfnamefont {Y.}~\bibnamefont {Xu}}, \bibinfo {author} {\bibfnamefont
  {B.}~\bibnamefont {Bradlyn}},\ and\ \bibinfo {author} {\bibfnamefont {B.~A.}\
  \bibnamefont {Bernevig}},\ }\bibfield  {title} {\bibinfo {title} {Magnetic
  topological quantum chemistry},\ }\href
  {https://doi.org/10.1038/s41467-021-26241-8} {\bibfield  {journal} {\bibinfo
  {journal} {Nature Communications}\ }\textbf {\bibinfo {volume} {12}},\
  \bibinfo {pages} {5965} (\bibinfo {year} {2021})}\BibitemShut {NoStop}%
\bibitem [{\citenamefont {Tang}\ \emph
  {et~al.}(2019{\natexlab{a}})\citenamefont {Tang}, \citenamefont {Po},
  \citenamefont {Vishwanath},\ and\ \citenamefont {Wan}}]{tang2019efficient}%
  \BibitemOpen
  \bibfield  {author} {\bibinfo {author} {\bibfnamefont {F.}~\bibnamefont
  {Tang}}, \bibinfo {author} {\bibfnamefont {H.~C.}\ \bibnamefont {Po}},
  \bibinfo {author} {\bibfnamefont {A.}~\bibnamefont {Vishwanath}},\ and\
  \bibinfo {author} {\bibfnamefont {X.}~\bibnamefont {Wan}},\ }\bibfield
  {title} {\bibinfo {title} {Efficient topological materials discovery using
  symmetry indicators},\ }\href {https://doi.org/10.1038/s41567-019-0418-7}
  {\bibfield  {journal} {\bibinfo  {journal} {Nature Physics}\ }\textbf
  {\bibinfo {volume} {15}},\ \bibinfo {pages} {470} (\bibinfo {year}
  {2019}{\natexlab{a}})}\BibitemShut {NoStop}%
\bibitem [{\citenamefont {Tang}\ \emph
  {et~al.}(2019{\natexlab{b}})\citenamefont {Tang}, \citenamefont {Po},
  \citenamefont {Vishwanath},\ and\ \citenamefont
  {Wan}}]{tang2019comprehensive}%
  \BibitemOpen
  \bibfield  {author} {\bibinfo {author} {\bibfnamefont {F.}~\bibnamefont
  {Tang}}, \bibinfo {author} {\bibfnamefont {H.~C.}\ \bibnamefont {Po}},
  \bibinfo {author} {\bibfnamefont {A.}~\bibnamefont {Vishwanath}},\ and\
  \bibinfo {author} {\bibfnamefont {X.}~\bibnamefont {Wan}},\ }\bibfield
  {title} {\bibinfo {title} {Comprehensive search for topological materials
  using symmetry indicators},\ }\href
  {https://doi.org/10.1038/s41586-019-0937-5} {\bibfield  {journal} {\bibinfo
  {journal} {Nature}\ }\textbf {\bibinfo {volume} {566}},\ \bibinfo {pages}
  {486} (\bibinfo {year} {2019}{\natexlab{b}})}\BibitemShut {NoStop}%
\bibitem [{\citenamefont {Xu}\ \emph {et~al.}(2019)\citenamefont {Xu},
  \citenamefont {Song}, \citenamefont {Wang}, \citenamefont {Weng},\ and\
  \citenamefont {Dai}}]{xu2019higherorder}%
  \BibitemOpen
  \bibfield  {author} {\bibinfo {author} {\bibfnamefont {Y.}~\bibnamefont
  {Xu}}, \bibinfo {author} {\bibfnamefont {Z.}~\bibnamefont {Song}}, \bibinfo
  {author} {\bibfnamefont {Z.}~\bibnamefont {Wang}}, \bibinfo {author}
  {\bibfnamefont {H.}~\bibnamefont {Weng}},\ and\ \bibinfo {author}
  {\bibfnamefont {X.}~\bibnamefont {Dai}},\ }\bibfield  {title} {\bibinfo
  {title} {Higher-{{Order Topology}} of the {{Axion Insulator EuIn}} 2 {{As}}
  2},\ }\href@noop {} {\bibfield  {journal} {\bibinfo  {journal} {Physical
  review letters}\ }\textbf {\bibinfo {volume} {122}},\ \bibinfo {pages}
  {256402} (\bibinfo {year} {2019})}\BibitemShut {NoStop}%
\bibitem [{\citenamefont {Xiao}\ \emph {et~al.}(2018)\citenamefont {Xiao},
  \citenamefont {Jiang}, \citenamefont {Shin}, \citenamefont {Wang},
  \citenamefont {Wang}, \citenamefont {Zhao}, \citenamefont {Liu},
  \citenamefont {Wu}, \citenamefont {Chan}, \citenamefont {Samarth},\ and\
  \citenamefont {Chang}}]{xiao2018realization}%
  \BibitemOpen
  \bibfield  {author} {\bibinfo {author} {\bibfnamefont {D.}~\bibnamefont
  {Xiao}}, \bibinfo {author} {\bibfnamefont {J.}~\bibnamefont {Jiang}},
  \bibinfo {author} {\bibfnamefont {J.-H.}\ \bibnamefont {Shin}}, \bibinfo
  {author} {\bibfnamefont {W.}~\bibnamefont {Wang}}, \bibinfo {author}
  {\bibfnamefont {F.}~\bibnamefont {Wang}}, \bibinfo {author} {\bibfnamefont
  {Y.-F.}\ \bibnamefont {Zhao}}, \bibinfo {author} {\bibfnamefont
  {C.}~\bibnamefont {Liu}}, \bibinfo {author} {\bibfnamefont {W.}~\bibnamefont
  {Wu}}, \bibinfo {author} {\bibfnamefont {M.~H.~W.}\ \bibnamefont {Chan}},
  \bibinfo {author} {\bibfnamefont {N.}~\bibnamefont {Samarth}},\ and\ \bibinfo
  {author} {\bibfnamefont {C.-Z.}\ \bibnamefont {Chang}},\ }\bibfield  {title}
  {\bibinfo {title} {Realization of the {{Axion Insulator State}} in {{Quantum
  Anomalous Hall Sandwich Heterostructures}}},\ }\href
  {https://doi.org/10.1103/PhysRevLett.120.056801} {\bibfield  {journal}
  {\bibinfo  {journal} {Physical Review Letters}\ }\textbf {\bibinfo {volume}
  {120}},\ \bibinfo {pages} {056801} (\bibinfo {year} {2018})}\BibitemShut
  {NoStop}%
\bibitem [{\citenamefont {Liu}\ \emph {et~al.}(2020)\citenamefont {Liu},
  \citenamefont {Wang}, \citenamefont {Li}, \citenamefont {Wu}, \citenamefont
  {Li}, \citenamefont {Li}, \citenamefont {He}, \citenamefont {Xu},
  \citenamefont {Zhang},\ and\ \citenamefont {Wang}}]{liu2020robust}%
  \BibitemOpen
  \bibfield  {author} {\bibinfo {author} {\bibfnamefont {C.}~\bibnamefont
  {Liu}}, \bibinfo {author} {\bibfnamefont {Y.}~\bibnamefont {Wang}}, \bibinfo
  {author} {\bibfnamefont {H.}~\bibnamefont {Li}}, \bibinfo {author}
  {\bibfnamefont {Y.}~\bibnamefont {Wu}}, \bibinfo {author} {\bibfnamefont
  {Y.}~\bibnamefont {Li}}, \bibinfo {author} {\bibfnamefont {J.}~\bibnamefont
  {Li}}, \bibinfo {author} {\bibfnamefont {K.}~\bibnamefont {He}}, \bibinfo
  {author} {\bibfnamefont {Y.}~\bibnamefont {Xu}}, \bibinfo {author}
  {\bibfnamefont {J.}~\bibnamefont {Zhang}},\ and\ \bibinfo {author}
  {\bibfnamefont {Y.}~\bibnamefont {Wang}},\ }\bibfield  {title} {\bibinfo
  {title} {Robust axion insulator and {{Chern}} insulator phases in a
  two-dimensional antiferromagnetic topological insulator},\ }\bibfield
  {journal} {\bibinfo  {journal} {Nature Materials}\ }\href
  {https://doi.org/10.1038/s41563-019-0573-3} {10.1038/s41563-019-0573-3}
  (\bibinfo {year} {2020})\BibitemShut {NoStop}%
\bibitem [{\citenamefont {Jo}\ \emph {et~al.}(2020)\citenamefont {Jo},
  \citenamefont {Wang}, \citenamefont {Slager}, \citenamefont {Yan},
  \citenamefont {Wu}, \citenamefont {Lee}, \citenamefont {Schrunk},
  \citenamefont {Vishwanath},\ and\ \citenamefont
  {Kaminski}}]{jo2020intrinsic}%
  \BibitemOpen
  \bibfield  {author} {\bibinfo {author} {\bibfnamefont {N.~H.}\ \bibnamefont
  {Jo}}, \bibinfo {author} {\bibfnamefont {L.-L.}\ \bibnamefont {Wang}},
  \bibinfo {author} {\bibfnamefont {R.-J.}\ \bibnamefont {Slager}}, \bibinfo
  {author} {\bibfnamefont {J.}~\bibnamefont {Yan}}, \bibinfo {author}
  {\bibfnamefont {Y.}~\bibnamefont {Wu}}, \bibinfo {author} {\bibfnamefont
  {K.}~\bibnamefont {Lee}}, \bibinfo {author} {\bibfnamefont {B.}~\bibnamefont
  {Schrunk}}, \bibinfo {author} {\bibfnamefont {A.}~\bibnamefont
  {Vishwanath}},\ and\ \bibinfo {author} {\bibfnamefont {A.}~\bibnamefont
  {Kaminski}},\ }\bibfield  {title} {\bibinfo {title} {Intrinsic axion
  insulating behavior in antiferromagnetic {{MnBi}} 6 {{Te}} 10},\ }\href@noop
  {} {\bibfield  {journal} {\bibinfo  {journal} {Physical Review B}\ }\textbf
  {\bibinfo {volume} {102}},\ \bibinfo {pages} {045130} (\bibinfo {year}
  {2020})}\BibitemShut {NoStop}%
\bibitem [{\citenamefont {Gao}\ \emph {et~al.}(2021)\citenamefont {Gao},
  \citenamefont {Liu}, \citenamefont {Hu}, \citenamefont {Qiu}, \citenamefont
  {Tzschaschel}, \citenamefont {Ghosh}, \citenamefont {Ho}, \citenamefont
  {B{\'e}rub{\'e}}, \citenamefont {Chen}, \citenamefont {Sun}, \citenamefont
  {Zhang}, \citenamefont {Zhang}, \citenamefont {Wang}, \citenamefont {Wang},
  \citenamefont {Huang}, \citenamefont {Felser}, \citenamefont {Agarwal},
  \citenamefont {Ding}, \citenamefont {Tien}, \citenamefont {Akey},
  \citenamefont {Gardener}, \citenamefont {Singh}, \citenamefont {Watanabe},
  \citenamefont {Taniguchi}, \citenamefont {Burch}, \citenamefont {Bell},
  \citenamefont {Zhou}, \citenamefont {Gao}, \citenamefont {Lu}, \citenamefont
  {Bansil}, \citenamefont {Lin}, \citenamefont {Chang}, \citenamefont {Fu},
  \citenamefont {Ma}, \citenamefont {Ni},\ and\ \citenamefont
  {Xu}}]{gao2021layer}%
  \BibitemOpen
  \bibfield  {author} {\bibinfo {author} {\bibfnamefont {A.}~\bibnamefont
  {Gao}}, \bibinfo {author} {\bibfnamefont {Y.-F.}\ \bibnamefont {Liu}},
  \bibinfo {author} {\bibfnamefont {C.}~\bibnamefont {Hu}}, \bibinfo {author}
  {\bibfnamefont {J.-X.}\ \bibnamefont {Qiu}}, \bibinfo {author} {\bibfnamefont
  {C.}~\bibnamefont {Tzschaschel}}, \bibinfo {author} {\bibfnamefont
  {B.}~\bibnamefont {Ghosh}}, \bibinfo {author} {\bibfnamefont {S.-C.}\
  \bibnamefont {Ho}}, \bibinfo {author} {\bibfnamefont {D.}~\bibnamefont
  {B{\'e}rub{\'e}}}, \bibinfo {author} {\bibfnamefont {R.}~\bibnamefont
  {Chen}}, \bibinfo {author} {\bibfnamefont {H.}~\bibnamefont {Sun}}, \bibinfo
  {author} {\bibfnamefont {Z.}~\bibnamefont {Zhang}}, \bibinfo {author}
  {\bibfnamefont {X.-Y.}\ \bibnamefont {Zhang}}, \bibinfo {author}
  {\bibfnamefont {Y.-X.}\ \bibnamefont {Wang}}, \bibinfo {author}
  {\bibfnamefont {N.}~\bibnamefont {Wang}}, \bibinfo {author} {\bibfnamefont
  {Z.}~\bibnamefont {Huang}}, \bibinfo {author} {\bibfnamefont
  {C.}~\bibnamefont {Felser}}, \bibinfo {author} {\bibfnamefont
  {A.}~\bibnamefont {Agarwal}}, \bibinfo {author} {\bibfnamefont
  {T.}~\bibnamefont {Ding}}, \bibinfo {author} {\bibfnamefont {H.-J.}\
  \bibnamefont {Tien}}, \bibinfo {author} {\bibfnamefont {A.}~\bibnamefont
  {Akey}}, \bibinfo {author} {\bibfnamefont {J.}~\bibnamefont {Gardener}},
  \bibinfo {author} {\bibfnamefont {B.}~\bibnamefont {Singh}}, \bibinfo
  {author} {\bibfnamefont {K.}~\bibnamefont {Watanabe}}, \bibinfo {author}
  {\bibfnamefont {T.}~\bibnamefont {Taniguchi}}, \bibinfo {author}
  {\bibfnamefont {K.~S.}\ \bibnamefont {Burch}}, \bibinfo {author}
  {\bibfnamefont {D.~C.}\ \bibnamefont {Bell}}, \bibinfo {author}
  {\bibfnamefont {B.~B.}\ \bibnamefont {Zhou}}, \bibinfo {author}
  {\bibfnamefont {W.}~\bibnamefont {Gao}}, \bibinfo {author} {\bibfnamefont
  {H.-Z.}\ \bibnamefont {Lu}}, \bibinfo {author} {\bibfnamefont
  {A.}~\bibnamefont {Bansil}}, \bibinfo {author} {\bibfnamefont
  {H.}~\bibnamefont {Lin}}, \bibinfo {author} {\bibfnamefont {T.-R.}\
  \bibnamefont {Chang}}, \bibinfo {author} {\bibfnamefont {L.}~\bibnamefont
  {Fu}}, \bibinfo {author} {\bibfnamefont {Q.}~\bibnamefont {Ma}}, \bibinfo
  {author} {\bibfnamefont {N.}~\bibnamefont {Ni}},\ and\ \bibinfo {author}
  {\bibfnamefont {S.-Y.}\ \bibnamefont {Xu}},\ }\bibfield  {title} {\bibinfo
  {title} {Layer {{Hall}} effect in a {{2D}} topological axion
  antiferromagnet},\ }\href {https://doi.org/10.1038/s41586-021-03679-w}
  {\bibfield  {journal} {\bibinfo  {journal} {Nature}\ }\textbf {\bibinfo
  {volume} {595}},\ \bibinfo {pages} {521} (\bibinfo {year}
  {2021})}\BibitemShut {NoStop}%
\bibitem [{\citenamefont {Cerjan}\ \emph {et~al.}(2020)\citenamefont {Cerjan},
  \citenamefont {J{\"u}rgensen}, \citenamefont {Benalcazar}, \citenamefont
  {Mukherjee},\ and\ \citenamefont {Rechtsman}}]{cerjan2020observation}%
  \BibitemOpen
  \bibfield  {author} {\bibinfo {author} {\bibfnamefont {A.}~\bibnamefont
  {Cerjan}}, \bibinfo {author} {\bibfnamefont {M.}~\bibnamefont
  {J{\"u}rgensen}}, \bibinfo {author} {\bibfnamefont {W.~A.}\ \bibnamefont
  {Benalcazar}}, \bibinfo {author} {\bibfnamefont {S.}~\bibnamefont
  {Mukherjee}},\ and\ \bibinfo {author} {\bibfnamefont {M.~C.}\ \bibnamefont
  {Rechtsman}},\ }\bibfield  {title} {\bibinfo {title} {Observation of a
  higher-order topological bound state in the continuum},\ }\href@noop {}
  {\bibfield  {journal} {\bibinfo  {journal} {Physical review letters}\
  }\textbf {\bibinfo {volume} {125}},\ \bibinfo {pages} {213901} (\bibinfo
  {year} {2020})}\BibitemShut {NoStop}%
\bibitem [{\citenamefont {Grinberg}\ \emph {et~al.}(2020)\citenamefont
  {Grinberg}, \citenamefont {Lin}, \citenamefont {Harris}, \citenamefont
  {Benalcazar}, \citenamefont {Peterson}, \citenamefont {Hughes},\ and\
  \citenamefont {Bahl}}]{grinberg2020robusta}%
  \BibitemOpen
  \bibfield  {author} {\bibinfo {author} {\bibfnamefont {I.~H.}\ \bibnamefont
  {Grinberg}}, \bibinfo {author} {\bibfnamefont {M.}~\bibnamefont {Lin}},
  \bibinfo {author} {\bibfnamefont {C.}~\bibnamefont {Harris}}, \bibinfo
  {author} {\bibfnamefont {W.~A.}\ \bibnamefont {Benalcazar}}, \bibinfo
  {author} {\bibfnamefont {C.~W.}\ \bibnamefont {Peterson}}, \bibinfo {author}
  {\bibfnamefont {T.~L.}\ \bibnamefont {Hughes}},\ and\ \bibinfo {author}
  {\bibfnamefont {G.}~\bibnamefont {Bahl}},\ }\bibfield  {title} {\bibinfo
  {title} {Robust temporal pumping in a magneto-mechanical topological
  insulator},\ }\href@noop {} {\bibfield  {journal} {\bibinfo  {journal}
  {Nature communications}\ }\textbf {\bibinfo {volume} {11}},\ \bibinfo {pages}
  {1} (\bibinfo {year} {2020})}\BibitemShut {NoStop}%
\bibitem [{\citenamefont {Peterson}\ \emph {et~al.}(2020)\citenamefont
  {Peterson}, \citenamefont {Li}, \citenamefont {Benalcazar}, \citenamefont
  {Hughes},\ and\ \citenamefont {Bahl}}]{peterson2020fractional}%
  \BibitemOpen
  \bibfield  {author} {\bibinfo {author} {\bibfnamefont {C.~W.}\ \bibnamefont
  {Peterson}}, \bibinfo {author} {\bibfnamefont {T.}~\bibnamefont {Li}},
  \bibinfo {author} {\bibfnamefont {W.~A.}\ \bibnamefont {Benalcazar}},
  \bibinfo {author} {\bibfnamefont {T.~L.}\ \bibnamefont {Hughes}},\ and\
  \bibinfo {author} {\bibfnamefont {G.}~\bibnamefont {Bahl}},\ }\bibfield
  {title} {\bibinfo {title} {A fractional corner anomaly reveals higher-order
  topology},\ }\href@noop {} {\bibfield  {journal} {\bibinfo  {journal}
  {Science (New York, N.Y.)}\ }\textbf {\bibinfo {volume} {368}},\ \bibinfo
  {pages} {1114} (\bibinfo {year} {2020})}\BibitemShut {NoStop}%
\bibitem [{\citenamefont {Xie}\ \emph {et~al.}(2018)\citenamefont {Xie},
  \citenamefont {Wang}, \citenamefont {Wang}, \citenamefont {Zhu},
  \citenamefont {Jiang}, \citenamefont {Lu},\ and\ \citenamefont
  {Chen}}]{xie2018secondorder}%
  \BibitemOpen
  \bibfield  {author} {\bibinfo {author} {\bibfnamefont {B.-Y.}\ \bibnamefont
  {Xie}}, \bibinfo {author} {\bibfnamefont {H.-F.}\ \bibnamefont {Wang}},
  \bibinfo {author} {\bibfnamefont {H.-X.}\ \bibnamefont {Wang}}, \bibinfo
  {author} {\bibfnamefont {X.-Y.}\ \bibnamefont {Zhu}}, \bibinfo {author}
  {\bibfnamefont {J.-H.}\ \bibnamefont {Jiang}}, \bibinfo {author}
  {\bibfnamefont {M.-H.}\ \bibnamefont {Lu}},\ and\ \bibinfo {author}
  {\bibfnamefont {Y.-F.}\ \bibnamefont {Chen}},\ }\bibfield  {title} {\bibinfo
  {title} {Second-order photonic topological insulator with corner states},\
  }\href@noop {} {\bibfield  {journal} {\bibinfo  {journal} {Physical Review
  B}\ }\textbf {\bibinfo {volume} {98}},\ \bibinfo {pages} {205147} (\bibinfo
  {year} {2018})}\BibitemShut {NoStop}%
\bibitem [{\citenamefont {Ota}\ \emph {et~al.}(2019)\citenamefont {Ota},
  \citenamefont {Liu}, \citenamefont {Katsumi}, \citenamefont {Watanabe},
  \citenamefont {Wakabayashi}, \citenamefont {Arakawa},\ and\ \citenamefont
  {Iwamoto}}]{ota2019photonic}%
  \BibitemOpen
  \bibfield  {author} {\bibinfo {author} {\bibfnamefont {Y.}~\bibnamefont
  {Ota}}, \bibinfo {author} {\bibfnamefont {F.}~\bibnamefont {Liu}}, \bibinfo
  {author} {\bibfnamefont {R.}~\bibnamefont {Katsumi}}, \bibinfo {author}
  {\bibfnamefont {K.}~\bibnamefont {Watanabe}}, \bibinfo {author}
  {\bibfnamefont {K.}~\bibnamefont {Wakabayashi}}, \bibinfo {author}
  {\bibfnamefont {Y.}~\bibnamefont {Arakawa}},\ and\ \bibinfo {author}
  {\bibfnamefont {S.}~\bibnamefont {Iwamoto}},\ }\bibfield  {title} {\bibinfo
  {title} {Photonic crystal nanocavity based on a topological corner state},\
  }\href@noop {} {\bibfield  {journal} {\bibinfo  {journal} {Optica}\ }\textbf
  {\bibinfo {volume} {6}},\ \bibinfo {pages} {786} (\bibinfo {year}
  {2019})}\BibitemShut {NoStop}%
\bibitem [{\citenamefont {El~Hassan}\ \emph {et~al.}(2019)\citenamefont
  {El~Hassan}, \citenamefont {Kunst}, \citenamefont {Moritz}, \citenamefont
  {Andler}, \citenamefont {Bergholtz},\ and\ \citenamefont
  {Bourennane}}]{elhassan2019corner}%
  \BibitemOpen
  \bibfield  {author} {\bibinfo {author} {\bibfnamefont {A.}~\bibnamefont
  {El~Hassan}}, \bibinfo {author} {\bibfnamefont {F.~K.}\ \bibnamefont
  {Kunst}}, \bibinfo {author} {\bibfnamefont {A.}~\bibnamefont {Moritz}},
  \bibinfo {author} {\bibfnamefont {G.}~\bibnamefont {Andler}}, \bibinfo
  {author} {\bibfnamefont {E.~J.}\ \bibnamefont {Bergholtz}},\ and\ \bibinfo
  {author} {\bibfnamefont {M.}~\bibnamefont {Bourennane}},\ }\bibfield  {title}
  {\bibinfo {title} {Corner states of light in photonic waveguides},\
  }\href@noop {} {\bibfield  {journal} {\bibinfo  {journal} {Nature Photonics}\
  }\textbf {\bibinfo {volume} {13}},\ \bibinfo {pages} {697} (\bibinfo {year}
  {2019})}\BibitemShut {NoStop}%
\bibitem [{\citenamefont {Chen}\ \emph {et~al.}(2019)\citenamefont {Chen},
  \citenamefont {Deng}, \citenamefont {Shi}, \citenamefont {Zhao},
  \citenamefont {Chen},\ and\ \citenamefont {Dong}}]{chen2019direct}%
  \BibitemOpen
  \bibfield  {author} {\bibinfo {author} {\bibfnamefont {X.-D.}\ \bibnamefont
  {Chen}}, \bibinfo {author} {\bibfnamefont {W.-M.}\ \bibnamefont {Deng}},
  \bibinfo {author} {\bibfnamefont {F.-L.}\ \bibnamefont {Shi}}, \bibinfo
  {author} {\bibfnamefont {F.-L.}\ \bibnamefont {Zhao}}, \bibinfo {author}
  {\bibfnamefont {M.}~\bibnamefont {Chen}},\ and\ \bibinfo {author}
  {\bibfnamefont {J.-W.}\ \bibnamefont {Dong}},\ }\bibfield  {title} {\bibinfo
  {title} {Direct observation of corner states in second-order topological
  photonic crystal slabs},\ }\href@noop {} {\bibfield  {journal} {\bibinfo
  {journal} {Physical Review Letters}\ }\textbf {\bibinfo {volume} {122}},\
  \bibinfo {pages} {233902} (\bibinfo {year} {2019})}\BibitemShut {NoStop}%
\bibitem [{\citenamefont {Zhang}\ \emph {et~al.}(2020)\citenamefont {Zhang},
  \citenamefont {Yang}, \citenamefont {Lin}, \citenamefont {Qin}, \citenamefont
  {Chen}, \citenamefont {Gao}, \citenamefont {Li}, \citenamefont {Jiang},
  \citenamefont {Zhang},\ and\ \citenamefont {Chen}}]{zhang2020higherorder}%
  \BibitemOpen
  \bibfield  {author} {\bibinfo {author} {\bibfnamefont {L.}~\bibnamefont
  {Zhang}}, \bibinfo {author} {\bibfnamefont {Y.}~\bibnamefont {Yang}},
  \bibinfo {author} {\bibfnamefont {Z.-K.}\ \bibnamefont {Lin}}, \bibinfo
  {author} {\bibfnamefont {P.}~\bibnamefont {Qin}}, \bibinfo {author}
  {\bibfnamefont {Q.}~\bibnamefont {Chen}}, \bibinfo {author} {\bibfnamefont
  {F.}~\bibnamefont {Gao}}, \bibinfo {author} {\bibfnamefont {E.}~\bibnamefont
  {Li}}, \bibinfo {author} {\bibfnamefont {J.-H.}\ \bibnamefont {Jiang}},
  \bibinfo {author} {\bibfnamefont {B.}~\bibnamefont {Zhang}},\ and\ \bibinfo
  {author} {\bibfnamefont {H.}~\bibnamefont {Chen}},\ }\bibfield  {title}
  {\bibinfo {title} {Higher-order topological states in surface-wave photonic
  crystals},\ }\href@noop {} {\bibfield  {journal} {\bibinfo  {journal}
  {Advanced Science}\ }\textbf {\bibinfo {volume} {7}},\ \bibinfo {pages}
  {1902724} (\bibinfo {year} {2020})}\BibitemShut {NoStop}%
\bibitem [{\citenamefont {Xie}\ \emph {et~al.}(2019)\citenamefont {Xie},
  \citenamefont {Su}, \citenamefont {Wang}, \citenamefont {Su}, \citenamefont
  {Shen}, \citenamefont {Zhan}, \citenamefont {Lu}, \citenamefont {Wang},\ and\
  \citenamefont {Chen}}]{xie2019visualization}%
  \BibitemOpen
  \bibfield  {author} {\bibinfo {author} {\bibfnamefont {B.-Y.}\ \bibnamefont
  {Xie}}, \bibinfo {author} {\bibfnamefont {G.-X.}\ \bibnamefont {Su}},
  \bibinfo {author} {\bibfnamefont {H.-F.}\ \bibnamefont {Wang}}, \bibinfo
  {author} {\bibfnamefont {H.}~\bibnamefont {Su}}, \bibinfo {author}
  {\bibfnamefont {X.-P.}\ \bibnamefont {Shen}}, \bibinfo {author}
  {\bibfnamefont {P.}~\bibnamefont {Zhan}}, \bibinfo {author} {\bibfnamefont
  {M.-H.}\ \bibnamefont {Lu}}, \bibinfo {author} {\bibfnamefont {Z.-L.}\
  \bibnamefont {Wang}},\ and\ \bibinfo {author} {\bibfnamefont {Y.-F.}\
  \bibnamefont {Chen}},\ }\bibfield  {title} {\bibinfo {title} {Visualization
  of higher-order topological insulating phases in two-dimensional dielectric
  photonic crystals},\ }\href@noop {} {\bibfield  {journal} {\bibinfo
  {journal} {Physical review letters}\ }\textbf {\bibinfo {volume} {122}},\
  \bibinfo {pages} {233903} (\bibinfo {year} {2019})}\BibitemShut {NoStop}%
\bibitem [{\citenamefont {Wang}\ \emph {et~al.}(2021)\citenamefont {Wang},
  \citenamefont {Liang}, \citenamefont {Jiang}, \citenamefont {Hu},
  \citenamefont {Lu},\ and\ \citenamefont {Jiang}}]{wang2021higherorder}%
  \BibitemOpen
  \bibfield  {author} {\bibinfo {author} {\bibfnamefont {H.-X.}\ \bibnamefont
  {Wang}}, \bibinfo {author} {\bibfnamefont {L.}~\bibnamefont {Liang}},
  \bibinfo {author} {\bibfnamefont {B.}~\bibnamefont {Jiang}}, \bibinfo
  {author} {\bibfnamefont {J.}~\bibnamefont {Hu}}, \bibinfo {author}
  {\bibfnamefont {X.}~\bibnamefont {Lu}},\ and\ \bibinfo {author}
  {\bibfnamefont {J.-H.}\ \bibnamefont {Jiang}},\ }\bibfield  {title} {\bibinfo
  {title} {Higher-order topological phases in tunable {{C}} 3 symmetric
  photonic crystals},\ }\href@noop {} {\bibfield  {journal} {\bibinfo
  {journal} {Photonics Research}\ }\textbf {\bibinfo {volume} {9}},\ \bibinfo
  {pages} {1854} (\bibinfo {year} {2021})}\BibitemShut {NoStop}%
\bibitem [{\citenamefont {Li}\ \emph {et~al.}(2020{\natexlab{a}})\citenamefont
  {Li}, \citenamefont {Zhirihin}, \citenamefont {Gorlach}, \citenamefont {Ni},
  \citenamefont {Filonov}, \citenamefont {Slobozhanyuk}, \citenamefont
  {Al{\`u}},\ and\ \citenamefont {Khanikaev}}]{li2020higherorder}%
  \BibitemOpen
  \bibfield  {author} {\bibinfo {author} {\bibfnamefont {M.}~\bibnamefont
  {Li}}, \bibinfo {author} {\bibfnamefont {D.}~\bibnamefont {Zhirihin}},
  \bibinfo {author} {\bibfnamefont {M.}~\bibnamefont {Gorlach}}, \bibinfo
  {author} {\bibfnamefont {X.}~\bibnamefont {Ni}}, \bibinfo {author}
  {\bibfnamefont {D.}~\bibnamefont {Filonov}}, \bibinfo {author} {\bibfnamefont
  {A.}~\bibnamefont {Slobozhanyuk}}, \bibinfo {author} {\bibfnamefont
  {A.}~\bibnamefont {Al{\`u}}},\ and\ \bibinfo {author} {\bibfnamefont {A.~B.}\
  \bibnamefont {Khanikaev}},\ }\bibfield  {title} {\bibinfo {title}
  {Higher-order topological states in photonic kagome crystals with long-range
  interactions},\ }\href@noop {} {\bibfield  {journal} {\bibinfo  {journal}
  {Nature Photonics}\ }\textbf {\bibinfo {volume} {14}},\ \bibinfo {pages} {89}
  (\bibinfo {year} {2020}{\natexlab{a}})}\BibitemShut {NoStop}%
\bibitem [{\citenamefont {Kim}\ \emph {et~al.}(2020)\citenamefont {Kim},
  \citenamefont {Jacob},\ and\ \citenamefont {Rho}}]{kim2020recent}%
  \BibitemOpen
  \bibfield  {author} {\bibinfo {author} {\bibfnamefont {M.}~\bibnamefont
  {Kim}}, \bibinfo {author} {\bibfnamefont {Z.}~\bibnamefont {Jacob}},\ and\
  \bibinfo {author} {\bibfnamefont {J.}~\bibnamefont {Rho}},\ }\bibfield
  {title} {\bibinfo {title} {Recent advances in {{2D}}, {{3D}} and higher-order
  topological photonics},\ }\href@noop {} {\bibfield  {journal} {\bibinfo
  {journal} {Light: Science \& Applications}\ }\textbf {\bibinfo {volume}
  {9}},\ \bibinfo {pages} {1} (\bibinfo {year} {2020})}\BibitemShut {NoStop}%
\bibitem [{\citenamefont {Proctor}\ \emph {et~al.}(2020)\citenamefont
  {Proctor}, \citenamefont {Huidobro}, \citenamefont {{radlyn}}, \citenamefont
  {{de Paz}}, \citenamefont {Vergniory}, \citenamefont {Bercioux},\ and\
  \citenamefont {{Garc{\'i}a-Etxarri}}}]{proctor2020robustness}%
  \BibitemOpen
  \bibfield  {author} {\bibinfo {author} {\bibfnamefont {M.}~\bibnamefont
  {Proctor}}, \bibinfo {author} {\bibfnamefont {P.~A.}\ \bibnamefont
  {Huidobro}}, \bibinfo {author} {\bibfnamefont {B.}~\bibnamefont {{radlyn}}},
  \bibinfo {author} {\bibfnamefont {M.~B.}\ \bibnamefont {{de Paz}}}, \bibinfo
  {author} {\bibfnamefont {M.~G.}\ \bibnamefont {Vergniory}}, \bibinfo {author}
  {\bibfnamefont {D.}~\bibnamefont {Bercioux}},\ and\ \bibinfo {author}
  {\bibfnamefont {A.}~\bibnamefont {{Garc{\'i}a-Etxarri}}},\ }\bibfield
  {title} {\bibinfo {title} {Robustness of topological corner modes in photonic
  crystals},\ }\href@noop {} {\bibfield  {journal} {\bibinfo  {journal}
  {Physical Review Research}\ }\textbf {\bibinfo {volume} {2}},\ \bibinfo
  {pages} {042038} (\bibinfo {year} {2020})}\BibitemShut {NoStop}%
\bibitem [{\citenamefont {Po}\ \emph {et~al.}(2018)\citenamefont {Po},
  \citenamefont {Watanabe},\ and\ \citenamefont {Vishwanath}}]{Po2017}%
  \BibitemOpen
  \bibfield  {author} {\bibinfo {author} {\bibfnamefont {H.~C.}\ \bibnamefont
  {Po}}, \bibinfo {author} {\bibfnamefont {H.}~\bibnamefont {Watanabe}},\ and\
  \bibinfo {author} {\bibfnamefont {A.}~\bibnamefont {Vishwanath}},\ }\bibfield
   {title} {\bibinfo {title} {Fragile topology and wannier obstructions},\
  }\href {https://doi.org/10.1103/PhysRevLett.121.126402} {\bibfield  {journal}
  {\bibinfo  {journal} {Physical Review Letters}\ }\textbf {\bibinfo {volume}
  {121}},\ \bibinfo {pages} {126402} (\bibinfo {year} {2018})}\BibitemShut
  {NoStop}%
\bibitem [{\citenamefont {Wieder}\ and\ \citenamefont
  {Bernevig}(2018)}]{wieder2018axion}%
  \BibitemOpen
  \bibfield  {author} {\bibinfo {author} {\bibfnamefont {B.~J.}\ \bibnamefont
  {Wieder}}\ and\ \bibinfo {author} {\bibfnamefont {B.~A.}\ \bibnamefont
  {Bernevig}},\ }\bibfield  {title} {\bibinfo {title} {The {{Axion Insulator}}
  as a {{Pump}} of {{Fragile Topology}}},\ }\href@noop {} {\bibfield  {journal}
  {\bibinfo  {journal} {arXiv preprint arXiv:1810.02373}\ } (\bibinfo {year}
  {2018})},\ \Eprint {https://arxiv.org/abs/1810.02373} {arXiv:1810.02373}
  \BibitemShut {NoStop}%
\bibitem [{\citenamefont {Wieder}\ \emph {et~al.}(2020)\citenamefont {Wieder},
  \citenamefont {Wang}, \citenamefont {Cano}, \citenamefont {Dai},
  \citenamefont {Schoop}, \citenamefont {Bradlyn},\ and\ \citenamefont
  {Bernevig}}]{wieder2020strong}%
  \BibitemOpen
  \bibfield  {author} {\bibinfo {author} {\bibfnamefont {B.~J.}\ \bibnamefont
  {Wieder}}, \bibinfo {author} {\bibfnamefont {Z.}~\bibnamefont {Wang}},
  \bibinfo {author} {\bibfnamefont {J.}~\bibnamefont {Cano}}, \bibinfo {author}
  {\bibfnamefont {X.}~\bibnamefont {Dai}}, \bibinfo {author} {\bibfnamefont
  {L.~M.}\ \bibnamefont {Schoop}}, \bibinfo {author} {\bibfnamefont
  {B.}~\bibnamefont {Bradlyn}},\ and\ \bibinfo {author} {\bibfnamefont {B.~A.}\
  \bibnamefont {Bernevig}},\ }\bibfield  {title} {\bibinfo {title} {Strong and
  fragile topological dirac semimetals with higher-order fermi arcs},\
  }\href@noop {} {\bibfield  {journal} {\bibinfo  {journal} {Nature
  communications}\ }\textbf {\bibinfo {volume} {11}},\ \bibinfo {pages} {1}
  (\bibinfo {year} {2020})}\BibitemShut {NoStop}%
\bibitem [{\citenamefont {Fang}\ and\ \citenamefont
  {Cano}(2021)}]{fang2021filling}%
  \BibitemOpen
  \bibfield  {author} {\bibinfo {author} {\bibfnamefont {Y.}~\bibnamefont
  {Fang}}\ and\ \bibinfo {author} {\bibfnamefont {J.}~\bibnamefont {Cano}},\
  }\bibfield  {title} {\bibinfo {title} {Filling anomaly for general two-and
  three-dimensional c 4 symmetric lattices},\ }\href@noop {} {\bibfield
  {journal} {\bibinfo  {journal} {Physical Review B}\ }\textbf {\bibinfo
  {volume} {103}},\ \bibinfo {pages} {165109} (\bibinfo {year}
  {2021})}\BibitemShut {NoStop}%
\bibitem [{\citenamefont {Benalcazar}\ \emph {et~al.}(2019)\citenamefont
  {Benalcazar}, \citenamefont {Li},\ and\ \citenamefont
  {Hughes}}]{benalcazar2019quantization}%
  \BibitemOpen
  \bibfield  {author} {\bibinfo {author} {\bibfnamefont {W.~A.}\ \bibnamefont
  {Benalcazar}}, \bibinfo {author} {\bibfnamefont {T.}~\bibnamefont {Li}},\
  and\ \bibinfo {author} {\bibfnamefont {T.~L.}\ \bibnamefont {Hughes}},\
  }\bibfield  {title} {\bibinfo {title} {Quantization of fractional corner
  charge in c n-symmetric higher-order topological crystalline insulators},\
  }\href@noop {} {\bibfield  {journal} {\bibinfo  {journal} {Physical Review
  B}\ }\textbf {\bibinfo {volume} {99}},\ \bibinfo {pages} {245151} (\bibinfo
  {year} {2019})}\BibitemShut {NoStop}%
\bibitem [{\citenamefont {Song}\ \emph {et~al.}(2017)\citenamefont {Song},
  \citenamefont {Fang},\ and\ \citenamefont {Fang}}]{song2017ensuremath2}%
  \BibitemOpen
  \bibfield  {author} {\bibinfo {author} {\bibfnamefont {Z.}~\bibnamefont
  {Song}}, \bibinfo {author} {\bibfnamefont {Z.}~\bibnamefont {Fang}},\ and\
  \bibinfo {author} {\bibfnamefont {C.}~\bibnamefont {Fang}},\ }\bibfield
  {title} {\bibinfo {title} {\$({{D}}\textbackslash
  ensuremath-2)\$-{{Dimensional Edge States}} of {{Rotation Symmetry Protected
  Topological States}}},\ }\href
  {https://doi.org/10.1103/PhysRevLett.119.246402} {\bibfield  {journal}
  {\bibinfo  {journal} {Physical Review Letters}\ }\textbf {\bibinfo {volume}
  {119}},\ \bibinfo {pages} {246402} (\bibinfo {year} {2017})}\BibitemShut
  {NoStop}%
\bibitem [{\citenamefont {Hwang}\ \emph {et~al.}(2019)\citenamefont {Hwang},
  \citenamefont {Ahn},\ and\ \citenamefont {Yang}}]{hwang2019fragile}%
  \BibitemOpen
  \bibfield  {author} {\bibinfo {author} {\bibfnamefont {Y.}~\bibnamefont
  {Hwang}}, \bibinfo {author} {\bibfnamefont {J.}~\bibnamefont {Ahn}},\ and\
  \bibinfo {author} {\bibfnamefont {B.-J.}\ \bibnamefont {Yang}},\ }\bibfield
  {title} {\bibinfo {title} {Fragile topology protected by inversion symmetry:
  Diagnosis, bulk-boundary correspondence, and wilson loop},\ }\href@noop {}
  {\bibfield  {journal} {\bibinfo  {journal} {Physical Review B}\ }\textbf
  {\bibinfo {volume} {100}},\ \bibinfo {pages} {205126} (\bibinfo {year}
  {2019})}\BibitemShut {NoStop}%
\bibitem [{\citenamefont {Bouhon}\ \emph {et~al.}(2019)\citenamefont {Bouhon},
  \citenamefont {{Black-Schaffer}},\ and\ \citenamefont
  {Slager}}]{bouhon2019wilson}%
  \BibitemOpen
  \bibfield  {author} {\bibinfo {author} {\bibfnamefont {A.}~\bibnamefont
  {Bouhon}}, \bibinfo {author} {\bibfnamefont {A.~M.}\ \bibnamefont
  {{Black-Schaffer}}},\ and\ \bibinfo {author} {\bibfnamefont {R.-J.}\
  \bibnamefont {Slager}},\ }\bibfield  {title} {\bibinfo {title} {Wilson loop
  approach to fragile topology of split elementary band representations and
  topological crystalline insulators with time-reversal symmetry},\ }\href@noop
  {} {\bibfield  {journal} {\bibinfo  {journal} {Physical Review B}\ }\textbf
  {\bibinfo {volume} {100}},\ \bibinfo {pages} {195135} (\bibinfo {year}
  {2019})}\BibitemShut {NoStop}%
\bibitem [{\citenamefont {Cano}\ \emph {et~al.}(2018)\citenamefont {Cano},
  \citenamefont {Bradlyn}, \citenamefont {Wang}, \citenamefont {Elcoro},
  \citenamefont {Vergniory}, \citenamefont {Felser}, \citenamefont {Aroyo},\
  and\ \citenamefont {Bernevig}}]{cano2018topology}%
  \BibitemOpen
  \bibfield  {author} {\bibinfo {author} {\bibfnamefont {J.}~\bibnamefont
  {Cano}}, \bibinfo {author} {\bibfnamefont {B.}~\bibnamefont {Bradlyn}},
  \bibinfo {author} {\bibfnamefont {Z.}~\bibnamefont {Wang}}, \bibinfo {author}
  {\bibfnamefont {L.}~\bibnamefont {Elcoro}}, \bibinfo {author} {\bibfnamefont
  {M.~G.}\ \bibnamefont {Vergniory}}, \bibinfo {author} {\bibfnamefont
  {C.}~\bibnamefont {Felser}}, \bibinfo {author} {\bibfnamefont {M.~I.}\
  \bibnamefont {Aroyo}},\ and\ \bibinfo {author} {\bibfnamefont {B.~A.}\
  \bibnamefont {Bernevig}},\ }\bibfield  {title} {\bibinfo {title} {Topology of
  disconnected elementary band representations},\ }\href@noop {} {\bibfield
  {journal} {\bibinfo  {journal} {Physical Review Letters}\ }\textbf {\bibinfo
  {volume} {120}},\ \bibinfo {pages} {266401} (\bibinfo {year}
  {2018})}\BibitemShut {NoStop}%
\bibitem [{\citenamefont {Bradlyn}\ \emph {et~al.}(2019)\citenamefont
  {Bradlyn}, \citenamefont {Wang}, \citenamefont {Cano},\ and\ \citenamefont
  {Bernevig}}]{bradlyn2019disconnected}%
  \BibitemOpen
  \bibfield  {author} {\bibinfo {author} {\bibfnamefont {B.}~\bibnamefont
  {Bradlyn}}, \bibinfo {author} {\bibfnamefont {Z.}~\bibnamefont {Wang}},
  \bibinfo {author} {\bibfnamefont {J.}~\bibnamefont {Cano}},\ and\ \bibinfo
  {author} {\bibfnamefont {B.~A.}\ \bibnamefont {Bernevig}},\ }\bibfield
  {title} {\bibinfo {title} {Disconnected elementary band representations,
  fragile topology, and {{Wilson}} loops as topological indices: {{An}} example
  on the triangular lattice},\ }\href@noop {} {\bibfield  {journal} {\bibinfo
  {journal} {Physical Review B}\ }\textbf {\bibinfo {volume} {99}},\ \bibinfo
  {pages} {045140} (\bibinfo {year} {2019})}\BibitemShut {NoStop}%
\bibitem [{\citenamefont {Liu}\ \emph {et~al.}(2019)\citenamefont {Liu},
  \citenamefont {Vishwanath},\ and\ \citenamefont {Khalaf}}]{liu2019shift}%
  \BibitemOpen
  \bibfield  {author} {\bibinfo {author} {\bibfnamefont {S.}~\bibnamefont
  {Liu}}, \bibinfo {author} {\bibfnamefont {A.}~\bibnamefont {Vishwanath}},\
  and\ \bibinfo {author} {\bibfnamefont {E.}~\bibnamefont {Khalaf}},\
  }\bibfield  {title} {\bibinfo {title} {Shift insulators:
  {{Rotation-protected}} two-dimensional topological crystalline insulators},\
  }\href@noop {} {\bibfield  {journal} {\bibinfo  {journal} {Physical Review
  X}\ }\textbf {\bibinfo {volume} {9}},\ \bibinfo {pages} {031003} (\bibinfo
  {year} {2019})}\BibitemShut {NoStop}%
\bibitem [{\citenamefont {Manjunath}\ and\ \citenamefont
  {Barkeshli}(2021)}]{manjunath2021crystalline}%
  \BibitemOpen
  \bibfield  {author} {\bibinfo {author} {\bibfnamefont {N.}~\bibnamefont
  {Manjunath}}\ and\ \bibinfo {author} {\bibfnamefont {M.}~\bibnamefont
  {Barkeshli}},\ }\bibfield  {title} {\bibinfo {title} {Crystalline gauge
  fields and quantized discrete geometric response for abelian topological
  phases with lattice symmetry},\ }\href@noop {} {\bibfield  {journal}
  {\bibinfo  {journal} {Physical Review Research}\ }\textbf {\bibinfo {volume}
  {3}},\ \bibinfo {pages} {013040} (\bibinfo {year} {2021})}\BibitemShut
  {NoStop}%
\bibitem [{\citenamefont {Li}\ \emph {et~al.}(2020{\natexlab{b}})\citenamefont
  {Li}, \citenamefont {Zhu}, \citenamefont {Benalcazar},\ and\ \citenamefont
  {Hughes}}]{li2020fractional}%
  \BibitemOpen
  \bibfield  {author} {\bibinfo {author} {\bibfnamefont {T.}~\bibnamefont
  {Li}}, \bibinfo {author} {\bibfnamefont {P.}~\bibnamefont {Zhu}}, \bibinfo
  {author} {\bibfnamefont {W.~A.}\ \bibnamefont {Benalcazar}},\ and\ \bibinfo
  {author} {\bibfnamefont {T.~L.}\ \bibnamefont {Hughes}},\ }\bibfield  {title}
  {\bibinfo {title} {Fractional disclination charge in two-dimensional c
  n-symmetric topological crystalline insulators},\ }\href@noop {} {\bibfield
  {journal} {\bibinfo  {journal} {Physical Review B}\ }\textbf {\bibinfo
  {volume} {101}},\ \bibinfo {pages} {115115} (\bibinfo {year}
  {2020}{\natexlab{b}})}\BibitemShut {NoStop}%
\bibitem [{\citenamefont {{May-Mann}}\ and\ \citenamefont
  {Hughes}(2022)}]{may2021crystalline}%
  \BibitemOpen
  \bibfield  {author} {\bibinfo {author} {\bibfnamefont {J.}~\bibnamefont
  {{May-Mann}}}\ and\ \bibinfo {author} {\bibfnamefont {T.~L.}\ \bibnamefont
  {Hughes}},\ }\bibfield  {title} {\bibinfo {title} {Crystalline responses for
  rotation-invariant higher-order topological insulators},\ }\href@noop {}
  {\bibfield  {journal} {\bibinfo  {journal} {Physical Review B}\ }\textbf
  {\bibinfo {volume} {106}},\ \bibinfo {pages} {L241113} (\bibinfo {year}
  {2022})}\BibitemShut {NoStop}%
\bibitem [{\citenamefont {Schindler}\ \emph {et~al.}(2022)\citenamefont
  {Schindler}, \citenamefont {Tsirkin}, \citenamefont {Neupert}, \citenamefont
  {Andrei~Bernevig},\ and\ \citenamefont {Wieder}}]{schindler2022topological}%
  \BibitemOpen
  \bibfield  {author} {\bibinfo {author} {\bibfnamefont {F.}~\bibnamefont
  {Schindler}}, \bibinfo {author} {\bibfnamefont {S.~S.}\ \bibnamefont
  {Tsirkin}}, \bibinfo {author} {\bibfnamefont {T.}~\bibnamefont {Neupert}},
  \bibinfo {author} {\bibfnamefont {B.}~\bibnamefont {Andrei~Bernevig}},\ and\
  \bibinfo {author} {\bibfnamefont {B.~J.}\ \bibnamefont {Wieder}},\ }\bibfield
   {title} {\bibinfo {title} {Topological zero-dimensional defect and flux
  states in three-dimensional insulators},\ }\href@noop {} {\bibfield
  {journal} {\bibinfo  {journal} {Nature communications}\ }\textbf {\bibinfo
  {volume} {13}},\ \bibinfo {pages} {5791} (\bibinfo {year}
  {2022})}\BibitemShut {NoStop}%
\bibitem [{\citenamefont {Geier}\ \emph {et~al.}(2021)\citenamefont {Geier},
  \citenamefont {Fulga},\ and\ \citenamefont {Lau}}]{geier2021bulk}%
  \BibitemOpen
  \bibfield  {author} {\bibinfo {author} {\bibfnamefont {M.}~\bibnamefont
  {Geier}}, \bibinfo {author} {\bibfnamefont {I.~C.}\ \bibnamefont {Fulga}},\
  and\ \bibinfo {author} {\bibfnamefont {A.}~\bibnamefont {Lau}},\ }\bibfield
  {title} {\bibinfo {title} {Bulk-boundary-defect correspondence at
  disclinations in rotation-symmetric topological insulators and
  superconductors},\ }\href@noop {} {\bibfield  {journal} {\bibinfo  {journal}
  {SciPost Physics}\ }\textbf {\bibinfo {volume} {10}},\ \bibinfo {pages} {092}
  (\bibinfo {year} {2021})}\BibitemShut {NoStop}%
\bibitem [{\citenamefont {Zhang}\ \emph {et~al.}(2022)\citenamefont {Zhang},
  \citenamefont {Manjunath}, \citenamefont {Nambiar},\ and\ \citenamefont
  {Barkeshli}}]{zhang2022fractional}%
  \BibitemOpen
  \bibfield  {author} {\bibinfo {author} {\bibfnamefont {Y.}~\bibnamefont
  {Zhang}}, \bibinfo {author} {\bibfnamefont {N.}~\bibnamefont {Manjunath}},
  \bibinfo {author} {\bibfnamefont {G.}~\bibnamefont {Nambiar}},\ and\ \bibinfo
  {author} {\bibfnamefont {M.}~\bibnamefont {Barkeshli}},\ }\bibfield  {title}
  {\bibinfo {title} {Fractional disclination charge and discrete shift in the
  {{Hofstadter}} butterfly},\ }\href@noop {} {\bibfield  {journal} {\bibinfo
  {journal} {Physical Review Letters}\ }\textbf {\bibinfo {volume} {129}},\
  \bibinfo {pages} {275301} (\bibinfo {year} {2022})}\BibitemShut {NoStop}%
\bibitem [{\citenamefont {Wen}\ and\ \citenamefont {Zee}(1992)}]{wen1992shift}%
  \BibitemOpen
  \bibfield  {author} {\bibinfo {author} {\bibfnamefont {X.~G.}\ \bibnamefont
  {Wen}}\ and\ \bibinfo {author} {\bibfnamefont {A.}~\bibnamefont {Zee}},\
  }\bibfield  {title} {\bibinfo {title} {Shift and spin vector: New topological
  quantum numbers for the hall fluids},\ }\href@noop {} {\bibfield  {journal}
  {\bibinfo  {journal} {Physical review letters}\ }\textbf {\bibinfo {volume}
  {69}},\ \bibinfo {pages} {953} (\bibinfo {year} {1992})}\BibitemShut
  {NoStop}%
\bibitem [{\citenamefont {Read}\ and\ \citenamefont
  {Rezayi}(2011)}]{read2011hall}%
  \BibitemOpen
  \bibfield  {author} {\bibinfo {author} {\bibfnamefont {N.}~\bibnamefont
  {Read}}\ and\ \bibinfo {author} {\bibfnamefont {E.}~\bibnamefont {Rezayi}},\
  }\bibfield  {title} {\bibinfo {title} {Hall viscosity, orbital spin, and
  geometry: paired superfluids and quantum hall systems},\ }\href@noop {}
  {\bibfield  {journal} {\bibinfo  {journal} {Physical Review B}\ }\textbf
  {\bibinfo {volume} {84}},\ \bibinfo {pages} {085316} (\bibinfo {year}
  {2011})}\BibitemShut {NoStop}%
\bibitem [{\citenamefont {Bradlyn}\ and\ \citenamefont
  {Read}(2015{\natexlab{a}})}]{bradlyn2014low}%
  \BibitemOpen
  \bibfield  {author} {\bibinfo {author} {\bibfnamefont {B.}~\bibnamefont
  {Bradlyn}}\ and\ \bibinfo {author} {\bibfnamefont {N.}~\bibnamefont {Read}},\
  }\bibfield  {title} {\bibinfo {title} {Low-energy effective theory in the
  bulk for transport in a topological phase},\ }\href@noop {} {\bibfield
  {journal} {\bibinfo  {journal} {Phys. Rev. B}\ }\textbf {\bibinfo {volume}
  {91}},\ \bibinfo {pages} {125303} (\bibinfo {year}
  {2015}{\natexlab{a}})}\BibitemShut {NoStop}%
\bibitem [{\citenamefont {Abanov}\ and\ \citenamefont
  {Gromov}(2014)}]{Gromov20141}%
  \BibitemOpen
  \bibfield  {author} {\bibinfo {author} {\bibfnamefont {A.}~\bibnamefont
  {Abanov}}\ and\ \bibinfo {author} {\bibfnamefont {A.}~\bibnamefont
  {Gromov}},\ }\href {https://doi.org/10.1103/PhysRevB.90.014435} {\bibfield
  {journal} {\bibinfo  {journal} {Phys. Rev. B}\ }\textbf {\bibinfo {volume}
  {90}},\ \bibinfo {pages} {014435} (\bibinfo {year} {2014})}\BibitemShut
  {NoStop}%
\bibitem [{\citenamefont {Gromov}\ and\ \citenamefont
  {Abanov}(2014)}]{Abanov2014}%
  \BibitemOpen
  \bibfield  {author} {\bibinfo {author} {\bibfnamefont {A.}~\bibnamefont
  {Gromov}}\ and\ \bibinfo {author} {\bibfnamefont {A.}~\bibnamefont
  {Abanov}},\ }\href {https://doi.org/10.1103/PhysRevLett.113.266802}
  {\bibfield  {journal} {\bibinfo  {journal} {Phys. Rev. Lett.}\ }\textbf
  {\bibinfo {volume} {113}},\ \bibinfo {pages} {266802} (\bibinfo {year}
  {2014})}\BibitemShut {NoStop}%
\bibitem [{\citenamefont {Bradlyn}\ and\ \citenamefont
  {Read}(2015{\natexlab{b}})}]{bradlyn2015gcs}%
  \BibitemOpen
  \bibfield  {author} {\bibinfo {author} {\bibfnamefont {B.}~\bibnamefont
  {Bradlyn}}\ and\ \bibinfo {author} {\bibfnamefont {N.}~\bibnamefont {Read}},\
  }\bibfield  {title} {\bibinfo {title} {Topological central charge from berry
  curvature: Gravitational anomalies in trial wave functions for topological
  phases},\ }\href {https://doi.org/10.1103/PhysRevB.91.165306} {\bibfield
  {journal} {\bibinfo  {journal} {Phys. Rev. B}\ }\textbf {\bibinfo {volume}
  {91}},\ \bibinfo {pages} {165306} (\bibinfo {year}
  {2015}{\natexlab{b}})}\BibitemShut {NoStop}%
\bibitem [{\citenamefont {{Herzog-Arbeitman}}\ \emph
  {et~al.}(2022)\citenamefont {{Herzog-Arbeitman}}, \citenamefont {Bernevig},\
  and\ \citenamefont {Song}}]{herzog2022interacting}%
  \BibitemOpen
  \bibfield  {author} {\bibinfo {author} {\bibfnamefont {J.}~\bibnamefont
  {{Herzog-Arbeitman}}}, \bibinfo {author} {\bibfnamefont {B.~A.}\ \bibnamefont
  {Bernevig}},\ and\ \bibinfo {author} {\bibfnamefont {Z.-D.}\ \bibnamefont
  {Song}},\ }\bibfield  {title} {\bibinfo {title} {Interacting topological
  quantum chemistry in {{2D}}: {{Many-body}} real space invariants},\
  }\href@noop {} {\bibfield  {journal} {\bibinfo  {journal} {arXiv preprint
  arXiv:2212.00030}\ } (\bibinfo {year} {2022})},\ \Eprint
  {https://arxiv.org/abs/2212.00030} {arXiv:2212.00030} \BibitemShut {NoStop}%
\bibitem [{\citenamefont {Gromov}\ \emph {et~al.}(2016)\citenamefont {Gromov},
  \citenamefont {Jensen},\ and\ \citenamefont {Abanov}}]{gromov2016boundary}%
  \BibitemOpen
  \bibfield  {author} {\bibinfo {author} {\bibfnamefont {A.}~\bibnamefont
  {Gromov}}, \bibinfo {author} {\bibfnamefont {K.}~\bibnamefont {Jensen}},\
  and\ \bibinfo {author} {\bibfnamefont {A.~G.}\ \bibnamefont {Abanov}},\
  }\bibfield  {title} {\bibinfo {title} {Boundary effective action for quantum
  hall states},\ }\href@noop {} {\bibfield  {journal} {\bibinfo  {journal}
  {Physical review letters}\ }\textbf {\bibinfo {volume} {116}},\ \bibinfo
  {pages} {126802} (\bibinfo {year} {2016})}\BibitemShut {NoStop}%
\bibitem [{\citenamefont {Callan~Jr}\ and\ \citenamefont
  {Harvey}(1985)}]{callan1985anomalies}%
  \BibitemOpen
  \bibfield  {author} {\bibinfo {author} {\bibfnamefont {C.~G.}\ \bibnamefont
  {Callan~Jr}}\ and\ \bibinfo {author} {\bibfnamefont {J.~A.}\ \bibnamefont
  {Harvey}},\ }\bibfield  {title} {\bibinfo {title} {Anomalies and fermion zero
  modes on strings and domain walls},\ }\href@noop {} {\bibfield  {journal}
  {\bibinfo  {journal} {Nuclear Physics B}\ }\textbf {\bibinfo {volume}
  {250}},\ \bibinfo {pages} {427} (\bibinfo {year} {1985})}\BibitemShut
  {NoStop}%
\bibitem [{\citenamefont {Bardeen}\ and\ \citenamefont
  {Zumino}(1984)}]{bardeen1984consistent}%
  \BibitemOpen
  \bibfield  {author} {\bibinfo {author} {\bibfnamefont {W.~A.}\ \bibnamefont
  {Bardeen}}\ and\ \bibinfo {author} {\bibfnamefont {B.}~\bibnamefont
  {Zumino}},\ }\bibfield  {title} {\bibinfo {title} {Consistent and covariant
  anomalies in gauge and gravitational theories},\ }\href@noop {} {\bibfield
  {journal} {\bibinfo  {journal} {Nuclear Physics B}\ }\textbf {\bibinfo
  {volume} {244}},\ \bibinfo {pages} {421} (\bibinfo {year}
  {1984})}\BibitemShut {NoStop}%
\bibitem [{\citenamefont {Bradley}\ and\ \citenamefont
  {Cracknell}(1972)}]{bradley1972mathematical}%
  \BibitemOpen
  \bibfield  {author} {\bibinfo {author} {\bibfnamefont {C.}~\bibnamefont
  {Bradley}}\ and\ \bibinfo {author} {\bibfnamefont {A.}~\bibnamefont
  {Cracknell}},\ }\href {https://books.google.es/books?id=OKXvAAAAMAAJ} {\emph
  {\bibinfo {title} {The mathematical theory of symmetry in solids:
  representation theory for point groups and space groups}}}\ (\bibinfo
  {publisher} {Clarendon Press},\ \bibinfo {year} {1972})\BibitemShut {NoStop}%
\bibitem [{\citenamefont {Gromov}\ and\ \citenamefont
  {Abanov}(2015)}]{gromov2015thermal}%
  \BibitemOpen
  \bibfield  {author} {\bibinfo {author} {\bibfnamefont {A.}~\bibnamefont
  {Gromov}}\ and\ \bibinfo {author} {\bibfnamefont {A.~G.}\ \bibnamefont
  {Abanov}},\ }\bibfield  {title} {\bibinfo {title} {Thermal {{Hall}} effect
  and geometry with torsion},\ }\href@noop {} {\bibfield  {journal} {\bibinfo
  {journal} {Physical review letters}\ }\textbf {\bibinfo {volume} {114}},\
  \bibinfo {pages} {016802} (\bibinfo {year} {2015})}\BibitemShut {NoStop}%
\bibitem [{\citenamefont {Son}(2013)}]{son2013newton}%
  \BibitemOpen
  \bibfield  {author} {\bibinfo {author} {\bibfnamefont {D.~T.}\ \bibnamefont
  {Son}},\ }\bibfield  {title} {\bibinfo {title} {Newton-cartan geometry and
  the quantum hall effect},\ }\href@noop {} {\bibfield  {journal} {\bibinfo
  {journal} {arXiv preprint arXiv:1306.0638}\ } (\bibinfo {year}
  {2013})}\BibitemShut {NoStop}%
\bibitem [{\citenamefont {Bradlyn}\ \emph {et~al.}(2012)\citenamefont
  {Bradlyn}, \citenamefont {Goldstein},\ and\ \citenamefont
  {Read}}]{bradlyn2012kubo}%
  \BibitemOpen
  \bibfield  {author} {\bibinfo {author} {\bibfnamefont {B.}~\bibnamefont
  {Bradlyn}}, \bibinfo {author} {\bibfnamefont {M.}~\bibnamefont {Goldstein}},\
  and\ \bibinfo {author} {\bibfnamefont {N.}~\bibnamefont {Read}},\ }\bibfield
  {title} {\bibinfo {title} {Kubo formulas for viscosity: Hall viscosity, ward
  identities, and the relation with conductivity},\ }\href@noop {} {\bibfield
  {journal} {\bibinfo  {journal} {Physical Review B}\ }\textbf {\bibinfo
  {volume} {86}},\ \bibinfo {pages} {245309} (\bibinfo {year}
  {2012})}\BibitemShut {NoStop}%
\bibitem [{\citenamefont {Hoyos}(2014)}]{hoyos2014hall}%
  \BibitemOpen
  \bibfield  {author} {\bibinfo {author} {\bibfnamefont {C.}~\bibnamefont
  {Hoyos}},\ }\bibfield  {title} {\bibinfo {title} {Hall viscosity, topological
  states and effective theories},\ }\href@noop {} {\bibfield  {journal}
  {\bibinfo  {journal} {International Journal of Modern Physics B}\ }\textbf
  {\bibinfo {volume} {28}},\ \bibinfo {pages} {1430007} (\bibinfo {year}
  {2014})}\BibitemShut {NoStop}%
\bibitem [{\citenamefont {Hoyos}\ and\ \citenamefont {Son}(2012)}]{Hoyos2012}%
  \BibitemOpen
  \bibfield  {author} {\bibinfo {author} {\bibfnamefont {C.}~\bibnamefont
  {Hoyos}}\ and\ \bibinfo {author} {\bibfnamefont {D.~T.}\ \bibnamefont
  {Son}},\ }\href {https://doi.org/10.1103/PhysRevLett.108.066805} {\bibfield
  {journal} {\bibinfo  {journal} {Phys. Rev. Lett.}\ }\textbf {\bibinfo
  {volume} {108}},\ \bibinfo {pages} {066805} (\bibinfo {year}
  {2012})}\BibitemShut {NoStop}%
\bibitem [{\citenamefont {Rao}\ and\ \citenamefont
  {Bradlyn}(2021)}]{rao2021resolving}%
  \BibitemOpen
  \bibfield  {author} {\bibinfo {author} {\bibfnamefont {P.}~\bibnamefont
  {Rao}}\ and\ \bibinfo {author} {\bibfnamefont {B.}~\bibnamefont {Bradlyn}},\
  }\bibfield  {title} {\bibinfo {title} {Resolving hall and dissipative
  viscosity ambiguities via boundary effects},\ }\href@noop {} {\bibfield
  {journal} {\bibinfo  {journal} {arXiv preprint arXiv:2112.04545}\ } (\bibinfo
  {year} {2021})}\BibitemShut {NoStop}%
\bibitem [{\citenamefont {Abanov}\ and\ \citenamefont
  {Monteiro}(2019)}]{abanov2019free}%
  \BibitemOpen
  \bibfield  {author} {\bibinfo {author} {\bibfnamefont {A.~G.}\ \bibnamefont
  {Abanov}}\ and\ \bibinfo {author} {\bibfnamefont {G.~M.}\ \bibnamefont
  {Monteiro}},\ }\bibfield  {title} {\bibinfo {title} {Free-surface variational
  principle for an incompressible fluid with odd viscosity},\ }\href@noop {}
  {\bibfield  {journal} {\bibinfo  {journal} {Physical review letters}\
  }\textbf {\bibinfo {volume} {122}},\ \bibinfo {pages} {154501} (\bibinfo
  {year} {2019})}\BibitemShut {NoStop}%
\bibitem [{\citenamefont {Ganeshan}\ and\ \citenamefont
  {Abanov}(2017)}]{ganeshan2017odd}%
  \BibitemOpen
  \bibfield  {author} {\bibinfo {author} {\bibfnamefont {S.}~\bibnamefont
  {Ganeshan}}\ and\ \bibinfo {author} {\bibfnamefont {A.~G.}\ \bibnamefont
  {Abanov}},\ }\bibfield  {title} {\bibinfo {title} {Odd viscosity in
  two-dimensional incompressible fluids},\ }\href@noop {} {\bibfield  {journal}
  {\bibinfo  {journal} {Physical review fluids}\ }\textbf {\bibinfo {volume}
  {2}},\ \bibinfo {pages} {094101} (\bibinfo {year} {2017})}\BibitemShut
  {NoStop}%
\bibitem [{\citenamefont {Eguchi}\ \emph {et~al.}(1980)\citenamefont {Eguchi},
  \citenamefont {Gilkey},\ and\ \citenamefont
  {Hanson}}]{eguchi1980gravitation}%
  \BibitemOpen
  \bibfield  {author} {\bibinfo {author} {\bibfnamefont {T.}~\bibnamefont
  {Eguchi}}, \bibinfo {author} {\bibfnamefont {P.~B.}\ \bibnamefont {Gilkey}},\
  and\ \bibinfo {author} {\bibfnamefont {A.~J.}\ \bibnamefont {Hanson}},\
  }\bibfield  {title} {\bibinfo {title} {Gravitation, gauge theories and
  differential geometry},\ }\href@noop {} {\bibfield  {journal} {\bibinfo
  {journal} {Physics reports}\ }\textbf {\bibinfo {volume} {66}},\ \bibinfo
  {pages} {213} (\bibinfo {year} {1980})}\BibitemShut {NoStop}%
\bibitem [{\citenamefont {Fradkin}(2013)}]{fradkin2013field}%
  \BibitemOpen
  \bibfield  {author} {\bibinfo {author} {\bibfnamefont {E.}~\bibnamefont
  {Fradkin}},\ }\href@noop {} {\emph {\bibinfo {title} {Field theories of
  condensed matter physics}}}\ (\bibinfo  {publisher} {Cambridge University
  Press, Cambridge},\ \bibinfo {year} {2013})\BibitemShut {NoStop}%
\bibitem [{\citenamefont {Von~Delft}\ and\ \citenamefont
  {Schoeller}(1998)}]{von1998bosonization}%
  \BibitemOpen
  \bibfield  {author} {\bibinfo {author} {\bibfnamefont {J.}~\bibnamefont
  {Von~Delft}}\ and\ \bibinfo {author} {\bibfnamefont {H.}~\bibnamefont
  {Schoeller}},\ }\bibfield  {title} {\bibinfo {title} {Bosonization for
  beginners—refermionization for experts},\ }\href@noop {} {\bibfield
  {journal} {\bibinfo  {journal} {Annalen der Physik}\ }\textbf {\bibinfo
  {volume} {7}},\ \bibinfo {pages} {225} (\bibinfo {year} {1998})}\BibitemShut
  {NoStop}%
\bibitem [{\citenamefont {Kane}\ and\ \citenamefont
  {Fisher}(1992{\natexlab{a}})}]{kane1992transport}%
  \BibitemOpen
  \bibfield  {author} {\bibinfo {author} {\bibfnamefont {C.}~\bibnamefont
  {Kane}}\ and\ \bibinfo {author} {\bibfnamefont {M.~P.}\ \bibnamefont
  {Fisher}},\ }\bibfield  {title} {\bibinfo {title} {Transport in a one-channel
  luttinger liquid},\ }\href@noop {} {\bibfield  {journal} {\bibinfo  {journal}
  {Physical review letters}\ }\textbf {\bibinfo {volume} {68}},\ \bibinfo
  {pages} {1220} (\bibinfo {year} {1992}{\natexlab{a}})}\BibitemShut {NoStop}%
\bibitem [{\citenamefont {Kane}\ and\ \citenamefont
  {Fisher}(1992{\natexlab{b}})}]{kane1992transmission}%
  \BibitemOpen
  \bibfield  {author} {\bibinfo {author} {\bibfnamefont {C.}~\bibnamefont
  {Kane}}\ and\ \bibinfo {author} {\bibfnamefont {M.~P.}\ \bibnamefont
  {Fisher}},\ }\bibfield  {title} {\bibinfo {title} {Transmission through
  barriers and resonant tunneling in an interacting one-dimensional electron
  gas},\ }\href@noop {} {\bibfield  {journal} {\bibinfo  {journal} {Physical
  Review B}\ }\textbf {\bibinfo {volume} {46}},\ \bibinfo {pages} {15233}
  (\bibinfo {year} {1992}{\natexlab{b}})}\BibitemShut {NoStop}%
\bibitem [{\citenamefont {Goldstone}\ and\ \citenamefont
  {Wilczek}(1981)}]{goldstone1981fractional}%
  \BibitemOpen
  \bibfield  {author} {\bibinfo {author} {\bibfnamefont {J.}~\bibnamefont
  {Goldstone}}\ and\ \bibinfo {author} {\bibfnamefont {F.}~\bibnamefont
  {Wilczek}},\ }\bibfield  {title} {\bibinfo {title} {Fractional quantum
  numbers on solitons},\ }\href@noop {} {\bibfield  {journal} {\bibinfo
  {journal} {Physical Review Letters}\ }\textbf {\bibinfo {volume} {47}},\
  \bibinfo {pages} {986} (\bibinfo {year} {1981})}\BibitemShut {NoStop}%
\bibitem [{\citenamefont {S{\'e}n{\'e}chal}(2004)}]{senechal2004introduction}%
  \BibitemOpen
  \bibfield  {author} {\bibinfo {author} {\bibfnamefont {D.}~\bibnamefont
  {S{\'e}n{\'e}chal}},\ }\bibfield  {title} {\bibinfo {title} {An introduction
  to bosonization},\ }\href@noop {} {\bibfield  {journal} {\bibinfo  {journal}
  {Theoretical Methods for Strongly Correlated Electrons}\ ,\ \bibinfo {pages}
  {139}} (\bibinfo {year} {2004})}\BibitemShut {NoStop}%
\bibitem [{\citenamefont {Fujikawa}\ \emph {et~al.}(2004)\citenamefont
  {Fujikawa}, \citenamefont {Fujikawa}, \citenamefont {Suzuki} \emph
  {et~al.}}]{fujikawa2004path}%
  \BibitemOpen
  \bibfield  {author} {\bibinfo {author} {\bibfnamefont {K.}~\bibnamefont
  {Fujikawa}}, \bibinfo {author} {\bibfnamefont {K.}~\bibnamefont {Fujikawa}},
  \bibinfo {author} {\bibfnamefont {H.}~\bibnamefont {Suzuki}}, \emph
  {et~al.},\ }\href@noop {} {\emph {\bibinfo {title} {Path Integrals and
  Quantum Anomalies}}},\ \bibinfo {number} {122}\ (\bibinfo  {publisher}
  {{Oxford University Press on Demand}},\ \bibinfo {year} {2004})\BibitemShut
  {NoStop}%
\bibitem [{\citenamefont {{van Miert}}\ and\ \citenamefont
  {Ortix}(2020)}]{van2020topological}%
  \BibitemOpen
  \bibfield  {author} {\bibinfo {author} {\bibfnamefont {G.}~\bibnamefont {{van
  Miert}}}\ and\ \bibinfo {author} {\bibfnamefont {C.}~\bibnamefont {Ortix}},\
  }\bibfield  {title} {\bibinfo {title} {On the topological immunity of corner
  states in two-dimensional crystalline insulators},\ }\href@noop {} {\bibfield
   {journal} {\bibinfo  {journal} {npj Quantum Materials}\ }\textbf {\bibinfo
  {volume} {5}},\ \bibinfo {pages} {63} (\bibinfo {year} {2020})}\BibitemShut
  {NoStop}%
\bibitem [{\citenamefont {Kempkes}\ \emph {et~al.}(2019)\citenamefont
  {Kempkes}, \citenamefont {Slot}, \citenamefont {{van Den Broeke}},
  \citenamefont {Capiod}, \citenamefont {Benalcazar}, \citenamefont
  {Vanmaekelbergh}, \citenamefont {Bercioux}, \citenamefont {Swart},\ and\
  \citenamefont {Morais~Smith}}]{kempkes2019robust}%
  \BibitemOpen
  \bibfield  {author} {\bibinfo {author} {\bibfnamefont {S.}~\bibnamefont
  {Kempkes}}, \bibinfo {author} {\bibfnamefont {M.}~\bibnamefont {Slot}},
  \bibinfo {author} {\bibfnamefont {J.}~\bibnamefont {{van Den Broeke}}},
  \bibinfo {author} {\bibfnamefont {P.}~\bibnamefont {Capiod}}, \bibinfo
  {author} {\bibfnamefont {W.}~\bibnamefont {Benalcazar}}, \bibinfo {author}
  {\bibfnamefont {D.}~\bibnamefont {Vanmaekelbergh}}, \bibinfo {author}
  {\bibfnamefont {D.}~\bibnamefont {Bercioux}}, \bibinfo {author}
  {\bibfnamefont {I.}~\bibnamefont {Swart}},\ and\ \bibinfo {author}
  {\bibfnamefont {C.}~\bibnamefont {Morais~Smith}},\ }\bibfield  {title}
  {\bibinfo {title} {Robust zero-energy modes in an electronic higher-order
  topological insulator},\ }\href@noop {} {\bibfield  {journal} {\bibinfo
  {journal} {Nature Materials}\ }\textbf {\bibinfo {volume} {18}},\ \bibinfo
  {pages} {1292} (\bibinfo {year} {2019})}\BibitemShut {NoStop}%
\bibitem [{\citenamefont {Xue}\ \emph {et~al.}(2019)\citenamefont {Xue},
  \citenamefont {Yang}, \citenamefont {Gao}, \citenamefont {Chong},\ and\
  \citenamefont {Zhang}}]{xue2019acoustic}%
  \BibitemOpen
  \bibfield  {author} {\bibinfo {author} {\bibfnamefont {H.}~\bibnamefont
  {Xue}}, \bibinfo {author} {\bibfnamefont {Y.}~\bibnamefont {Yang}}, \bibinfo
  {author} {\bibfnamefont {F.}~\bibnamefont {Gao}}, \bibinfo {author}
  {\bibfnamefont {Y.}~\bibnamefont {Chong}},\ and\ \bibinfo {author}
  {\bibfnamefont {B.}~\bibnamefont {Zhang}},\ }\bibfield  {title} {\bibinfo
  {title} {Acoustic higher-order topological insulator on a kagome lattice},\
  }\href@noop {} {\bibfield  {journal} {\bibinfo  {journal} {Nature materials}\
  }\textbf {\bibinfo {volume} {18}},\ \bibinfo {pages} {108} (\bibinfo {year}
  {2019})}\BibitemShut {NoStop}%
\bibitem [{\citenamefont {Ni}\ \emph {et~al.}(2019)\citenamefont {Ni},
  \citenamefont {Weiner}, \citenamefont {Alu},\ and\ \citenamefont
  {Khanikaev}}]{ni2019observation}%
  \BibitemOpen
  \bibfield  {author} {\bibinfo {author} {\bibfnamefont {X.}~\bibnamefont
  {Ni}}, \bibinfo {author} {\bibfnamefont {M.}~\bibnamefont {Weiner}}, \bibinfo
  {author} {\bibfnamefont {A.}~\bibnamefont {Alu}},\ and\ \bibinfo {author}
  {\bibfnamefont {A.~B.}\ \bibnamefont {Khanikaev}},\ }\bibfield  {title}
  {\bibinfo {title} {Observation of higher-order topological acoustic states
  protected by generalized chiral symmetry},\ }\href@noop {} {\bibfield
  {journal} {\bibinfo  {journal} {Nature materials}\ }\textbf {\bibinfo
  {volume} {18}},\ \bibinfo {pages} {113} (\bibinfo {year} {2019})}\BibitemShut
  {NoStop}%
\bibitem [{\citenamefont {Huang}\ \emph {et~al.}(2022)\citenamefont {Huang},
  \citenamefont {Hsieh},\ and\ \citenamefont {Yu}}]{huang2022effective}%
  \BibitemOpen
  \bibfield  {author} {\bibinfo {author} {\bibfnamefont {S.-J.}\ \bibnamefont
  {Huang}}, \bibinfo {author} {\bibfnamefont {C.-T.}\ \bibnamefont {Hsieh}},\
  and\ \bibinfo {author} {\bibfnamefont {J.}~\bibnamefont {Yu}},\ }\bibfield
  {title} {\bibinfo {title} {Effective field theories of topological
  crystalline insulators and topological crystals},\ }\href@noop {} {\bibfield
  {journal} {\bibinfo  {journal} {Physical Review B}\ }\textbf {\bibinfo
  {volume} {105}},\ \bibinfo {pages} {045112} (\bibinfo {year}
  {2022})}\BibitemShut {NoStop}%
\bibitem [{\citenamefont {Gioia}\ \emph {et~al.}(2021)\citenamefont {Gioia},
  \citenamefont {Wang},\ and\ \citenamefont {Burkov}}]{gioia2021unquantized}%
  \BibitemOpen
  \bibfield  {author} {\bibinfo {author} {\bibfnamefont {L.}~\bibnamefont
  {Gioia}}, \bibinfo {author} {\bibfnamefont {C.}~\bibnamefont {Wang}},\ and\
  \bibinfo {author} {\bibfnamefont {A.}~\bibnamefont {Burkov}},\ }\bibfield
  {title} {\bibinfo {title} {Unquantized anomalies in topological semimetals},\
  }\href@noop {} {\bibfield  {journal} {\bibinfo  {journal} {Physical Review
  Research}\ }\textbf {\bibinfo {volume} {3}},\ \bibinfo {pages} {043067}
  (\bibinfo {year} {2021})}\BibitemShut {NoStop}%
\bibitem [{\citenamefont {Else}\ \emph {et~al.}(2021)\citenamefont {Else},
  \citenamefont {Huang}, \citenamefont {Prem},\ and\ \citenamefont
  {Gromov}}]{else2021quantum}%
  \BibitemOpen
  \bibfield  {author} {\bibinfo {author} {\bibfnamefont {D.~V.}\ \bibnamefont
  {Else}}, \bibinfo {author} {\bibfnamefont {S.-J.}\ \bibnamefont {Huang}},
  \bibinfo {author} {\bibfnamefont {A.}~\bibnamefont {Prem}},\ and\ \bibinfo
  {author} {\bibfnamefont {A.}~\bibnamefont {Gromov}},\ }\bibfield  {title}
  {\bibinfo {title} {Quantum many-body topology of quasicrystals},\ }\href@noop
  {} {\bibfield  {journal} {\bibinfo  {journal} {Physical Review X}\ }\textbf
  {\bibinfo {volume} {11}},\ \bibinfo {pages} {041051} (\bibinfo {year}
  {2021})}\BibitemShut {NoStop}%
\bibitem [{\citenamefont {Else}(2021)}]{else2021topological}%
  \BibitemOpen
  \bibfield  {author} {\bibinfo {author} {\bibfnamefont {D.~V.}\ \bibnamefont
  {Else}},\ }\bibfield  {title} {\bibinfo {title} {Topological {{Goldstone}}
  phases of matter},\ }\href@noop {} {\bibfield  {journal} {\bibinfo  {journal}
  {Physical Review B}\ }\textbf {\bibinfo {volume} {104}},\ \bibinfo {pages}
  {115129} (\bibinfo {year} {2021})}\BibitemShut {NoStop}%
\bibitem [{\citenamefont {Dubinkin}\ \emph {et~al.}(2021)\citenamefont
  {Dubinkin}, \citenamefont {Burnell},\ and\ \citenamefont
  {Hughes}}]{dubinkin2021higher}%
  \BibitemOpen
  \bibfield  {author} {\bibinfo {author} {\bibfnamefont {O.}~\bibnamefont
  {Dubinkin}}, \bibinfo {author} {\bibfnamefont {F.}~\bibnamefont {Burnell}},\
  and\ \bibinfo {author} {\bibfnamefont {T.~L.}\ \bibnamefont {Hughes}},\
  }\bibfield  {title} {\bibinfo {title} {Higher rank chiral fermions in 3d weyl
  semimetals},\ }\href@noop {} {\bibfield  {journal} {\bibinfo  {journal}
  {arXiv preprint arXiv:2102.08959}\ } (\bibinfo {year} {2021})},\ \Eprint
  {https://arxiv.org/abs/2102.08959} {arXiv:2102.08959} \BibitemShut {NoStop}%
\bibitem [{\citenamefont {Han}\ \emph {et~al.}(2019)\citenamefont {Han},
  \citenamefont {Wang},\ and\ \citenamefont {Ye}}]{han2019generalized}%
  \BibitemOpen
  \bibfield  {author} {\bibinfo {author} {\bibfnamefont {B.}~\bibnamefont
  {Han}}, \bibinfo {author} {\bibfnamefont {H.}~\bibnamefont {Wang}},\ and\
  \bibinfo {author} {\bibfnamefont {P.}~\bibnamefont {Ye}},\ }\bibfield
  {title} {\bibinfo {title} {Generalized wen-zee terms},\ }\href@noop {}
  {\bibfield  {journal} {\bibinfo  {journal} {Physical Review B}\ }\textbf
  {\bibinfo {volume} {99}},\ \bibinfo {pages} {205120} (\bibinfo {year}
  {2019})}\BibitemShut {NoStop}%
\bibitem [{\citenamefont {Stone}(2012)}]{stone2012gravitational}%
  \BibitemOpen
  \bibfield  {author} {\bibinfo {author} {\bibfnamefont {M.}~\bibnamefont
  {Stone}},\ }\bibfield  {title} {\bibinfo {title} {Gravitational anomalies and
  thermal {{Hall}} effect in topological insulators},\ }\href@noop {}
  {\bibfield  {journal} {\bibinfo  {journal} {Physical Review B}\ }\textbf
  {\bibinfo {volume} {85}},\ \bibinfo {pages} {184503} (\bibinfo {year}
  {2012})}\BibitemShut {NoStop}%
\bibitem [{\citenamefont {Read}\ and\ \citenamefont
  {Green}(2000)}]{read2000paired}%
  \BibitemOpen
  \bibfield  {author} {\bibinfo {author} {\bibfnamefont {N.}~\bibnamefont
  {Read}}\ and\ \bibinfo {author} {\bibfnamefont {D.}~\bibnamefont {Green}},\
  }\bibfield  {title} {\bibinfo {title} {Paired states of fermions in two
  dimensions with breaking of parity and time-reversal symmetries and the
  fractional quantum hall effect},\ }\href@noop {} {\bibfield  {journal}
  {\bibinfo  {journal} {Physical Review B}\ }\textbf {\bibinfo {volume} {61}},\
  \bibinfo {pages} {10267} (\bibinfo {year} {2000})}\BibitemShut {NoStop}%
\bibitem [{\citenamefont {Cappelli}\ and\ \citenamefont
  {Randellini}(2016)}]{cappelli2016multipole}%
  \BibitemOpen
  \bibfield  {author} {\bibinfo {author} {\bibfnamefont {A.}~\bibnamefont
  {Cappelli}}\ and\ \bibinfo {author} {\bibfnamefont {E.}~\bibnamefont
  {Randellini}},\ }\bibfield  {title} {\bibinfo {title} {Multipole expansion in
  the quantum hall effect},\ }\href@noop {} {\bibfield  {journal} {\bibinfo
  {journal} {Journal of High Energy Physics}\ }\textbf {\bibinfo {volume}
  {2016}},\ \bibinfo {pages} {1} (\bibinfo {year} {2016})}\BibitemShut
  {NoStop}%
\bibitem [{\citenamefont {Levin}(2013)}]{levin2013protected}%
  \BibitemOpen
  \bibfield  {author} {\bibinfo {author} {\bibfnamefont {M.}~\bibnamefont
  {Levin}},\ }\bibfield  {title} {\bibinfo {title} {Protected edge modes
  without symmetry},\ }\href@noop {} {\bibfield  {journal} {\bibinfo  {journal}
  {Physical Review X}\ }\textbf {\bibinfo {volume} {3}},\ \bibinfo {pages}
  {021009} (\bibinfo {year} {2013})}\BibitemShut {NoStop}%
\bibitem [{\citenamefont {Lin}\ \emph {et~al.}(2022)\citenamefont {Lin},
  \citenamefont {Palumbo}, \citenamefont {Guo}, \citenamefont {Blackburn},
  \citenamefont {Shoemaker}, \citenamefont {Mahmood}, \citenamefont {Wang},
  \citenamefont {Fiete}, \citenamefont {Wieder},\ and\ \citenamefont
  {Bradlyn}}]{lin2022spin}%
  \BibitemOpen
  \bibfield  {author} {\bibinfo {author} {\bibfnamefont {K.-S.}\ \bibnamefont
  {Lin}}, \bibinfo {author} {\bibfnamefont {G.}~\bibnamefont {Palumbo}},
  \bibinfo {author} {\bibfnamefont {Z.}~\bibnamefont {Guo}}, \bibinfo {author}
  {\bibfnamefont {J.}~\bibnamefont {Blackburn}}, \bibinfo {author}
  {\bibfnamefont {D.~P.}\ \bibnamefont {Shoemaker}}, \bibinfo {author}
  {\bibfnamefont {F.}~\bibnamefont {Mahmood}}, \bibinfo {author} {\bibfnamefont
  {Z.}~\bibnamefont {Wang}}, \bibinfo {author} {\bibfnamefont {G.~A.}\
  \bibnamefont {Fiete}}, \bibinfo {author} {\bibfnamefont {B.~J.}\ \bibnamefont
  {Wieder}},\ and\ \bibinfo {author} {\bibfnamefont {B.}~\bibnamefont
  {Bradlyn}},\ }\bibfield  {title} {\bibinfo {title} {Spin-resolved topology
  and partial axion angles in three-dimensional insulators},\ }\href@noop {}
  {\bibfield  {journal} {\bibinfo  {journal} {arXiv preprint arXiv:2207.10099}\
  } (\bibinfo {year} {2022})},\ \Eprint {https://arxiv.org/abs/2207.10099}
  {arXiv:2207.10099} \BibitemShut {NoStop}%
\bibitem [{\citenamefont {Lange}\ \emph {et~al.}(2022)\citenamefont {Lange},
  \citenamefont {Bouhon},\ and\ \citenamefont {Slager}}]{lange2022projected}%
  \BibitemOpen
  \bibfield  {author} {\bibinfo {author} {\bibfnamefont {G.~F.}\ \bibnamefont
  {Lange}}, \bibinfo {author} {\bibfnamefont {A.}~\bibnamefont {Bouhon}},\ and\
  \bibinfo {author} {\bibfnamefont {R.-J.}\ \bibnamefont {Slager}},\ }\bibfield
   {title} {\bibinfo {title} {Projected spin texture as a bulk indicator of
  fragile topology},\ }\href@noop {} {\bibfield  {journal} {\bibinfo  {journal}
  {arXiv preprint arXiv:2211.05137}\ } (\bibinfo {year} {2022})},\ \Eprint
  {https://arxiv.org/abs/2211.05137} {arXiv:2211.05137} \BibitemShut {NoStop}%
\bibitem [{\citenamefont {Hwang}\ \emph {et~al.}(2022)\citenamefont {Hwang},
  \citenamefont {Qian}, \citenamefont {Kang}, \citenamefont {Lee},
  \citenamefont {Ryu}, \citenamefont {Choi},\ and\ \citenamefont
  {Yang}}]{hwang2022magnetic}%
  \BibitemOpen
  \bibfield  {author} {\bibinfo {author} {\bibfnamefont {Y.}~\bibnamefont
  {Hwang}}, \bibinfo {author} {\bibfnamefont {Y.}~\bibnamefont {Qian}},
  \bibinfo {author} {\bibfnamefont {J.}~\bibnamefont {Kang}}, \bibinfo {author}
  {\bibfnamefont {J.}~\bibnamefont {Lee}}, \bibinfo {author} {\bibfnamefont
  {D.}~\bibnamefont {Ryu}}, \bibinfo {author} {\bibfnamefont {H.~C.}\
  \bibnamefont {Choi}},\ and\ \bibinfo {author} {\bibfnamefont {B.-J.}\
  \bibnamefont {Yang}},\ }\bibfield  {title} {\bibinfo {title} {Magnetic
  wallpaper {{Dirac}} fermions and topological magnetic {{Dirac}} insulators},\
  }\href@noop {} {\bibfield  {journal} {\bibinfo  {journal} {arXiv preprint
  arXiv:2210.10740}\ } (\bibinfo {year} {2022})},\ \Eprint
  {https://arxiv.org/abs/2210.10740} {arXiv:2210.10740} \BibitemShut {NoStop}%
\bibitem [{\citenamefont {Xu}\ \emph {et~al.}(2021{\natexlab{a}})\citenamefont
  {Xu}, \citenamefont {Elcoro}, \citenamefont {Song}, \citenamefont
  {Vergniory}, \citenamefont {Felser}, \citenamefont {Parkin}, \citenamefont
  {Regnault}, \citenamefont {Ma{\~n}es},\ and\ \citenamefont
  {Bernevig}}]{xu2021filling}%
  \BibitemOpen
  \bibfield  {author} {\bibinfo {author} {\bibfnamefont {Y.}~\bibnamefont
  {Xu}}, \bibinfo {author} {\bibfnamefont {L.}~\bibnamefont {Elcoro}}, \bibinfo
  {author} {\bibfnamefont {Z.-D.}\ \bibnamefont {Song}}, \bibinfo {author}
  {\bibfnamefont {M.}~\bibnamefont {Vergniory}}, \bibinfo {author}
  {\bibfnamefont {C.}~\bibnamefont {Felser}}, \bibinfo {author} {\bibfnamefont
  {S.~S.}\ \bibnamefont {Parkin}}, \bibinfo {author} {\bibfnamefont
  {N.}~\bibnamefont {Regnault}}, \bibinfo {author} {\bibfnamefont {J.~L.}\
  \bibnamefont {Ma{\~n}es}},\ and\ \bibinfo {author} {\bibfnamefont {B.~A.}\
  \bibnamefont {Bernevig}},\ }\bibfield  {title} {\bibinfo {title}
  {Filling-enforced obstructed atomic insulators},\ }\href@noop {} {\bibfield
  {journal} {\bibinfo  {journal} {arXiv preprint arXiv:2106.10276}\ } (\bibinfo
  {year} {2021}{\natexlab{a}})},\ \Eprint {https://arxiv.org/abs/2106.10276}
  {arXiv:2106.10276} \BibitemShut {NoStop}%
\bibitem [{\citenamefont {Xu}\ \emph {et~al.}(2021{\natexlab{b}})\citenamefont
  {Xu}, \citenamefont {Elcoro}, \citenamefont {Li}, \citenamefont {Song},
  \citenamefont {Regnault}, \citenamefont {Yang}, \citenamefont {Sun},
  \citenamefont {Parkin}, \citenamefont {Felser},\ and\ \citenamefont
  {Bernevig}}]{xu2021three}%
  \BibitemOpen
  \bibfield  {author} {\bibinfo {author} {\bibfnamefont {Y.}~\bibnamefont
  {Xu}}, \bibinfo {author} {\bibfnamefont {L.}~\bibnamefont {Elcoro}}, \bibinfo
  {author} {\bibfnamefont {G.}~\bibnamefont {Li}}, \bibinfo {author}
  {\bibfnamefont {Z.-D.}\ \bibnamefont {Song}}, \bibinfo {author}
  {\bibfnamefont {N.}~\bibnamefont {Regnault}}, \bibinfo {author}
  {\bibfnamefont {Q.}~\bibnamefont {Yang}}, \bibinfo {author} {\bibfnamefont
  {Y.}~\bibnamefont {Sun}}, \bibinfo {author} {\bibfnamefont {S.}~\bibnamefont
  {Parkin}}, \bibinfo {author} {\bibfnamefont {C.}~\bibnamefont {Felser}},\
  and\ \bibinfo {author} {\bibfnamefont {B.~A.}\ \bibnamefont {Bernevig}},\
  }\bibfield  {title} {\bibinfo {title} {Three-dimensional real space
  invariants, obstructed atomic insulators and a new principle for active
  catalytic sites},\ }\href@noop {} {\bibfield  {journal} {\bibinfo  {journal}
  {arXiv preprint arXiv:2111.02433}\ } (\bibinfo {year}
  {2021}{\natexlab{b}})},\ \Eprint {https://arxiv.org/abs/2111.02433}
  {arXiv:2111.02433} \BibitemShut {NoStop}%
\bibitem [{\citenamefont {Nie}\ \emph {et~al.}(2021)\citenamefont {Nie},
  \citenamefont {Qian}, \citenamefont {Gao}, \citenamefont {Fang},
  \citenamefont {Weng},\ and\ \citenamefont {Wang}}]{nie2021application}%
  \BibitemOpen
  \bibfield  {author} {\bibinfo {author} {\bibfnamefont {S.}~\bibnamefont
  {Nie}}, \bibinfo {author} {\bibfnamefont {Y.}~\bibnamefont {Qian}}, \bibinfo
  {author} {\bibfnamefont {J.}~\bibnamefont {Gao}}, \bibinfo {author}
  {\bibfnamefont {Z.}~\bibnamefont {Fang}}, \bibinfo {author} {\bibfnamefont
  {H.}~\bibnamefont {Weng}},\ and\ \bibinfo {author} {\bibfnamefont
  {Z.}~\bibnamefont {Wang}},\ }\bibfield  {title} {\bibinfo {title}
  {Application of topological quantum chemistry in electrides},\ }\href@noop {}
  {\bibfield  {journal} {\bibinfo  {journal} {Physical Review B}\ }\textbf
  {\bibinfo {volume} {103}},\ \bibinfo {pages} {205133} (\bibinfo {year}
  {2021})}\BibitemShut {NoStop}%
\bibitem [{\citenamefont {{May-Mann}}\ \emph {et~al.}(2022)\citenamefont
  {{May-Mann}}, \citenamefont {Hirsbrunner}, \citenamefont {Cao},\ and\
  \citenamefont {Hughes}}]{may2022topological}%
  \BibitemOpen
  \bibfield  {author} {\bibinfo {author} {\bibfnamefont {J.}~\bibnamefont
  {{May-Mann}}}, \bibinfo {author} {\bibfnamefont {M.~R.}\ \bibnamefont
  {Hirsbrunner}}, \bibinfo {author} {\bibfnamefont {X.}~\bibnamefont {Cao}},\
  and\ \bibinfo {author} {\bibfnamefont {T.~L.}\ \bibnamefont {Hughes}},\
  }\bibfield  {title} {\bibinfo {title} {Topological field theories of
  three-dimensional rotation symmetric insulators: {{Coupling}} curvature and
  electromagnetism},\ }\href@noop {} {\bibfield  {journal} {\bibinfo  {journal}
  {arXiv preprint arXiv:2209.00026}\ } (\bibinfo {year} {2022})},\ \Eprint
  {https://arxiv.org/abs/2209.00026} {arXiv:2209.00026} \BibitemShut {NoStop}%
\end{thebibliography}%
\end{document}